\documentclass[10pt,a4paper]{article}
\usepackage{mymacro}

\newcommand{\bbQ}{\mathds{Q}}

\newcommand{\cro}{\times}
\newcommand{\uncr}{\mbox{$\shortparallel$}}

\newcommand{\SU}{\mathrm{SU}}
\newcommand{\COp}[1]{\CO^{(#1)}}

\newcommand{\pic}[3][1]{\raisebox{#2}{\includegraphics[scale=#1]{Graphics/Wick-#3.pdf}}}

\title{Bootstrapping mixed correlators in ${\boldsymbol{\mathcal{N}=4}}$  Super Yang-Mills}
\author[a]{Agnese Bissi,}
\author[a,b]{Andrea Manenti\lsp}
\author[b,c]{and Alessandro Vichi\lsp}
\affiliation[a]{Department of Physics and Astronomy, Uppsala University, Box 516, SE-751 20 Uppsala, Sweden}
\affiliation[b]{Institute of Physics, École Polytechnique Fédérale de Lausanne, CH-1015 Lausanne, Switzerland}
\affiliation[c]{Department of Physics \emph{E. Fermi}, Pisa University, Largo B. Pontecorvo 3, 56127 Pisa, Italy}

\emailAdd{agnese.bissi@physics.uu.se}
\emailAdd{andrea.manenti@epfl.ch}
\emailAdd{alessandro.vichi@unipi.it}

\preprint{UUITP-46/20}

\abstract{We perform a numerical bootstrap study of the mixed correlator system containing the half-BPS operators of dimension two and three in $\mathcal N = 4$ Super Yang-Mills. This setup improves on previous works in the literature that only considered single correlators of one or the other operator. We obtain upper bounds on the leading twist in a given representation of the R-symmetry by imposing gaps on the twist of all operators rather than the dimension of a single one. As a result we find a tension between the large $N$ supergravity predictions and the numerical finite $N$ results already at $N\sim 100$.  A few possible solutions are discussed: the extremal spectrum suggests that at large but finite $N$, in addition to the double trace operators, there exists a second tower of states with smaller dimension.
We also obtain new bounds on the dimension of operators which were not accessible with a single correlator setup.
Finally we consider bounds on the OPE coefficients of various operators.
The results obtained for the OPE coefficient of the lightest scalar singlet show evidences of a two dimensional conformal manifold.
}

\begin{document}
\maketitle

\newpage
\section{Summary of results}

Super Yang-Mills with $\mathrm{SU}(N)$ gauge group and $\mathcal N=4$ supersymmetry in four dimensions is perhaps one of the most studied Superconformal Field Theories (SCFTs) of the last decades. Despite much progress has been made and the theory is well understood in certain limits, such as large $N$, large 't~Hooft coupling $\lambda$ or small gauge coupling $g_{\mathrm{YM}}$, there is still much that we do not know about other regions of the parameter space. It is precisely in these setups, when the absence of perturbative expansion of any kind makes the theory intrinsically strongly coupled, that the conformal bootstrap can play a major role (see for instance \cite{Poland:2018epd} for a review). 

Numerical studies of 4d $\mathcal N=4$ SCFTs were initiated in \cite{Beem:2013qxa,Beem:2016wfs} which applied numerical bootstrap techniques to study the four-point function of the lowest scalar transforming in the $[0,2,0]$ of $\mathrm{SU}(4)_R$, and sitting in half-BPS supermultiplet of the stress tensor. These works found evidences that upper bounds on operator dimensions at infinite values of $N$ are saturated by the supergravity solution. Furthermore, the extrapolation of the bounds to infinite numerical power was consistent with the leading $1/N$ supergravity corrections.

At small values of $N$ instead the bounds did not seem to correspond to any known result. On the other hand, the allowed region in the space of the smallest spin-0,2,4 unprotected operator dimension assumed a peculiar cubic-like shape, leading to the conjecture that, for each given finite $N$, there exists a value of the gauge coupling that simultaneously maximizes these operator dimensions. It was also suggested that this value would correspond to one of the self-dual points.

In \cite{Alday:2014qfa} the numerical study was further extended to the correlation function of the first graviton Kaluza Klein (KK) state, namely a half-BPS operator transforming in the $[0,3,0]$ representation of the $\mathrm{SU}(4)_R$ R-symmetry. This analysis allowed to access new unprotected operators but it was not competitive to the previous one due to the more complicated numerical setup.

In this work we combine the two previous studies in a full mixed system of correlation functions involving the stress tensor multiplet $\COp2$ and the first graviton KK mode  $\COp3$. Our setup includes all four-point functions of the form $\langle2222\rangle$, $\langle3333\rangle$ and $\langle2233\rangle$. Compared to previous studies, this system has two advantages: \emph{i)} it allows to access even more unprotected operators through the Operator Product Expansion (OPE) $\COp2\times\COp3$ and \emph{ii)} it links the correlation functions in a nontrivial way due to the contribution of protected operators obtained by solving the Ward identities.\footnote{The contribution of protected operators is fixed also requiring the absence of higher spin currents, negative norm states and consistency of various channels. We discuss this in detail in Section~\ref{sec:multiplets}.}

We should also stress that the present work focuses on Super Yang-Mills with $\mathrm{SU}(N)$ gauge group rather than considering the most general $\mathcal N=4$ SCFT.  As we review in the next section, superconformal symmetry fixes certain three-point function coefficients among short and semi-short multiplets in terms of few quantities, such as the central charge and few others.\footnote{See equations (\ref{eq:color2}), (\ref{eq:color3}), and (\ref{eq:color23})} Whenever the underlying theory is a gauge theory, it is possible to express these constants in terms of the rank of the gauge group $N$. In this work we chose to focus on $\mathrm{SU}(N)$ gauge theory and thus substituted the exact dependence on $N$. It would be interesting to use the technology developed in \cite{Chester:2019ifh} to scan the parameter space of all possible values of these constants, similarly to what has been done in three dimensions \cite{Agmon:2019imm} and search for isolated allowed regions that might signal the presence of new $\mathcal N=4$ SCFTs with no Lagrangian description.

Let us review and discuss the main results obtained in this work. Additional details and all plots can be found in Section~\ref{sec:results}. The first part, Sections~\ref{sec:generalities}, \ref{sec:multiplets} and \ref{sec:crossing}, is instead dedicated to set up the formalism, while a few known analytical results are reviewed in Section~\ref{sec:knownanalytic}. 

As a consistency check we reproduced some of the bounds first obtained in \cite{Beem:2016wfs, Alday:2014qfa} and improved on some of those by pushing the numerics. We also present bounds on the unprotected sector that was not accessible by the previous studies, namely operators transforming in the $[0,1,0]$ representation of $\mathrm{SU}(4)_R$. All such bounds on operator dimensions present a rather common structure: at large values of $N$, they are a monotonically increasing function of $N$;\footnote{This is true for $N\geq 3$ which is the smallest values consistent with the existence of $\COp3$.} moreover they asymptote to the dimension of the lightest double trace of the two external operators that are merged in a given OPE channel. In the majority of the cases this behavior is expected and consistent with our understanding of the large $N$ limit of the theory. The AdS/CFT correspondence predicts that the large 't Hooft coupling, large $N$ limit the theory is dual to a weakly coupled supergravity theory on $AdS_5\times S^5$. One can then compute the dimension of the lightest double trace operator appearing in a given OPE. These are reviewed in Section~\ref{sec:knownanalytic} and at zeroth order in $1/N$ they have the form
\eqn{
[\COp2 \COp2] \, : \,\Delta \sim 4\,, \qquad [\COp2 \COp3] \,:\,  \Delta \sim 5\,, \qquad [\COp3 \COp3] \,:\,  \Delta \sim 6\,.\nonumber
}[]
These values seem to drive the behavior observed in Figure~\ref{fig:BisectionSingletMixed}. On the other hand the supergravity dual predicts that the double trace $[\COp2 \COp2]$ appears in the OPE $\COp3 \times \COp3$ as well, with a $O(1/N^2)$ suppressed OPE coefficient. This operator is only absent in the strict planar limit, but should instead be present at any other finite value of $N$. The numerical bounds obtained by bootstrapping the  $\langle3333\rangle$ correlation function alone are instead not strong enough to impose the presence of an operator below the $[\COp3 \COp3]$ double trace. This lack of constraining power is a  common feature of many numerical bootstrap analyses where the external operator has dimension considerably away from the unitarity bound.

Next we considered a different approach. Instead of imposing gaps on single operators, we assumed a twist gap for a given sector, where the twist of an operator $\CO$ is defined as $\tau_\CO=\Delta_\CO -\ell_\CO$.  More specifically, for all the operators in a given representation $r$ of $\mathrm{SU}(4)_R$, we imposed
\eqn{
\tau_\CO \geq \tau_\text{gap} \quad \text{for any} \quad \CO \in r \,.
}[]
With this assumption, all the operators in a given sector are lifted together. The main motivation to this analysis comes from recent works on conformal Regge theory and lightcone bootstrap 
\cite{
Cornalba:2007fs,
Komargodski:2012ek,Fitzpatrick:2012yx,
Kaviraj:2015cxa,Alday:2015eya,
Simmons-Duffin:2016wlq,Alday:2016mxe,Alday:2016jfr,Alday:2016njk,Aharony:2016dwx,
Li:2017lmh,Costa:2017twz,Simmons-Duffin:2017nub,Caron-Huot:2017vep,Alday:2017vkk,
Liu:2020tpf,Caron-Huot:2020adz,Caron-Huot:2020nem,Caron-Huot:2020ouj}, 
which showed that CFT operators organize in families, or Regge trajectories. Hence, a twist gap corresponds to imposing that the leading Regge trajectory lies above a certain line. Moreover, if the leading Regge trajectory is a concave function $\tau(\ell)$, then the twist gap corresponds to a bound on the lowest spin operator on the trajectory and can be compared with explicit predictions.

\begin{figure}[ht]
\subfloat[Plot in linear scale.]
{\includegraphics[scale=.65]{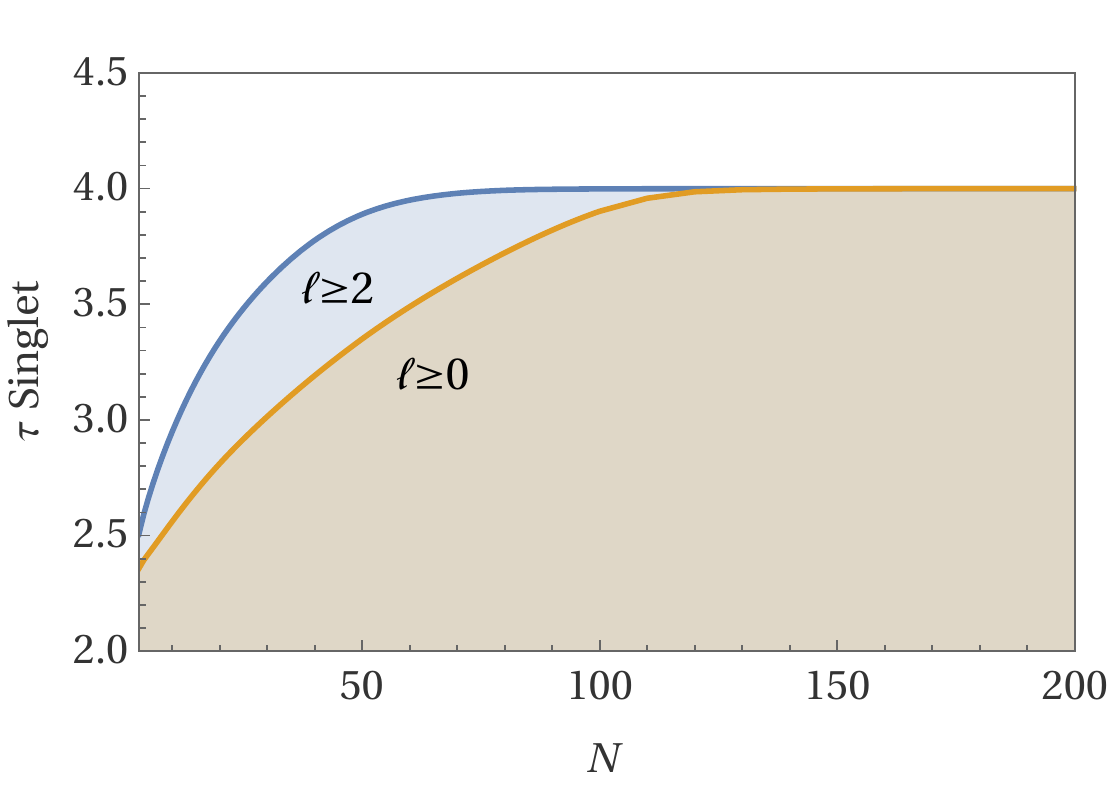}}
\;
\subfloat[Plot in log scale. The $y$ axis is $4-\tau$.]
{\includegraphics[scale=.65]{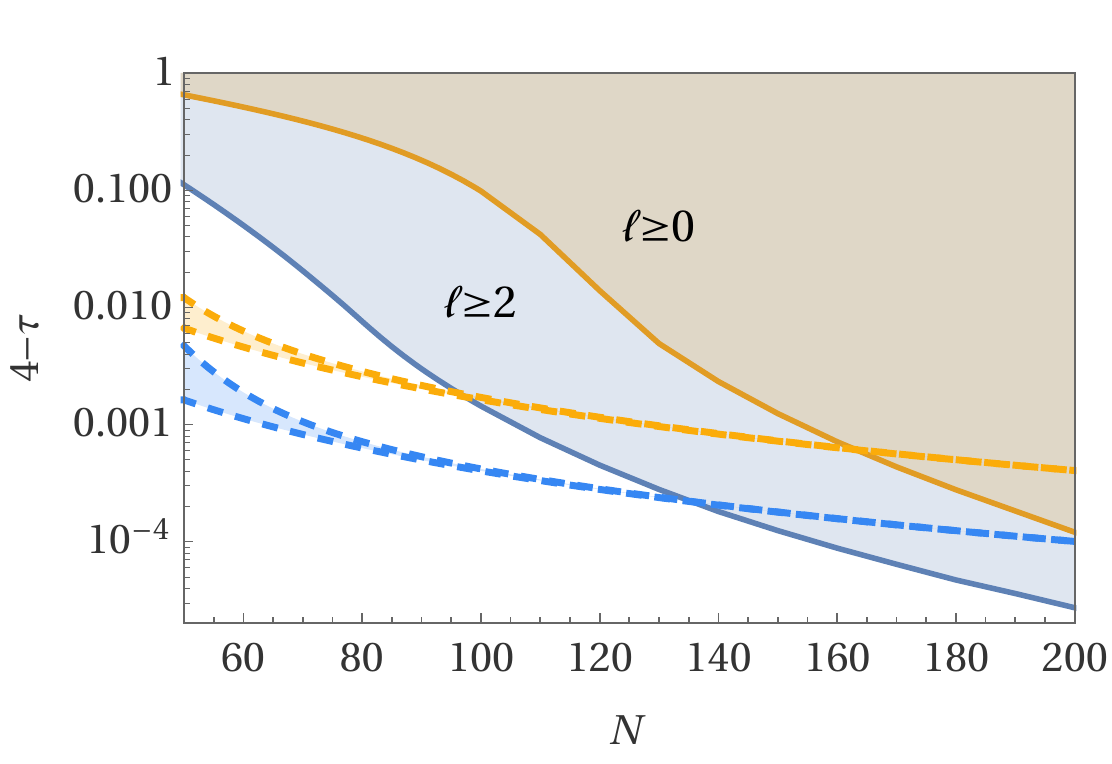}}

\caption{Bounds obtained by gapping the twist of all singlets (in yellow) and by gapping the twist of only the singlets with spin $\ell \geq 2$ (in blue). The shaded region is allowed. These plots have been obtained with $\Lambda = 27$. The dashed lines correspond to the supergravity prediction for $\ell=0$ (yellow) and $\ell=2$ (blue), which can be found in~\eqref{localizationgamma}, with $\lambda \in [c/10,c]$.}
\label{fig:BisectionReggeSingletComparisonINTRO}
\end{figure}

We first performed this analysis using the $\langle 2222\rangle$ or $\langle 3333\rangle$ correlation functions alone and did not find substantial improvements over the previous dimension bounds, except a faster convergence of the numerics. However, when we considered the full mixed system, we found a surprising and intriguing result. The plot is shown in \figurename~\ref{fig:BisectionReggeSingletComparisonINTRO}. The main aspect to focus on is how the bound approaches the value of $\tau=4$. However, before making an interpretation of the result, we should understand the analytical predictions to which it can be compared.

Let us start discussing the most natural expectation. As we will see, this picture creates a tension and must be revisited in light of our results. In the limit of large $N$ and large $\lambda$, the dual theory is described by weakly coupled supergravity. In this regime it makes sense to classify the operators as single trace, double trace and, more generally, multi trace. The dimension of the unprotected double traces grows with $\lambda$ and so they decouple. Therefore, the leading Regge trajectory is given by the double trace operators of the schematic form
\eqn{
[\COp2 \COp2]_{0,\ell} \sim \COp2 \partial^\ell \COp2 \,.
}[eq:doubletrace]
The computation of their anomalous dimension can be done in the gravity side by computing Witten diagrams. However gravity is the low energy description of a more fundamental string theory and it is valid up to a certain scale, which is the mass of the massive string modes. By integrating them out one introduces a cutoff in the low energy theory and this generates a whole series of higher dimensional operators, suppressed by the string scale, i.e. inverse powers of the t'Hooft coupling. Therefore, as it is the case for any effective theory, going beyond tree level forces us to introduce more and more of these higher dimensional operators in order to cancel loop divergences. This ultimately will give ambiguities for the anomalous dimensions of the operators under study that cannot be resolved without addressing the full UV complete string theory.

Recently however, field theory techniques \cite{Arutyunov:2000ku,DHoker:1999mic,Goncalves:2014ffa,Binder:2019jwn,Chester:2019pvm,Aprile:2017bgs,Alday:2018pdi,Chester:2020dja} have been exploited to compute subleading orders in the anomalous dimension of the double trace \eqref{eq:doubletrace}, which we report in \eqref{localizationgamma}, in an expansion in the 't Hooft coupling and the central charge $c=(N^2-1)/4$.
A few comments about this result are in order: First, beyond the leading order, the analyticity in spin down to $\ell=0$ is broken: at order $1/N^4$ analyticity extends only to $\ell\geq 2$.\footnote{Note the pole at $\ell=1$ in \eqref{gammal}.} This is a consequence of the higher dimensional operator $R^4$ that contributes to the scalar double trace anomalous dimension and spoils the Regge behavior at this order. Similar things happen at higher orders for any spin.
Second, equation \eqref{localizationgamma} contains terms that naively diverge in the limit $\lambda\rightarrow \infty$ with $N$ fixed. This is due to the fact that the supergravity cutoff cannot be removed without reintroducing divergencies and ambiguities. Indeed, strictly speaking, this limit cannot be taken because one can show that the boundedness of the double discontinuity of the correlator requires $\lambda \leq c$~\cite{Alday:2017vkk}. Interestingly, if one takes the upper limit $\lambda\sim c$, some terms that seem to diverge with $\lambda$ give rise to a contribution at the same order as a lower loop level.

Despite all the above caveats, one can observe that the inclusion of $1/\lambda$ corrections is parametrically small, at least for the terms that have been computed. Hence it makes sense to compare our bounds with the leading order prediction
\eqn{
\tau_{[\COp2 \COp2]_{0,0}} = 4 - \frac{16}{N^2} +\frac{\kappa_0}{N^4}+O\!\left(\frac1{N^6}\right)\,,
}[sugraprediction]
where $\kappa_0$ parametrizes the contribution from the $R^4$ operator discussed before. Here we face an apparent tension: if we compare the above prediction with the bound in \figurename~\ref{fig:BisectionReggeSingletComparisonINTRO} we observe a strong disagreement already for $N\lesssim 160$. We also computed twist bounds including only operators of spin $\ell\geq2$, while leaving the scalar unconstrained. Nevertheless, we still observe a disagreement between the supergravity prediction
\eqn{
\tau_{[\COp2 \COp2]_{0,2}} = 4 - \frac{4}{N^2} -\frac{45}{N^4} + \frac{\kappa_2}{N^6}+O\!\left(\frac1{N^8}\right)\,,
}[sugrapredictionL2]
and the numerical bound, albeit for somewhat smaller $N$. As it is clear from \eqref{sugrapredictionL2}, $[\COp2\COp2]_{0,2}$ suffers from the same problems as the scalar double trace, but they start appearing at order $1/N^6$.

The observed tension comes from insisting that the double trace operators have the smallest twist at large $N$. Let us revise this assumption.

Generically it is a complicated matter to define which operator has the smallest twist for a given value of the coupling and $N$. In particular, there are two families of candidates: the Konishi-like operators, schematically  $\mathcal{O}^{K}_\ell\sim \Tr (\varphi^i \partial^\ell \varphi^i)$, whose dimension is $\Delta_{K}=\ell+2+ f(\lambda, N)$ and the above-mentioned double trace operators with dimension $4+\ell + g(\lambda, N)$. 
We know that in the planar limit $f(\lambda, N)\sim \lambda^{1/4}$, thus the Konishi-like operators become heavy and the leading twist operators are the double trace ones. Oppositely, for fixed $N$ and small enough $\lambda$ the leading twist operators are the $\mathcal{O}^{K}_\ell$. 

While at infinite $N$ the two trajectories are allowed to cross without mixing, at finite $N$ no level-crossing is possible. Nevertheless,  the expectation at large $N$ is that, by following the evolution of the Konishi-like operators as the coupling $\lambda$ grows, one will find a combination of the two candidates that remains light and one that presumably becomes heavy.  Contrarily, at small $N$ as one moves in coupling space, both towers remain with twist below four.\footnote{See also discussion in section 4.3 of \cite{Beem:2016wfs}} We do not know for which values of $N$ the transition between the two regimes takes place. 

\emph{The bounds we presented suggest that there exists an extended intermediate range of $N$ which is closer to the small $N$, strong coupling regime, namely with two towers of operators with twist smaller than four}. Only when $N$ is parametrically very large one tower will decouple; the low dimension spectrum will then contain a single tower of states that mimic the double trace operators. Notice that the bounds shown have not converged yet: in particular Figure~\ref{fig:BisectionReggeSingletMixeda} shows that in principle the regime of validity of supergravity could be restricted to even higher values of $N$.

In order to get further evidences of this scenario, we extracted the extremal spectrum of the crossing solution living on the boundary shown in Figure \ref{fig:BisectionReggeSingletMixeda}  using the extremal functional method \cite{ElShowk:2012hu, Simmons-Duffin:2016wlq}. The results are shown in Figure \ref{fig:spectraMixed} and they seem to suggest that for $N\lesssim 100$ there exist two towers\footnote{Due to trucation in $\Lambda$ and numerical precision these towers appear incomplete.} of operators: one corresponding to the double trace operators $[\COp2 \COp2]_{0,\ell}$ and one corresponding to operators with twist at the boundary. As $N$ increases these two towers merge into a single one, with twist approximately 4. Thus, the extremal spectrum seems to favor this scenario, although we do not claim that it is the correct one. 

Other less likely scenarios could also explain the observed tension:

\begin{enumerate}

\item In the large $\lambda$, large $N$ limit only double trace operators remain, however the large $N$ expansion is asymptotic: in order to compare our bound with the supergravity prediction at finite $N$ one should first resum the perturbative series. Moreover, as shown in equation (\ref{localizationgamma}), a given order in $1/N$ receives contribution to higher orders, once $1/\lambda$ corrections are included. It could be that all those higher order terms of the perturbative expansion, despite they seem negligible at $N\sim 10^2$, once resummed they do give a finite contribution and make the leading Regge trajectory consistent with the bound.

\item The anomalous dimension of the leading Regge trajectory could receive sizeable non-perturbative contributions that modify the prediction of the large $N$ expansion. We believe this possibility is quite unlikely.

\item As we discuss in the next sections, the four-point functions of half-BPS operators receive contribution from both short and long multiplets. We computed the contribution of the protected operators using free theory \cite{Dolan:2004iy} and chiral algebra techniques \cite{Beem:2013sza}. Our results agree with the several other works in the literature \cite{Rayson:2017mma,Aprile:2017xsp,Caron-Huot:2018kta}. Still, it could be that the setup used in the present work misses something. For instance there could be operators whose dimension is dynamically protected: in this case our bound would be trivially satisfied. 
\end{enumerate}

The above scenarios are quite peculiar and it will be fundamental to gather additional evidences to support one or the other, and eventually formulate alternative explanations.

All in all, we discovered a tension between the common lore about the validity of large $N$ expansion and non-perturbative bootstrap results. It will be interesting to understand the source of the discrepancy, which hopefully will lead to a better understanding of this class of SCFTs.

\vspace{1em}

In this work we also explored bounds on OPE coefficients in various sector. We first computed upper bounds on OPE coefficients and then lower bounds by assuming a gap to the next operators in the same sector.

As a first example we considered the lowest dimensional scalar $\CO$ in the $[0,0,0]$ representation. Because this operator appears both in the $\COp2 \times \COp2$ and $\COp3 \times \COp3$ OPE, there are two independent OPE coefficients associated to it. We explored bounds on the norm of the vector of OPE coefficients as a function of their ratio, parametried by an angle $\theta$, and the dimension of $\CO$. For concreteness we fixed $N=20$ and assumed a gap to the next scalar singlet $\Delta\geq 4.2$. As a function of $\theta$ and $\Delta_\CO$ the upper and lower bounds describe a somewhat thin three-dimensional region, see Figure \ref{fig:OPEMinMaxSinglet}. Although this is not a conclusive evidence, this analysis points in the direction of a two dimensional allowed region, rather than a three dimensional one. This is in agreement with the expectation that the conformal manifold of $\mathcal N=4$ SYM is also two dimensional, since it is parametrized by the gauge coupling $g_{\mathrm{YM}}$ and the $\vartheta$-angle\footnote{This is the coupling in front of the $F\tilde F$ term in the Lagrangian, not to be confused with the angle parametrizing the ratio of OPE coefficients defined above.}. It is tempting to conjecture that in the limit of infinite numerical power, the upper and lower bound will coincide and the allowed region will shrink to a two dimensional surface. If this is the case, then specifying say $\theta$ and $\Delta_\CO$ we would be able to completely fix all other CFT data.

A similar analysis was carried out also for the lowest scalar in the $[0,1,0]$ and $[0,2,0]$ representation and it is shown in Figure \ref{fig:OPEMinMax}. In that case we do not scan over two parameters since there is a single OPE coefficient.

\section{Set up}
\label{sec:generalities}
In this section we keep the discussion as general as possible and we will specialize later to the case at hand. 
Our object of study is the four-point function of non identical half-BPS operators $\COp{p}$ in four dimensional $\mathcal{N}=4$ SCFT. These are scalar operators, of protected dimension $\Delta=p$ and transforming in the $[0,p,0]$ representation of the $\SU(4)$ R-symmetry group. They can be written as
\begin{equation}
\COp{p}(x,t)=t_{i_1}\cdots t_{i_p}\mathcal{O}^{i_1\ldots i_p}(x)\,,
\end{equation}
where $\mathcal{O}^{i_1\ldots i_p}$ is a symmetric traceless tensor of $\mathrm{SO}(6)$ and $t$ is a null, complex six dimensional vector which encodes the R-symmetry polarization. Supersymmetry guarantees that both two- and three-point functions of half-BPS operators are protected. Four-point functions are nontrivial and, due to superconformal symmetry, can be written as
\eqna{
&\langle \COp{p_1}(x_1,t_1)\COp{p_2}(x_2,t_2)\COp{p_3}(x_3,t_3)\COp{p_4}(x_4,t_4)\rangle =\\
&
\left(\frac{t_1\cdot t_2}{x_{12}^2}\right)^{\Sigma-p_4}
\left(\frac{t_3\cdot t_4}{x_{34}^2}\right)^{p_3}
\left(\frac{t_1\cdot t_4}{x_{14}^2}\right)^{p_1+p_4-\Sigma}
\left(\frac{t_2\cdot t_4}{x_{24}^2}\right)^{p_2+p_4-\Sigma}\times\\&
\mathcal{F}^{(p_1,p_2,p_3,p_4)}(u,v;\sigma,\tau)\,,
}[]
where $\Sigma = \tfrac12(p_1+p_2+p_3+p_4)$ and the function $\mathcal{F}^{(p_1,p_2,p_3,p_4)}(u,v;\sigma,\tau)$ contains dynamical information. We have introduced space-time cross ratios 
\begin{align}
&u =z \bar{z} =\frac{x_{12}^2 x_{34}^2}{x_{13}^2x_{24}^2}\;, &v =(1-z)(1-\bar{z})= \frac{x_{14}^2 x_{23}^2}{x_{13}^2x_{24}^2}\,,\label{eq:vars}
\end{align}
as well as harmonic cross ratios
\begin{align}
&\sigma =\alpha \bar{\alpha}= \frac{t_1 \cdot t_3\, t_2 \cdot t_4}{t_1 \cdot t_2\, t_3 \cdot t_4}\; &\tau =(1-\alpha)(1-\bar{\alpha})= \frac{t_1 \cdot t_4\,t_2 \cdot t_3}{t_1 \cdot t_2\,t_3 \cdot t_4}\,.
\end{align}
For brevity we will often replace superscripts $(p,p,p,p)$ by $(p)$. The function $\mathcal{F}^{(p_1,p_2,p_3,p_4)}(u,v;\sigma,\tau)$ is a polynomial in $\sigma$ and $\tau$ of degree $p\equiv \min(\Sigma-p_4,p_3)$. In the following we will consider a specific ordering such that
$p_1,p_2,p_3 \leq p_4$.
The function $\mathcal{F}^{(p_1,p_2,p_3,p_4)}(u,v;\sigma,\tau)$ admits a conformal partial wave decomposition in both harmonic and space-time cross ratios
\eqn{ \label{cpwex}
\mathcal{F}^{(p_1,p_2,p_3,p_4)}(u,v;\sigma,\tau) = \sum_{\Delta,\ell} \sum_{0\leq m\leq n \leq p} c_{\Delta,\ell}^{[n,m]}\, Y_{nm}^{(a,b)}(\sigma,\tau)\,G_{\Delta}^{(\ell)}(u,v;p_{21},p_{43})\,,
}[]
where $p_{ij} \equiv p_i-p_j$, $a = \frac{1}{2}(p_{43}-p_{21})$, $b = \frac{1}{2}(p_{43}+p_{21})$. The labels $(n,m)$ parametrize the $\SU(4)$ representations under which intermediate operators belonging to the OPE of $\COp{p_1}\times \COp{p_2}$ transform, and the OPE reads (for $p_1\leq p_2$)
\eqn{
[0,p_1,0] \otimes [0,p_2,0] \sim \bigoplus_{0\leq m\leq n \leq p_1} [n-m,2m + p_2-p_1,n-m]\,,
}[]
There are $\frac{(p_1+1)(p_1+2)}{2}$ representations, each of them is associated to a specific combination of Jacobi polynomials $P_n^{(a,b)}$
\eqn{
Y_{nm}^{(a,b)}(y,\yb) = \frac{P_{n+1}^{(a,b)}(y)P_m^{(a,b)}(\yb)-P_{m}^{(a,b)}(y)P_{n+1}^{(a,b)}(\yb)}{y-\yb}\,,
}[]
where $y=2 \alpha-1$ and $\yb=2 \bar{\alpha}-1$
The remaining part of \eqref{cpwex} depends only on the cross ratios $u$ and $v$ and it is the usual four dimensional conformal block
\eqn{
G_{\Delta}^{(\ell)}(z,\zb;p_{21},p_{43}) = \frac{(z\zb)^{\frac{1}{2}(\Delta-\ell-p_{43})}}{z-\zb}\left(\left(-\lifrac{1}{2}z\right)^\ell z\, F_{p_{21},p_{43}}\lnsp\left[\lifrac{\Delta+\ell}{2},z\right]F_{p_{21},p_{43}}\lnsp\left[\lifrac{\Delta-\ell-2}{2},\zb\right]-(z\leftrightarrow \zb)\right)\,,\label{confblockdef}
}[]
with
\eqn{
F_{p_{21},p_{43}}[\lambda,z] = \F\left(\lambda+\lifrac{p_{21}}{2},\lambda-\lifrac{p_{43}}{2};2\lambda;z\right)\,,\label{confblockdef2}
}[]
$\F$ being the hypergeometric function. The structure of the conformal partial wave decomposition \eqref{cpwex} is dictated only by conformal invariance. However we would like to consider the implication of the full $\mathcal{N}=4$ superconformal symmetry. In particular we will use two consequences of superconformal symmetry. The first is the fact that each conformal primary operator belongs to a superconformal multiplet which relates specific ranges of conformal dimension $\Delta$ and spin $\ell$. Such supermultiplets have highest weight states which are called superconformal primaries. The challenge is then to reorganize the conformal partial wave into a superconformal partial wave. The second is the presence of protected operators in the OPE $\COp{p_1}\times \COp{p_2}$. As already mentioned, this fact simplifies extremely the superconformal block decomposition, since both dimensions and OPE coefficients of such operators are protected from quantum corrections and are explicitly known. In analogy with the conformal case, the most efficient method to combine these constraints is to find solutions of the superconformal Ward identities. We will now review briefly the logic, which was first developed in \cite{Nirschl:2004pa}. The Ward identities on the correlator $\langle\mathcal{O}^{(p_1)}\cdots\mathcal{O}^{(p_4)}\rangle$ imply
\eqn{
\left.\mathcal{F}^{(p_1,p_2,p_3,p_4)}(z,\zb;\alpha,\alphab)\right|_{\zb=1/\alphab} = f(z,\alpha)\,,\label{eq:WI}
}[]
together with the other combinations of $\alpha$ or $\alphab$ and $z$ or $\zb$. By consistency $f\left(z,z\right)$ must be a constant, which we denote by $k$. If one regards $z,\alpha$ as ``chiral'' variables and $\zb,\alphab$ as ``antichiral'', this identity suggests that the four-point function, when restricted to the region $\zb=1/\alphab$, acquires a chiral structure. This is indeed a feature of all four dimensional $\CN=2$ theories \cite{Beem:2013sza}. Such a chiral structure is related to an exactly solvable subsector of the theory, dual to a non-unitary two dimensional SCFT, which in principle allows us to compute the exact form of $f(z,\alpha)$. From the knowledge of $f(z,\alpha)$ we can compute all OPE coefficients of the protected (or \emph{short}) operators, which is precisely what we need in order to write down the bootstrap equations. This can be done for any $p_i$ and the general solution is of the form 
\eqna{
&\mathcal{F}^{(p_1,p_2,p_3,p_4)} = \mathcal{F}^{(p_1,p_2,p_3,p_4)}_{\hat{f}}(z,\zb,\alpha,\alphab)
 + s(z,\zb,\alpha,\alphab) \,\mathcal{H}^{(p_1,p_2,p_3,p_4)}(z,\zb,\alpha,\alphab)\,.\\
 }[eq:Hdef]
The function $\mathcal{F}^{(p_1,p_2,p_3,p_4)}_{\hat{f}}$ contains information only from the protected sector of the OPE and it is written as 
 \eqna{
&\mathcal{F}^{(p_1,p_2,p_3,p_4)}_{\hat{f}}(z,\zb,y,\yb) =\\& \quad-k + 	\frac{(\alphab z-1)(\alpha\zb-1)\left(f(z,\alpha)+f(\zb,\alphab)\right)-(\alpha z-1)(\alphab \zb -1)\left(f(z,\alphab)+f(\zb,\alpha)\right)}{(z-\zb)(\alpha-\alphab)}\,,
}[]
where the superscript $(p_1,p_2,p_3,p_4)$ on $f$ are implicit and
\eqna{
f(z,\alpha) &= k + \left(\alpha-\frac1z\right)\,\hat{f}(z,\alpha)\,.
}[eq:fhatdef]
The remaining part, which we call $\mathcal{H}^{(p_1,p_2,p_3,p_4)}$, contains the contribution to the OPE of unprotected operators. It is multiplied by the function $s(z,\zb,\alpha,\alphab)$, which must be always zero when $z = 1/\alpha$ (or $\zb = 1/\alphab$, $z = 1/\alphab$ and $\zb = 1/\alpha$  ). We define the function $s(z,\zb,\alpha,\alphab)$ to have this form
\eqna{
s(z,\zb,\alpha,\alphab) &=(\alpha z - 1)(\alphab z - 1)(\alpha \zb - 1)(\alphab \zb - 1)\,.\label{eq:sdef}
}[]
Notice that $s(z,\zb,\alpha,\alphab)$ is a polynomial of degree two in $\sigma$ and $\tau$, thus $\mathcal{H}^{(p_1,p_2,p_3,p_4)}$ is a polynomial of degree $p-2$ where $p = \mathrm{min}(\Sigma- p_4,p_3)$.
This allows writing a superconformal partial wave decomposition involving $\mathcal{H}^{(p_1,p_2,p_3,p_4)}$
\eqna{
\mathcal{H}^{(p_1,p_2,p_3,p_4)}(u,v;\sigma,\tau) &= \sum_{0\leq m\leq n\leq p-2} \sum_{\Delta,\ell} a_{\Delta,\ell}^{[n,m]}\,Y_{nm}^{(a,b)}(\sigma,\tau)\, \mathcal{G}_{\Delta}^{(\ell)}(u,v;p_{21},p_{43})\,,}[supercpwex]
where $\mathcal{G}_{\Delta}^{(\ell)}(u,v;p_{21},p_{43})=u^{-2}G_{\Delta+4}^{(\ell)}(u,v;p_{21},p_{43})$ is the superconformal block which repackages the contributions of superdescendants of a given superprimary of dimension $\Delta$, spin $\ell$ and transforming under the $[m,n]$ representation.\footnote{We will often abbreviate $[m,n] \equiv [n-m,2m+p_{21},n-m]$.\label{foot:mn}} This decomposition might seem in contradiction with \eqref{cpwex}, since the possible representation are reduced to $\frac{p_1(p_1-1)}{2}$. The key is that the sum in \eqref{supercpwex} runs over superconformal primary operators, and those will transform under fewer representations of $\SU(4)$ R-symmetry. In particular for the case $p_i = p = 2$, $\mathcal{H}^{(2,2,2,2)}$ does not depend on the polarizations and this means that all superconformal primaries of the exchanged long multiplets are singlets.
\par
There is a decomposition for the function $\hat{f}$ which reads
\eqna{
\hat{f}^{(p_1,p_2,p_3,p_4)}(x,y)&=\sum_{n=0}^{p-1}\sum_{
\left\lbrace
\substack{\ell \geq -1\;n\; \mathrm{even}\\
\ell \geq 0\;n\; \mathrm{odd}\;\;\;\,}\right.
} b_{n,\ell} \,P_n^{(a,b)}(y)\,g_\ell^{(p_{21},p_{43})}(x) \,,
}[eq:hatfex]
where $P_n^{(a,b)}(y)$ are the Jacobi polynomials and $g_\ell^{(p_{21},p_{43})}(x) $ are defined as
\eqn{
g_\ell^{(p_{21},p_{43})}(x) = (-x)^{\ell+2}F_{p_{21},p_{43}}\left[\ell+\tfrac{p_{43}}2+2,x\right]\,.
}[eq:hatfblock]
\par
The four-point function is constrained by crossing symmetry as
\eqna{
 \mathcal{F}^{(p_1,p_2,p_3,p_4)}(u,v,\sigma, \tau)&=\left(\frac{u}{v}\right)^{\Sigma-p_4}\tau^{\Sigma-p_4} \mathcal{F}^{(p_3,p_2,p_1,p_4)}\left(v,u,\frac{\sigma}{\tau},\frac{1}{ \tau}\right)\\
 &=u^{p_3}\sigma^{p_3}\mathcal{F}^{(p_4,p_2,p_3,p_1)}\left(\frac{1}{u},\frac{v}{u},\frac{1}{\sigma},\frac{\tau}{\sigma}\right).
 }[]
Similar relations hold for the unprotected function $\mathcal{H}^{(p_1,p_2,p_3,p_4)}$. Let us consider only the $(13)$ permutation. First we define the crossing of the $\CF_{\hat{f}}$ functions as follows
\eqn{
\Delta^{(p_1,p_2,p_3,p_4)} \equiv  \frac1{s(u,v,\sigma,\tau)}\left[\left(\frac{u}{v}\right)^{\Sigma-p_4}\tau^{\Sigma-p_4} \mathcal{F}_{\hat{f}}^{(p_3,p_2,p_1,p_4)}\left(v,u,\frac{\sigma}{\tau},\frac{1}{ \tau}\right)-\mathcal{F}_{\hat{f}}^{(p_1,p_2,p_3,p_4)}(u,v,\sigma, \tau)\right].
}[]
Then the crossing equation for the unprotected function reads
\eqn{
\mathcal{H}^{(p_1,p_2,p_3,p_4)}(u,v,\sigma, \tau)=\left(\frac{u}{v}\right)^{\Sigma-p_4}\tau^{\Sigma-p_4-2}\, \mathcal{H}^{(p_3,p_2,p_1,p_4)}\left(v,u,\frac{\sigma}{\tau},\frac{1}{\tau}\right)+\Delta^{(p_1,p_2,p_3,p_4)}\,.
}[]

\section{Multiplet recombination}
\label{sec:multiplets}
Thanks to $\CN=4$ supersymmetry, the contributions of the short multiplets to the OPE $\COp{p}\times\COp{q}$ can be exactly computed, namely both the dimension and the OPE coefficient are fixed. In order to fully exploit the constraints of maximal supersymmetry, it is important to input this information into the bootstrap problem. This will be the goal of the present section. Since the short multiplets are protected, we can compute their contribution to the four-point function by considering the theory at exactly zero coupling $g_{\mathrm{YM}} = 0$. An alternative derivation which makes use of the chiral algebra~\cite{Beem:2013sza} is presented in \appendixname~ \ref{app:chiral}. However, this is not enough: the result from the free theory is ambiguous due to the impossibility of distinguishing between long operators at the unitarity threshold and direct sums of short operators. What typically happens is that if some short multiplets can make up a long multiplet when combined, then they will do so and gain an anomalous dimension when the coupling is turned on. This phenomenon is usually referred to as ``multiplet recombination.'' However, there are some situations where this does not happen and some short multiplets that could recombine actually remain protected. Multiplets that exhibit this behavior are called ``dynamically protected''~\cite{Bianchi:2006ti, Doobary:2015gia}. After carrying out the analysis of the free theory we will address this issue.

\subsection{Short multiplets from the free theory}

The free theory result is obtained by doing Wick contractions. A field $\COp{p}$ is realized as the normal ordered product
\eqn{
\CO^{i_1\ldots i_p}(x) = \,:\lnsp\Tr \varphi^{\{i_1}\cdots \varphi^{i_p\}}:\,,
}[]
where the curly brackets indicate a projection to the symmetric traceless tensor. A contraction of $p$ lines is denoted as
\eqn{
\pic[1]{-2.5pt}{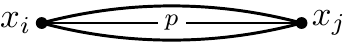} = \left(\frac{t_i\cdot t_j}{x_{ij^2}}\right)^p\,.
}[]
For simplicity let us use the shorthand
\eqn{
s = \sigma u\,,\qquad t = \tfrac{\tau u}v\,.
}
The free theory result is readily obtained as
\fourseqn{
&\begin{aligned}
\CF^{(2,2,2,2)} &= a_1(1+s^2+t^2) + a_2 (s + t + st) = \\
&= a_1\Bigg(\;\pic{-16pt}{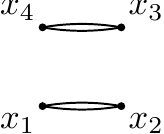} +\; \pic{-9.2pt}{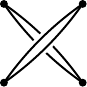}+\;\pic{-9.2pt}{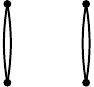}\Bigg) + a_2 \Bigg(\pic{-9.2pt}{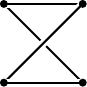}+\;\pic{-9.2pt}{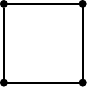}+\;\pic{-9.2pt}{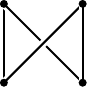}\Bigg)\,.
\end{aligned}
}[eq:F2freetheory]{
&\begin{aligned}
\CF^{(3,3,3,3)} &= b_1(1+s^3+t^3) + b_2 (s + s^2+t+t^2+s^2t+ st^2) + b_3 st = \\
&= b_1\Bigg(\;\pic{-10.5pt}{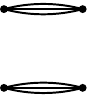} +\; \pic{-9.2pt}{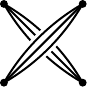}+\;\pic{-9.2pt}{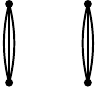}\Bigg) \\&+ b_2 \Bigg(\pic{-10pt}{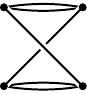}+\;\pic{-9.2pt}{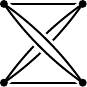}+\;\pic{-9.2pt}{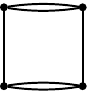}+\;\pic{-9.2pt}{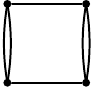}+\;\pic{-9.2pt}{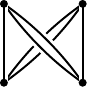}+\;\pic{-9.2pt}{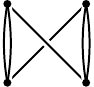}\Bigg)+b_3\;\pic{-9.2pt}{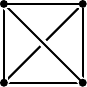}\,.
\end{aligned}
}[]{
&\begin{aligned}
\CF^{(2,2,3,3)} &= c_1  + c_2(s^2+t^2) + c_3(s+t) + c_4\,st \\&=
c_1\;\pic{-10pt}{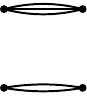} + c_2\Bigg(\pic{-9.2pt}{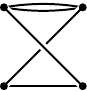}+\;\pic{-9.2pt}{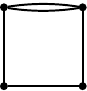}\Bigg) + c_3 \Bigg(\pic{-9.2pt}{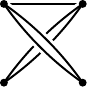}+\;\pic{-9.2pt}{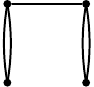}\Bigg)+c_4\;\pic{-9.2pt}{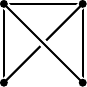}
\end{aligned}
}[]{
&\CF^{(2,3,2,3)} = c_1\, s^2 + c_2(1+t^2) + c_3(s^2+st) + c_4\,t \,.
}[][]
The various coefficients $a_i,\,b_i$ and $c_i$ are fixed by the gauge group $G$. For $\CF^{(2,2,2,2)}$ we have
\eqn{
a_1 = 1\,,\qquad a_2 = \frac{4}{\mathrm{dim}(G)}\,.
}[eq:color2]
For $\mathrm{SU}(N)$, $\mathrm{dim}(G) = N^2-1$. The other correlators instead exist only when the cubic Casimir $d_{abc}= 2 \,\Tr(\{t_a,t_b\}t_c)$ is non vanishing. The case of $\mathrm{SU}(N)$, which is the one we are interested in, has $d_{abc} \neq 0$ and we have
\twoseqn{
b_1 &= 1\,,&b_2 &= \frac{9}{N^2-1}\,,&b_3 &= \frac{18\lsp(N^2-12)}{(N^2-4)(N^2-1)}\,,
}[eq:color3]{
c_1 &= 1\,, &c_2 &= 0\,,&c_3 = \frac12 \llsp c_4 &= \frac{6}{N^2-1}\,.
}[eq:color23]
From the free theory expressions for $\CF^{(p_1,p_2,p_3,p_4)}$ we can deduce $\CH^{(p_1,p_2,p_3,p_4)}$ and $\hat{f}^{(p_1,p_2,p_3,p_4)}$, which we will not report here for brevity. The latter will maintain its form in the interacting theory, whereas the former will change. However some contributions in $\CH$ must remain fixed even in the interacting theory. They are divided in two classes
\begin{enumerate}
\item {\bf Multiplets below threshold}: by expanding $\hat{f}$ in conformal blocks we will discover some operators with conformal dimension below the unitarity bound. Their contribution can be canceled by introducing other non-unitary contributions to $\CH$. These are of course already present in the free theory result, but they must remain unchanged even at nonzero coupling.
\item {\bf Multiplets at threshold}: these are the multiplet that give rise to the free theory ambiguity discussed at the beginning of this section.
\end{enumerate}

Before studying both cases we need to expand the functions $\CH$ and $\hat{f}$ in superconformal blocks, as in \eqref{supercpwex} and \eqref{eq:hatfex}. This can be done using the results of \cite{Dolan:2004iy,Rayson:2017mma}. In \appendixname~\ref{app:expansion} we will work out explicitly the case $(2,3,2,3)$ and present the results for all $b_{n,\ell}$ coefficients. For brevity we will not report the coefficients $a^{[m,n]}_{\Delta,\ell}$.

\subsection{Multiplets below threshold}
The contributions to $\hat{f}$ that correspond to operators below unitarity are $b_{n,\ell}$ for $0 < n \leq p$.\footnote{Recall $p = \mathrm{min}(\tfrac12(p_1+p_2+p_3-p_4),p_3)$.} These can be canceled by adding to $\CH$ blocks of twist $2t+p_{43}$, with $0\leq t \leq n$ transforming in the $[n-m,2m+p_{43},n-m]$ of $\mathrm{SU}(4)$. As anticipated, these contributions are not unitary because the lower bound on the twist for those long operators is
\eqn{
\tau \equiv \Delta - \ell \geq 2n + p_{43} + 2 > 2t + p_{43}\,.
}
The non unitary sector of $\CH^{(2)}$ reads
\eqn{
\CH^{(2)} \supset \sum_{\ell \geq 0\,\mathrm{even}} A_{0,\ell}\,\CG_{\ell}^{(\ell)}(u,v)\,,
}[]
where the value of $ A_{0,\ell}$ will be given shortly. Similarly for the other correlators one has
\twoseqn{
\CH^{(2,2,3,3)} &\supset \sum_{\ell \geq 0\,\mathrm{even}} A^{(2,2,3,3)}_{0,\ell}\,\CG_{\ell}^{(\ell)}(u,v)\,, \qquad \mbox{Similarly for $(2,3,2,3)$}\,,
}[]
{
\CH^{(3)} &\supset \sum_{\substack{0\leq m \leq n \leq 1 \\ t\leq n}}\sum_{\substack{\ell \geq 0 \\ \ell \equiv n+m \,\mathrm{mod}\,2}} A_{t,\ell}^{[n,m]}\,Y_{n,m}(\sigma,\tau)\,\CG_{2t+\ell}^{(\ell)}(u,v)\,.
}[]
The coefficients are fixed as follows
\eqna{
A_{0,\ell} &= 2^{\ell+1} b_{1,\ell}\;,\\
A^{(2,3,2,3)}_{0,\ell} &= 2^{\ell+1} b^{(2,3,2,3)}_{1,\ell}\;,\\
A^{(2,2,3,3)}_{0,\ell} &=2^{\ell+1} b^{(2,2,3,3)}_{1,\ell} \;,\\
A^{[0,0]}_{0,\ell} &= 2^{\ell+1} b^{(3)}_{1,\ell}\;,\\
A^{[1,0]}_{0,\ell} &= 2^{\ell+1} b^{(3)}_{2,\ell}\;,\quad&
A^{[1,0]}_{1,\ell} &= 0\;,\\
A^{[1,1]}_{0,\ell} &= 0\;,\quad&
A^{[1,1]}_{1,\ell} &= -\lifrac{1}{4} A^{[1,0]}_{0,\ell+1}\,,
}[]
where the $b_{n,\ell}$ coefficients are given in \tablename~\ref{tab:bnlcoeffs}.
\par
The introduction of such long multiplets in $\CH$ does not only have the effect of removing the unwanted contributions in $\hat{f}$, it also implies the presence of other protected multiplets. For example, the recombination rule
\eqn{
\CA^{\ell}_{[0,0]\ell} \simeq \CD_{[0,0]\ell} \oplus \CC_{[1,1]\ell-2}\,,
}[]
says that after the cancellation of the $\CD$ non unitary multiplet, a $\CC$ multiplet appears with coefficient given by $A_{0,\ell}$. All other cases can be read from \tablename~\ref{tab:recombination}.

\subsection{Multiplets at threshold: recombination}\label{sec:recombination}
As explained previously, the free theory presents an ambiguity. Namely it is impossible to distinguish a long operator at threshold from a sum of protected operators. This is due to representation theoretic recombination rules such as
\eqn{
\CA^{\ell+2}_{[0,0,0]\ell} \simeq \CC_{[0,0,0]\ell} \oplus \CC_{[1,0,1]\ell-1}\,.
}
We can make use of two criteria to resolve the ambiguity
\begin{enumerate}
\item In $\CH^{(2)}$, $\CH^{(3)}$ and $\CH^{(2,2,3,3)}$ we find $\CC_{[0,0,0]\ell}$ multiplets. These contain (for $\ell > 2$) higher spin conserved currents, which are known to constitute a decoupled free subsector of the theory~\cite{Maldacena:2011jn,Alba:2013yda}. We will thus impose that such multiplets are absent and resolve the ambiguity for all operators of twist $2$.
\item The knowledge of the number of protected operators in a given representation gives an useful equality that can be used to fix the ambiguity~\cite{Doobary:2015gia}. Assuming for example that there is only one protected operator in a given representation $\gamma \equiv (\Delta,\ell,[p,q,k])$. Then the following equality between OPE coefficients must hold
\eqn{
c^{(p_1,p_1,p_1,p_1)}_\gamma\,c^{(p_2,p_2,p_2,p_2)}_\gamma = \big(c^{(p_1,p_1,p_2,p_2)}_\gamma\big)^2\,,
}[]
simply because $c^{(p_1,p_1,p_2,p_2)}_\gamma = \lambda_{p_1p_1\COp\gamma}\lambda_{p_2p_2\COp\gamma}$. This identity with $p_1 = 2,\,p_2=3,\gamma = (\ell+4,\ell,[n-m,2m,n-m]), \,n>0$ can be used to fix the contributions of twist $4$ in $\CH^{(3)}$.
\end{enumerate}
The correlator $\CH^{(2,3,2,3)}$ contain multiplets of twist $3$ which are not fixed by the two criteria. We will thus not make any assumption on their OPE coefficient other than its positivity. The contribution from operators at threshold reads
\twoseqn{
\CH^{(2)} &\supset \sum_{\ell \geq 0\,\mathrm{even}} A_{1,\ell}\,\CG_{\ell+2}^{(\ell)}(u,v)\,, \qquad \mbox{Similarly for $(2,2,3,3)$ and $(2,3,2,3)$}\,,
}[]
{
\CH^{(3)} &\supset \sum_{0\leq m \leq n \leq 1}\sum_{\substack{\ell \geq 0 \\ \ell \equiv n+m \,\mathrm{mod}\,2}} A_{n+1,\ell}^{[n,m]}\,Y_{n,m}(\sigma,\tau)\,\CG_{2n+\ell+2}^{(\ell)}(u,v)\,.
}[]
And the coefficients are given by
\eqna{
A_{1,\ell} &= 2^\ell b_{0,\ell+1}\;,\\
A^{(2,3,2,3)}_{1,\ell} &= \tfrac23 2^\ell b^{(2,3,2,3)}_{\ell+1}\;,\\
A^{(2,2,3,3)}_{1,\ell} &=2^\ell b^{(2,2,3,3)}_{0,\ell+1}\;,\\
A^{[0,0]}_{1,\ell} &= 2^\ell b^{(3)}_{0,\ell+1}\;,\\
A^{[1,0]}_{2,\ell} &= -\lifrac{1}{3}2^\ell b^{(3)}_{0,\ell+2} + \tilde{A}^{[1,0]}_{\ell}\;,\\
A^{[1,1]}_{2,\ell} &= -\lifrac{1}{24}A^{[0,0]}_{0,\ell+2} + \tilde{A}^{[1,1]}_{\ell}\,.
}[eq:thresholdA]
where the $b_{n,\ell}$ coefficients are given in \tablename~\ref{tab:bnlcoeffs} and
\eqna{
\tilde{A}^{[1,0]}_{\ell} &= \frac{2^\ell}3\frac{\big(b^{(2,2,3,3)}_{0,\ell+2}\big)^2}{b^{(2)}_{\ell+2}} \,,\\
\tilde{A}^{[1,1]}_{\ell} &= \frac1{24} \frac{\big(A^{(2,2,3,3)}_{0,\ell+2}\big)^2}{A^{(2)}_{0,\ell+2}}\,.
}[]
The precise recombination rules that take place are written in \tablename~\ref{tab:recombination}. In the last two lines of \eqref{eq:thresholdA} the recombination rules would have set $\tilde{A}_\ell^{[n,m]}$ to zero so this is an example of a dynamically protected multiplet.

\renewcommand{\arraystretch}{1.4}
\begin{table}[t]
\centering
\begin{tabular}{|r|ll|}
\hline
Coeff. & Decomposition used & Multiplet associated \\
\Hhline
$A_{0,\ell}$ & $\mathcal{A}_{[0,0,0]\ell}^{\ell} \cong \mathcal{D}_{[0,0,0]\ell} \oplus \mathcal{C}_{[0,2,0]\ell-2}$ & $\mathcal{C}_{[0,2,0]\ell-2}$\quad ($\mathcal{B}_{[0,2,0]} \oplus \mathcal{B}_{[0,4,0]}$ if $\ell=0$) \\
$A_{1,\ell}$ & $\mathcal{A}_{[0,0,0]\ell}^{\ell+2} \cong \mathcal{C}_{[0,0,0]\ell} \oplus \mathcal{C}_{[1,0,1]\ell-1}$ & $\mathcal{C}_{[1,0,1]\ell-1}$\quad ($\mathcal{B}_{[2,0,2]}$ if $\ell=0$)\\
\hline
$A^{[0,0]}_{0,\ell}$ &$\mathcal{A}_{[0,0,0]\ell}^{\ell} \cong \mathcal{D}_{[0,0,0]\ell} \oplus \mathcal{C}_{[0,2,0]\ell-2}$ &$\mathcal{C}_{[0,2,0]\ell-2}$\quad ($\mathcal{B}_{[0,2,0]} \oplus \mathcal{B}_{[0,4,0]}$ if $\ell=0$) \\
$A^{[1,0]}_{0,\ell}$ &$\mathcal{A}_{[1,0,1]\ell}^{\ell} \cong \mathcal{D}_{[1,0,1]\ell} \oplus \mathcal{D}_{[0,2,0]\ell-1}$&\multirow{2}{*}{$\hspace{-.44cm}\left.\raisebox{.75cm}{}\right\rbrace$\hspace{-.07cm} $\mathcal{C}_{[0,4,0]\ell-2}$\quad ($\mathcal{B}_{[0,4,0]}\oplus \mathcal{B}_{[0,6,0]}$ if $\ell =0$)}\\
$A^{[1,1]}_{1,\ell}$ &$\mathcal{A}_{[0,2,0]\ell}^{\ell+2} \cong \mathcal{D}_{[0,2,0]\ell} \oplus \mathcal{C}_{[0,4,0]\ell-2}$&\\
$A^{[0,0]}_{1,\ell}$ &$\mathcal{A}_{[0,0,0]\ell}^{\ell+2} \cong \mathcal{C}_{[0,0,0]\ell} \oplus \mathcal{C}_{[1,0,1]\ell-1}$&$\mathcal{C}_{[1,0,1]\ell-1}$\quad($\mathcal{B}_{[2,0,2]}$ if $\ell=0$)\\
$A^{[1,0]}_{2,\ell}$ &$\mathcal{A}_{[1,0,1]\ell}^{\ell+4} \cong \mathcal{C}_{[1,0,1]\ell} \oplus \mathcal{C}_{[2,0,2]\ell-1}$&$\mathcal{C}_{[1,0,1]\ell} \oplus (\mathcal{C}_{[2,0,2]\ell-1}$\;or $\mathcal{B}_{[3,0,3]}$ if $\ell=0$)\\
$A^{[1,1]}_{2,\ell}$ & $\mathcal{A}_{[0,2,0]\ell}^{\ell+4} \cong \mathcal{C}_{[0,2,0]\ell} \oplus \mathcal{C}_{[1,2,1]\ell-1}$&$\mathcal{C}_{[0,2,0]\ell} \oplus (\mathcal{C}_{[1,2,1]\ell-1}$\;or $\mathcal{B}_{[2,2,2]}$ if $\ell=0$)\\
\hline
$A^{(2,3,2,3)}_{0,\ell}$&$\mathcal{A}_{[0,1,0]\ell}^{\ell+1} \cong \mathcal{D}_{[0,1,0]\ell} \oplus \mathcal{C}_{[0,3,0]\ell-2}$ & $\mathcal{C}_{[0,3,0]\ell-2}$\quad ($\mathcal{B}_{[0,3,0]} \oplus \mathcal{B}_{[0,5,0]}$ if $\ell=0$)\\
$A^{(2,3,2,3)}_{1,\ell}$&$\mathcal{A}_{[0,1,0]\ell}^{\ell+3} \cong \mathcal{C}_{[0,1,0]\ell} \oplus \mathcal{C}_{[1,1,1]\ell-1}$&$\mathcal{C}_{[1,1,1]\ell-1}$\quad($\mathcal{B}_{[2,1,2]}$ if $\ell=0$)\\
$A^{(2,2,3,3)}_{0,\ell}$&$\mathcal{A}_{[0,0,0]\ell}^{\ell} \cong \mathcal{D}_{[0,0,0]\ell} \oplus \mathcal{C}_{[0,2,0]\ell-2}$ & $\mathcal{C}_{[0,2,0]\ell-2}$\quad ($\mathcal{B}_{[0,2,0]} \oplus \mathcal{B}_{[0,4,0]}$ if $\ell=0$)\\
$A^{(2,2,3,3)}_{1,\ell}$& $\mathcal{A}_{[0,0,0]\ell}^{\ell+2} \cong \mathcal{C}_{[0,0,0]\ell} \oplus \mathcal{C}_{[1,0,1]\ell-1}$ & $\mathcal{C}_{[1,0,1]\ell-1}$\quad ($\mathcal{B}_{[2,0,2]}$ if $\ell=0$)\\
\hline
\end{tabular}

\caption{Summary of semishort multiplets arising from the recombination described in the previous paragraphs. Now the notation $[n,m] \equiv [n-m,2m+p_{43},n-m]$ has been expanded in order to avoid confusion. These results can be found in \cite{Dolan:2002zh} and Appendix D of \cite{Nirschl:2004pa}.\label{tab:recombination}}
\end{table}
\renewcommand{\arraystretch}{1}

\section{Numerical implementation}
\label{sec:crossing}

\subsection{Crossing equations}
The goal of this section is to write down the crossing equations for the set of correlators
\eqn{
\langle \COp2\COp2\COp2\COp2\rangle\,,\quad
\langle \COp2\COp2\COp3\COp3\rangle\,,\quad
\langle \COp2\COp3\COp2\COp3\rangle\,,\quad
\langle \COp3\COp3\COp3\COp3\rangle\,.\quad
}[]
This will translate then in a set of equations for the unprotected part of the correlators $\CH$. In this section we will use ``$\lnsp\cro$'' as a shorthand for $(2,3,2,3)$ and ``$\llsp\uncr\lsp$'' as a shorthand for $(2,2,3,3)$. In order to fully exploit the information about supersymmetry it is necessary to separate the contributions of long and of short multiplets. The latter will be given by the coefficients computed in the previous section while the former will appear with arbitrary coefficients $a_{\Delta,\ell}$. Concretely one can write\footnote{Note $Y_{00}^{(0,1)}=\frac{3}{2}$.}
\fourseqn{
\mathcal{H}^{(2)}(u,v) &= \sum_{\Delta,\ell} a_{\Delta,\ell}\, \mathcal{G}_{\Delta}^{(\ell)}(u,v) + \sum_{\substack{ \ell \geq 0\;\mathrm{even}\\t=0,1}} A_{t,\ell}\, \mathcal{G}_{\ell+2t}^{(\ell)}(u,v)\,,
}[]{
\mathcal{H}^{\uncr}(u,v) &= \sum_{\Delta,\ell} a^{\uncr}_{\Delta,\ell}\, \mathcal{G}_{\Delta}^{(\ell)}(u,v) + \sum_{\substack{ \ell \geq 0\;\mathrm{even}\\t=0,1}} A^{\uncr}_{t,\ell} \mathcal{G}_{\ell+2t}^{(\ell)}(u,v)\;,
}[]{
\mathcal{H}^{\cro}(u,v) &= \sum_{\Delta,\ell} a^{\cro}_{\Delta,\ell}\,Y_{00}^{(0,1)}\, \mathcal{G}_{\Delta}^{(\ell)}(u,v;1,1) + \sum_{\substack{\ell \geq 0\\t=0,1}} A^{\cro}_{t,\ell}\,Y_{00}^{(0,1)} \, \mathcal{G}_{\ell+2t+1}^{(\ell)}(u,v;1,1)\,,
}[]{
\mathcal{H}^{(3)}(u,v;\sigma,\tau) &= \sum_{\Delta,\ell}\sum_{0\leq m\leq n\leq 1} a^{[n,m]}_{\Delta,\ell}\,Y_{nm}(\sigma,\tau)\, \mathcal{G}_{\Delta}^{(\ell)}(u,v) + 
\notag\\&+\sum_{\substack{ \ell \geq 0\;\mathrm{even}\\t=0,1}} A^{[0,0]}_{t,\ell}\,Y_{00}(\sigma,\tau) \,\mathcal{G}_{\ell+2t}^{(\ell)}(u,v)
\notag\\&+
\sum_{\substack{ \ell > 0\;\mathrm{odd}\\t=0,2}} A^{[1,0]}_{t,\ell}\,Y_{10}(\sigma,\tau) \,\mathcal{G}_{\ell+2t}^{(\ell)}(u,v)
\notag\\&+
\sum_{\substack{ \ell \geq 0\;\mathrm{even}\\t=1,2}} A^{[1,1]}_{t,\ell}\,Y_{11}(\sigma,\tau) \,\mathcal{G}_{\ell+2t}^{(\ell)}(u,v)
\,.
}[]

The crossing equations for $\mathcal{H}$ are not be homogeneous due to the transformation property of $\mathcal{F}_{\hat{f}}$. In order to take this into account it will be convenient to define the following quantities
\fourseqn{
\Delta_{(2)}(u,v) &= (u-v)(a_2+a_1(u+v))\;,}[]{
\Delta_{\uncr}(u,v) &= c_1-c_2+\frac{u^2-1}{v^2}c_2+\frac{u c_3 -c_4}{v}\;,}[]{
\Delta_{\cro}(u,v) &= (u-v)(c_3+c_1(u+v))\;,}[]{
\Delta_{(3)}(u,v;\sigma,\tau) &=
b_3 \left(\tau  u^2-v^2\right)+b_1 \big(-\sigma  u^4-\tau  u^4+u^4+\sigma  u^3+2 \tau  u^3+2 \sigma  u^3 v+\tau  u^3 v \nonumber\\\nonumber&\quad  -u^3 v-u^3-2 \sigma  u v^3+\tau  u v^3-u v^3+\sigma  v^4-\tau  v^4+v^4-\sigma  v^3+\tau  v^3-2 v^3\big) \\\nonumber&\quad +b_2 \big(-\sigma  u^3+\tau  u^3+u^3+2 \tau  u^2+\sigma  u^2 v+\sigma  u+\tau  u-\sigma  u v^2-\tau  u v+u v \\&\quad-u+\sigma  v^3-\tau  v^3-v^3-2 v^2-\sigma  v+\tau  v-v\big)
\,.
}[][]
The following combinations of conformal blocks will appear in the crossing equations:
\eqna{
F^{(p)}_{\Delta,\ell}(u,v) &= u^p \mathcal{G}_{\Delta}^{(\ell)}(v,u) - 
v^p \mathcal{G}_{\Delta}^{(\ell)}(u,v)\,,\\
H^{(p)}_{\Delta,\ell}(u,v) &= u^p \mathcal{G}_{\Delta}^{(\ell)}(v,u) +
v^p \mathcal{G}_{\Delta}^{(\ell)}(u,v)\,, \\
I_{\Delta,\ell}(u,v) &= u^2 \mathcal{G}_{\Delta}^{(\ell)}(v,u;1,1) - 
v^2 \mathcal{G}_{\Delta}^{(\ell)}(u,v;1,1)\,,\\
F^{\cro,\pm}_{\Delta,\ell}(u,v) &= (-1)^{\ell} \left(u^2 \mathcal{G}_{\Delta}^{(\ell)}(v,u;1,-1) \pm v^2 \mathcal{G}_{\Delta}^{(\ell)}(u,v;1,-1)\right)\,,\\
F^{\uncr,+}_{\Delta,\ell}(u,v) &=(-1)^\ell H^{(2)}_{\Delta,\ell}(u,v)\,,\\
F^{\uncr,-}_{\Delta,\ell}(u,v) &=(-1)^\ell F^{(2)}_{\Delta,\ell}(u,v)\,.
}[]
In the $\langle 3333 \rangle$ case we have a $\sigma,\tau$ dependence. This means that we should first project onto $\SU(4)$ structures (i.e. the $Y_{nm}$ polynomials). Crossing acts linearly on them in the following way
\eqn{
\tau \mathcal{Y}_a\left(\lifrac{\sigma}{\tau},\lifrac{1}{\tau}\right) = M_{ab}\, \mathcal{Y}_b(\sigma,\tau)\;,\qquad \mbox{where}\quad\vec{\mathcal{Y}}= \{ Y_{00},Y_{10},Y_{11}\}\,.
}[]
The eigenvalues of $M_{ab}$ are $1$,$1$ and $-1$. The matrix that diagonalizes it is
\eqn{
P M^T P^{-1} = \mathrm{diag}(1,1,-1)\;,\qquad 
P =\left(\begin{array}{ccc}
-1 & 0 & -5 \\
0 & -1 & -1 \\
-1 & -3 & 4
\end{array}
\right)\,.
}[]
We also define the vector $\vec{\Delta}_{(3)}$ as $\Delta_{(3)} = \vec{\Delta}_{(3)} \cdot \vec{\mathcal{Y}}$. 
Let us first write the vectors associated to the short multiplets.
\fourseqn{
F^{(2)}_{\mathrm{short}}(u,v) &= \Delta_{(2)}(u,v) - \sum_{\substack{\ell\geq 0 \;\mathrm{even}\\t=0,1}} A_{t,\ell} F^{(2)}_{2t+\ell,\ell}(u,v)\,,
}[eq:2222]{
&\begin{aligned}\hspace{-2cm}\vec{F}^{(3)}_\mathrm{short}(u,v) &= 
P \cdot \vec{\Delta}_{(3)}-
\sum_{\substack{t=0,1\\ \ell\;\mathrm{even}}} A_{t,\ell}^{[0,0]}\left(\begin{array}{c}
- F^{(3)}_{2t+\ell,\ell} \\ 0 \\ H^{(3)}_{2t+\ell,\ell}
\end{array}\right)-
\sum_{\substack{t=0,2\\ \ell\;\mathrm{odd}}} A_{t,\ell}^{[1,0]}\left(\begin{array}{c}
0 \\ -F^{(3)}_{2t+\ell,\ell} \\ 3H^{(3)}_{2t+\ell,\ell}
\end{array}\right)\\&\quad-
\sum_{\substack{t=1,2\\ \ell\;\mathrm{even}}} A_{t,\ell}^{[1,1]}\left(\begin{array}{c}
-5 F^{(3)}_{2t+\ell,\ell} \\ -F^{(3)}_{2t+\ell,\ell} \\ -4H^{(3)}_{2t+\ell,\ell}
\end{array}\right)\,,
\end{aligned}
}[eq:3333]
{
I_{\mathrm{short}}(u,v) &=\Delta_{\cro}(u,v) - \sum_{\substack{t=0,1\\\ell\geq 0}} A^\cro_{t,\ell} I_{2t+\ell+1,\ell}(u,v) \,,
}[eq:cro]{
F^{\pm}_{\mathrm{short}} &= \Delta_{\uncr}\left(\lifrac{u}{v},\lifrac{1}{v}\right) \pm \Delta_{\uncr}\left(\lifrac{v}{u},\lifrac{1}{u}\right)-\sum_{\ell\geq 0} A_{0,\ell}^{\cro} F_{\ell+1,\ell}^{\cro,\pm} \pm \sum_{\substack{t=0,1\\ \ell\; \mathrm{even}}} A_{t,\ell}^{\uncr} F_{2t+\ell,\ell}^{\uncr,\pm}\,.
}[eq:uncr]
Now we can finally state the seven crossing equations in the following compact vector form.\footnote{We call $\lambda_{pp(\Delta,\ell)}$ the coefficient of $\COp{\Delta,\ell,[0,0,0]}$ in the OPE $\COp{p} \times \COp{p}$. We still allow for degeneracies in the spectrum, but we did not take this into account in \eqref{eq:crosseqfinal} simply for notational convenience.}
\eqna{
&
\sum_{\substack{\Delta\geq\ell+2\\\ell\;\mathrm{even}}} \begin{array}{cc}
\Big[\lambda_{22(\Delta,\ell)} & \lambda_{33(\Delta,\ell)}\Big]\\ \ \end{array}\vec{\mathcal{V}}_{\Delta,\ell}\left[\begin{array}{c}
\lambda_{22(\Delta,\ell)} \\ \lambda_{33(\Delta,\ell)}
\end{array}\right] +\sum_{\Delta\geq\ell+2,\ell} a_{\Delta,\ell}^{\cro}\,\vec{V}_{\Delta,\ell;\cro}+\\&+\sum_{\substack{\Delta\geq\ell+4\\\ell\;\mathrm{odd}}} a_{\Delta,\ell}^{[1,0]}\,\vec{V}_{\Delta,\ell;[1,0]}+\sum_{\substack{\Delta\geq\ell+4\\\ell\;\mathrm{even}}}a_{\Delta,\ell}^{[1,1]}\,\vec{V}_{\Delta,\ell;[1,1]} = \vec{V}_{\mathrm{short}}\,. 
}[eq:crosseqfinal]
In the above equation all vectors are seven dimensional and $\vec{\mathcal{V}}_{\Delta,\ell}$ is a vector of $2\times2$ matrices. The coefficients $a_{\Delta,\ell}^\cro,\,a_{\Delta,\ell}^{[m,n]}$ are positive. The explicit definitions of the vectors $\vec{V}$ and $\vec{\mathcal{V}}$ is
\eqn{
\begin{aligned}
\vec{\mathcal{V}}_{\Delta,\ell} &= \left[\begin{array}{cc}
\left(\begin{array}{c}
F_{\Delta,\ell}^{(2)}\\ 0\vspace{-5pt}\\\vdots\vspace{-2pt}\\0
\end{array}\right) & 
\left(\begin{array}{c}
0\\-\lifrac{1}{2} F_{\Delta,\ell}^{\uncr,+}\\\lifrac{1}{2} F_{\Delta,\ell}^{\uncr,-}
\\0\vspace{-5pt}\\\vdots\vspace{-2pt}\\0
\end{array}\right) \\
\left(\begin{array}{c}
0\\-\lifrac{1}{2} F_{\Delta,\ell}^{\uncr,+}\\\lifrac{1}{2} F_{\Delta,\ell}^{\uncr,-}
\\0\vspace{-5pt}\\\vdots\vspace{-2pt}\\0
\end{array}\right) &
\left(\begin{array}{c}
0\vspace{-5pt}\\\vdots\vspace{-2pt}\\0\\-F_{\Delta,\ell}^{(3)}\\0\\H_{\Delta,\ell}^{(3)}
\end{array}\right)
\end{array}\right]
\end{aligned}\,,
\begin{aligned}
\vec{V}_{\Delta,\ell;\cro} = \left(
\begin{array}{c}
0\\\lifrac{1}{2} F_{\Delta,\ell}^{\cro,+}\\\lifrac{1}{2} F_{\Delta,\ell}^{\cro,-}\\ I_{\Delta,\ell} \\0\\0\\0
\end{array}
\right)\,,\\
\vec{V}_{\Delta,\ell;[1,0]} =\left(
\begin{array}{c}
0\vspace{-5pt}\\\vdots\vspace{-2pt}\\0\\ -F^{(3)}_{\Delta,\ell} \\ 3H^{(3)}_{\Delta,\ell}
\end{array}
\right)\,,\\
\vec{V}_{\Delta,\ell;[1,1]} =\left(
\begin{array}{c}
0\vspace{-5pt}\\\vdots\vspace{-2pt}\\0\\-5 F^{(3)}_{\Delta,\ell} \\ -F^{(3)}_{\Delta,\ell} \\ -4H^{(3)}_{\Delta,\ell}
\end{array}
\right)\,,\\
\end{aligned}
}[]
\eqn{
\vec{V}_{\mathrm{short}} = \left(F_{\mathrm{short}}^{(2)},	\,F_{\mathrm{short}}^{+}\,F_{\mathrm{short}}^{-},\,I_{\mathrm{short}},\,F^{(3)}_{\mathrm{short},1} ,\,F^{(3)}_{\mathrm{short},2} ,F^{(3)}_{\mathrm{short},3} \right)^T\,.
}[eq:vshort]

\subsection{Numerical bootstrap techniques}
The crossing equations in \eqref{eq:crosseqfinal} may be studied by means of a numerical method called conformal bootstrap~\cite{Rattazzi:2008pe, Poland:2018epd}. We will briefly review it in this subsection. The following general introduction is not specific to supersymmetry and, in fact, applies to all conformal field theories.
\par
The first step is to discretize the equations. This is done by Taylor expanding the functions $F(z,\zb)$ that appear in the crossing equations around a fixed point, retaining only a finite number of coefficients. This still leaves an infinite sum over $\Delta$ and $\ell$. The sum is thus truncated to a maximal value of the spin \texttt{Lmax} and the Taylor coefficients $(\partial_z^n\partial_\zb^m F(z,\zb))\big|_{z=\zb=\frac12}$ are approximated by rational functions of $\Delta$~\cite{Poland:2011ey}. The details of the approximation are given in Appendix~\ref{app:numerical}. After the truncation we end up with a system of equation of the form\footnote{$\lambda_{\rho}$ represents the OPE coefficient of the representation $\rho$. E.g. $\lambda_{\Delta,\ell,[0,1,0]} = \sqrt{a^\cro_{\Delta,\ell}}$ and $\lambda_{\Delta,\ell,[0,0,0]}$ is the 2-vector $(\lambda_{22(\Delta,\ell)},\lambda_{33(\Delta,\ell)})$.}
\eqn{
\sum_\rho \sum_{i,j}(\lambda_\rho)_i\lsp(\mathcal{V}_{\rho,I})_{i,j}\lsp (\lambda_\rho)_j = V_I\,,\qquad	\forall\;I\,.
}[eq:truncated]
The index $\rho = (\Delta,\ell,R)$ runs over the finite sum over spins and R-symmetry representations and over the infinite sum over conformal dimensions. The index $I$ runs over all crossing equations (in our case seven) times the number of Taylor coefficients kept in the discretization of $F(z,\zb)$. Finally, the indices $i$ and $j$ run over the the space of independent OPE coefficients in the same representation. In our case they take either two values or just one. The right hand side of the equation is obtained from the vector $V_\mathrm{short}$ \eqref{eq:vshort}. The strategy of the conformal bootstrap is to try and to rule out possible candidate solutions to \eqref{eq:crosseqfinal} by showing that they lead to a contradiction. One typically starts from a set of assumptions that depend on a small number of parameters and tries to ``carve out'' regions in parameter space. The contradictions may be found with the following general strategy: suppose that there exists a linear functional $\alpha_I$ such that
\eqna{
\sum_I\alpha_I \,\mathcal{V}_{\rho,I} &\succeq 0\,,\qquad \forall\;\rho \in \{\mathrm{assumptions}\}\,,\\
\sum_I\alpha_I \,V_I &= -1\,.
}[]
This looks like an infinite set of constraints because we have not restricted the conformal dimensions $\Delta$ to a finite set, therefore $\rho$ may assume a continuum of values. But, in fact, the functions that appear in $\mathcal{V}$ are rational and their denominator is known. So one simply has to impose positivity on a polynomial in $\Delta$ over some interval, typically of the form $[\Delta_0,\infty)$. This can be done rigorously with a computer. If there exists such an $\alpha$, then, by contracting $\alpha_I$ with \eqref{eq:truncated}, one would obtain $0 \leq \sum_\rho \lambda_\rho^T(\alpha\cdot \mathcal{V}_\rho)\lambda_\rho =  -1$, namely a contradiction. The problem now consists in implementing a numerical algorithm that searches for a functional $\alpha$ with the property described above. If such a functional is found, then the assumptions made are inconsistent. To this end, we use the program \texttt{sdpb}\footnote{\href{https://github.com/davidsd/sdpb}{\texttt{github.com/davidsd/sdpb}}}~\cite{Simmons-Duffin:2015qma} which is a semidefinite program solver optimized for the conformal bootstrap. The parameters used and various details of the numerical implementation are given in Appendix~\ref{app:numerical}.
\par
According to the kind of assumptions that we make, the bootstrap problems that we need to solve can be quite different. We will now explain the various approaches used to make the plots that will be presented in the next section. First we start with the assumptions of the type
\begin{quote}\it
Given a fixed value of $N$, all operators in a representation $(\Delta,\ell,R)$ have conformal dimension $\Delta \geq \Delta_{\ell,R}$.
\end{quote}
Where $\Delta_{\ell,R}$ is a function that we can choose. In the most common case it is
\eqna{
\Delta_{\ell^*,R^*} &= \Delta^*\,,\\ \Delta_{\ell,R} &= \Delta^{\mathrm{unitarity}}_{\ell,R}\,,\qquad \forall\;\ell,R \neq \ell^*,R^*\,,
}[]
for a chosen representation $\ell^*,R^*$ and a real number $\Delta^*$. We defined $\Delta^{\mathrm{unitarity}}_{\ell,R}$ as the unitarity bound for the representation with spin $\ell$ and R-symmetry $R$.\footnote{Since we only consider unitary theories, the assumption $\Delta \geq \Delta^{\mathrm{unitarity}}$ is always the default one.} The goal in this kind of problem is to find an upper bound on $\Delta_{\ell,R}$, as a function of $N$. This can be done by a simple binary search. Namely one fixes two values $\Delta^\mathrm{max}$ and $\Delta^\mathrm{min}$, which are respectively disallowed and allowed. Then, iteratively, the interval is divided in half and either $\Delta^\mathrm{max}$ or $\Delta^\mathrm{min}$ is updated according to whether the middle point results allowed or disallowed by the bootstrap.
\par
Next, we would like to discuss a slightly different setup that allows us to find upper and lower bounds on OPE coefficients without having to make multiple runs of the semidefinite solver~\cite{Caracciolo:2009bx,Poland:2011ey}. Indeed \texttt{sdpb} can also maximize a given objective vector, subject to certain semidefinite positiveness conditions. Suppose we are interested in the OPE coefficients of a certain operator in the representation $\rho^* = (\Delta^*,\ell^*,R^*)$. The crossing equations can be rewritten to isolate the contribution of that operator
\eqn{
|\lambda_{\rho^*}|^2\,\hat{n}_\theta^T\cdot\vec{\mathcal{V}}_{\rho^*} \cdot \hat{n}_\theta + \sum_{\Delta,\ell,R \neq \rho^*} \lambda_\rho^T \cdot\vec{\mathcal{V}}_\rho \cdot \lambda_\rho = \vec{V}_\mathrm{short}\,.
}[]
If $R^*$ is not the singlet, then the unit vectors are trivial (i.e. $\hat{n}=1$).
Then we search for functionals $\alpha$ satisfying
\eqna{
\alpha[\hat{n}_\theta^T\cdot\vec{\mathcal{V}}_{\rho^*} \cdot \hat{n}_\theta] &= \pm 1 \equiv s\,,\\
\alpha[\vec{\mathcal{V}}_\rho] &\succeq 0\,,\quad\;\qquad\forall\;\rho \neq \rho^*\,,\\
\alpha[\vec{V}_\mathrm{short}] &= B\,,
}[]
and minimize the value of $B$. This yields an upper or lower bound on $|\lambda_{\rho^*}|^2$, namely 
\eqna{
|\lambda_{\rho^*}|^2 &\leq B \,,\qquad	&\mbox{if $s = 1$}\,,\\
|\lambda_{\rho^*}|^2 &\geq -B \,,\qquad	&\mbox{if $s = -1$}\,.
}[]
\par Finally we briefly introduce the extremal functional method~\cite{ElShowk:2012hu, Simmons-Duffin:2016wlq}. It is a technique that can be used to extract an approximation of the spectrum (conformal dimensions and OPE coefficients) of the theory that lives at the boundary of an allowed region in parameter space. First notice that the functional $\alpha$ takes its normalization from the condition $\alpha[V_{\mathrm{short}}]=1$. We can relax this to $\alpha[V_{\mathrm{short}}]> 0$ while still getting a contradiction when $\alpha$ is found. The boundary of this region in functional space is given by all $\alpha_\partial$ that satisfy
\eqn{
\alpha_\partial[\vec{\mathcal{V}}_\rho] \succeq 0\quad \forall\;\rho\,,\qquad \alpha_\partial[\vec{V}_{\mathrm{short}}] =  0\,.
}[]
This implies that the only terms that can contribute to the sum of a consistent theory at the boundary must satisfy
\eqn{
\alpha_\partial[\hat{n}_\theta^T\cdot\vec{\mathcal{V}}_\rho\cdot \hat{n}_\theta] = 0\,.
}[]
For simplicity, let us consider only the simpler case where all $\mathcal{V}_\rho$ are one-by-one matrices. The equality above together with the positivity constraint implies that the function $f_{\ell,R}$ defined by
\eqn{
f_{\ell,R}(\Delta) \equiv \alpha_\partial[\vec{\mathcal{V}}_{\Delta,\ell,R}]\,,
}[]
has even order zeros (typically double zeros) only on those values of $\Delta$ that belong to the physical spectrum of operators with spin $\ell$ and R-charge $R$. Naturally in a numerical computation we will only find a finite number of such zeros. Then, in order to find the OPE coefficients, we can truncate the crossing equation to that finite number of operators and solve the linear system of equations
\eqn{
\sum_{\Delta,\ell,R \in \{f_{\ell,R}(\Delta\hspace{-.1pt}) = 0\}}\vec{\mathcal{V}}_{(\Delta,\ell,R)} a_{(\Delta,\ell,R)} = \vec{V}_{\mathrm{short}}\,,
}[]
with $a_\rho = \lambda_\rho^2$. If we find a forbidden point in parameter space very close to the allowed region,\footnote{This can be done by either running a binary search to very high accuracy or by extremizing an OPE coefficient with the methods explained before.} the functional $\alpha$ that excludes it will be a very good approximation of $\alpha_\partial$. We can therefore use it to define the function $f_{\ell,R}(\Delta)$ and extract the spectrum. For the results to be meaningful one needs to both be very close to the boundary and to observe that the position of the boundary remains stable if the size of the numerics is increased.

\section{Known analytic results}
\label{sec:knownanalytic}

In this section we summarize some results for OPE coefficients and conformal dimensions at large $N$ obtained through the supergravity description of $\CN=4$ SYM. In the large $N$ limit the operators on the OPE $\COp{p_1}\times \COp{p_2}$ organize in double trace towers $[\COp{p_1}\COp{p_2}]_{n,\ell}$ whose conformal dimensions are $p_1+p_2+2n+\ell$ plus $1/N$ corrections. It should be stressed that, strictly speaking, the operators $[\COp{p_1}\COp{p_2}]_{n,\ell}$ are not simply given by $\COp{p_1}\partial^\ell\square^{n}\COp{p_2}$. Instead, at each order in the $1/N$ expansion, the operators with the same twist mix with each other. For example, $[\COp2\COp2]_{n,\ell}$ will be a linear combination of
\eqn{
\{\COp2\partial^\ell\square^{n}\COp2,\, \COp3\partial^\ell\square^{n-1}\COp3,\ldots, \COp{n+2} \partial^\ell \COp{n+2}\}\,.
}[]
Solving this mixing is important for deriving the higher order corrections to the anomalous dimension. This problem is addressed in~\cite{Alday:2017xua, Aprile:2017xsp,Aprile:2018efk,Aprile:2017qoy} and in the references specified later in this section. Here we will report only the averaged anomalous dimensions
\eqn{
\langle\gamma_\CO\rangle = \frac{1}{\sum_{\CO'} a_{\CO'}}\sum_{\CO'} a_{\CO'}\gamma_{\CO'}\,,
}
where the sum ranges over the operators $\CO'$ that have the same quantum numbers as $\CO$ at tree level.  We will omit the angle brackets when the operator is non degenerate.

\par
We first show the results in the supergravity limit, namely $\lambda=\infty$ and $N$ large. Then we present some corrections in $1/\lambda$ for the operator $[\COp2\COp2]$ at lowest twist.

\subsection[Double traces \texorpdfstring{$[\COp2\COp2]$}{[O2O2]}]{Double traces $\boldsymbol{[\COp2\COp2]}$}
Let us start by showing the results at first order	 in $1/N^2$ and at $\lambda$ strictly infinity. The anomalous dimensions at order $1/N^2$ of the long singlet operators with twist $2t$ and spin $\ell$ in the OPE $\COp2\times\COp2$ have been computed in~\cite{Dolan:2001tt}.\footnote{For $t=2$ the results first appeared in~\cite{DHoker:1999mic,Hoffmann:2000dx,Arutyunov:2000ku}.} For $t=2,3,\ldots$ and $\ell$ even one has
\eqn{
\langle\gamma_{t,\ell}\rangle =-\frac{4 (t-1) t (t+1) (t+2)}{(\ell+1) (\ell+2 t+2)}\frac{1}{N^2}
\,,
}[gammatl]
where the conformal dimension is given by $\Delta_{t,\ell} = 2t+\ell+\gamma_{t,\ell}$. The anomalous dimensions of the operators of twist four and spin $\ell\geq 2$ are available also up to order $1/N^4$. They were obtained independently in~\cite{Alday:2017xua, Aprile:2017bgs}
\eqn{
\gamma_{t=2,\ell}\big|_{1/N^4} = \left(\frac{2688 (\ell-7) (\ell+14)}{(\ell-1) (\ell+1)^2 (\ell+6)^2 (\ell+8)}-\frac{4608 (2 \ell+7)}{(\ell+1)^3 (\ell+6)^3}- \frac{96}{(\ell+1) (\ell+6)}\right)\frac1{N^4}\,.
}[gammal]
The OPE coefficients\footnote{The coefficients $a_{\Delta,\ell}$ are defined as in \supercpwex, ommitting the implied $[0,0]$ superscript in this case.} at order $1/N^0$, for $t=2,3,\ldots$ and $\ell$ even, are given by
\eqn{
a_{2t+\ell,\ell} =\frac{2^{\ell+1}  (t!)^2  ((\ell+t+1)!)^2}{(2 t)! (2 \ell+2t+2)!}(\ell+1)(\ell+2 t+2)\,,
}[opetwo]
while for $t=1$ at order $1/N^2$ we have
\eqn{
a_{\ell+2,\ell} = \frac{2^{\ell} ((\ell+2)!)^2 }{(2 \ell+4)!}\left((\ell+1) (\ell+4)-\frac{12}{N^2} \right) = A_{1,\ell}\,.
}[]
The OPE coefficient for $t=1$ is precisely the one described in \eqref{eq:thresholdA} and thus it cancels exactly the contribution from the short $\CC_{[0,0,0]\ell}$ multiplets. This means that at this order only the long multiplets with twist at least four gain anomalous dimension, whereas the ones of twist two are absent.
\subsection[Double traces \texorpdfstring{$[\COp2\COp3]$}{[O2O3]}]{Double traces $\boldsymbol{[\COp2\COp3]}$}

The double trace operators in the OPE $\COp2\times\COp3$ were studied in~\cite{Aprile:2017qoy}. The anomalous dimensions at order $1/N^2$ for all $\ell$ and $t=2,3,\ldots$ read
\eqn{
\big\langle\gamma_{t,\ell}^{(23)}\big\rangle = \left\lbrace
\begin{aligned}
&-\frac{4 (t-1) t (t+2) (t+3)}{(\ell+1) (\ell+t+2)}\frac{1}{N^2}&\qquad&\mbox{$\ell$ even}\,,\\
&-\frac{4 (t-1) t (t+2) (t+3)}{(\ell+t+2) (\ell+2 t+3)}\frac{1}{N^2}&\qquad&\mbox{$\ell$ odd}\,,
\end{aligned}
\right.
}[]
where now the dimension is given by $\Delta_{t,\ell} = 2t+\ell+1+\gamma_{t,\ell}^{(23)}$. For spin greater or equal to two and twist five ($t=2$) there are also results at order $1/N^4$
\eqn{
\gamma_{t=2,\ell}^{(23)}\big|_{1/N^4} = \left\lbrace
\begin{aligned}
&\frac{640(9\ell^4+68\ell^3-1151\ell^2-5738\ell-3688)}{(\ell-1)(\ell+1)^3(\ell+4)^3(\ell+8)}\frac{1}{N^4}&\qquad&\mbox{$\ell= 2,4,\ldots$}\,,\\
&\frac{640(9\ell^4+140\ell^3-487\ell^2-11262\ell-29400)}{\ell(\ell+4)^3(\ell+7)^3(\ell+9)}\frac{1}{N^4}&\qquad&\mbox{$\ell=3,5,\ldots$}\,,
\end{aligned}
\right.
}[]
The OPE coefficients at order $1/N^0$ for twist five or greater coincide with the free theory, as before. Namely, for $t=2,3,\ldots$
\eqn{
a_{2t+\ell+1,\ell}^{(2323)} = \frac{2^l  ((t+1)!)^2  ((\ell+t+2)!)^2}{3 (2 t+1)! (2 \ell+2t+3)!}(\ell+1)(\ell+2 t+3)\,.
}[opetwothree]
And the OPE coefficients at threshold exactly match the protected contribution at order $1/N^2$ so that the corresponding operators decouple 
\eqn{
a_{\ell+3,\ell}^{(2323)} = A_{1,\ell}^{(2323)}\,.
}[]

\subsection[Double traces \texorpdfstring{$[\COp3\COp3]$}{[O3O3]}]{Double traces $\boldsymbol{[\COp3\COp3]}$}
The anomalous dimensions of the analogous operators in the OPE $\COp3\times\COp3$ were computed in~\cite{Arutyunov:2002fh}. In this case the long multiplet can transform in the representations $[0,0],[1,0],[1,1]$.\footnote{See footnote~\ref{foot:mn}.} The first and last equations are for $\ell$ even while the second is for $\ell$ odd. In all cases $t=3,4,\ldots$
\threeseqn{
\big\langle\gamma^{[0,0]}_{t,\ell}\big\rangle &= -\frac{4 (t+2) (t+1) t (t-1)}{(\ell+1) (\ell+2 t+2)} \left(1+\frac{5 (t-2) (t+3)}{(\ell+t-1) (\ell+t+4)}\right)
\frac{1}{N^2}\,,
}[]{
\big\langle\gamma^{[1,0]}_{t,\ell}\big\rangle &= -\frac{4 (t+3) (t+1) t (t-2)}{(\ell+t) (\ell+t+3)}
\frac{1}{N^2}\,,
}[gamma101]{
\big\langle\gamma^{[1,1]}_{t,\ell}\big\rangle &= -\frac{2 (t+3) (t+2) (t-1) (t-2) ((2 \ell+3 t+4) (\ell+t+1)-\ell t)}{(\ell+1) (\ell+t+1) (\ell+t+2) (\ell+2 t+2)}
\frac{1}{N^2}\,,
}[gamma020][]
where $\Delta_{t,\ell} = 2t+\ell+\gamma_{t,\ell}$. The OPE coefficients of the operators at threshold exactly cancel the respective short contributions and the contribution of twist four in $[0,0]$ vanishes. This means that only the long operators with twist at least six gain anomalous dimensions while the others decouple, as in the previous case. Let us show the results for twist two and four at order $1/N^2$ and the results for higher twists at order $1/N^0$. The spin parity and the values of $t$ in the following equations are the same as above
\threeseqn{
&\begin{aligned}
a_{\ell+2}^{[0,0]} &=A_{1,\ell}^{[0,0]}\,,\\
a_{\ell+4}^{[0,0]} &=0\,,\\
a_{2t+\ell}^{[0,0]} &=
\frac{2^{\ell-2}  (t!)^2 ((\ell+t+1)!)^2}{3 (2 t)! (2 (\ell+t+1))!}(\ell+1) (t-2) (t+3) (\ell+t-1) (\ell+t+4) (\ell+2 t+2) 
\,,
\end{aligned}
}[]{
&\begin{aligned}
a_{\ell+4}^{[1,0]} &=A_{2,\ell}^{[1,0]}\,,\\
a_{2t+\ell}^{[1,0]} &=
\frac{2^{\ell-2}  (t!)^2  ((\ell+t+1)!)^2}{3 (2 t)! (2 (\ell+t+1))!}(\ell+1) (t-1) (t+2) (\ell+t) (\ell+t+3) (\ell+2 t+2)
\,,
\end{aligned}
}[]{
&\begin{aligned}
a_{\ell+4}^{[1,1]} &=A_{2,\ell}^{[1,1]}\,,\\
a_{2t+\ell}^{[1,1]} &=
\frac{2^{\ell-2}  (t!)^2  ((\ell+t+1)!)^2}{3 (2 t)! (2 (\ell+t+1))!}(\ell+1) t (t+1) (\ell+t+1) (\ell+t+2) (\ell+2 t+2)
\,.
\end{aligned}
}[opezerotwozero][]
Note that the $\tilde{A}_{\ell}^{[n,m]}$ contributions that appear in $A_{2,\ell}^{[n,m]}$ (see \eqref{eq:thresholdA}) may be neglected at this order.

\subsection[Corrections \texorpdfstring{$1/\lambda$}{1/lambda}]{Corrections $\boldsymbol{1/\lambda}$}

The $1/\lambda$ corrections to the CFT data of $[\COp2\COp2]_{t,\ell}$ at generic twist are not known.\footnote{The $1/\lambda$ correction to the $[\CL\CL]$ double trace operator was computed in~\cite{Goncalves:2014ffa}, where $\CL$ is the Lagrangian operator: a descendant of $\COp2$. In order to relate that result to what we need one would still need to solve a mixing problem.} However, for the lowest twist operators ($t=2$) one does not need to solve any mixing problem and thus it was possible to obtain several orders in $1/\lambda$ and $1/N$. The latest results can be found in~\cite{Chester:2020dja}, part of which had been already obtained in~\cite{Arutyunov:2000ku,DHoker:1999mic,Goncalves:2014ffa,Binder:2019jwn,Chester:2019pvm,Aprile:2017bgs,Alday:2018pdi}. We will write $\gamma_{2,\ell}$ as an expansion on $1/c$ rather than $1/N^2$. The parameter $c$ is the central charge and it is defined as $c = (N^2-1)/4$. The anomalous dimensions read
\eqna{
\gamma_{t=2,\ell} &=
\frac1c\left[-\frac{24}{(\ell+1)(\ell+6)} - \frac{4320\lsp \zeta(3)}{7 \lambda^{\frac32}}\delta_{\ell,0} - \frac{\zeta(5)}{\lambda^{\frac52}}\left(30600\lsp\delta_{\ell,0}+\frac{201600}{11}\delta_{\ell,2}\right)\right.\\&
\qquad\;\,\lsp \left.-\frac{\zeta(3)^2}{\lambda^3}\left(172800\lsp\delta_{\ell,0}+\frac{2419200}{11}\delta_{\ell,2}\right)+ O(\lambda^{-\frac72})\right]\\
&\quad\,+\frac1{c^2}\left[
-\frac{45\sqrt{\lambda}}{14}\delta_{\ell,0} + \frac{24\lsp(7\ell^4+74\ell^3-533\ell^2-4904\ell-3444)}{(\ell-1)(\ell+1)^3(\ell+6)^3(\ell+8)}-\frac{135}{7}\delta_{\ell,0}
\right]\\
&\quad\,+\frac{1}{c^3}\left[-\lambda^{\frac32}\left(\frac{85}{768}\delta_{\ell,0} + \frac{35}{528}\delta_{\ell,2}\right) + O(\lambda)\right] + \frac{1}{c^4}\left(\frac{5\lambda^3}{16896}\delta_{\ell,2}+O(\lambda)\right)+ O(c^{-5})\,,
}[localizationgamma]
where for convenience we have repeated the results \gammatl for $t=2$ and \gammal.
\par
In the expression above one can observe that at each order in the $1/c$ expansion after tree level there are contributions diverging as $\lambda\to\infty$ which have support on a finite number of spins. These terms are those that make the extrapolation to supergravity ambiguous because one would need to renormalize those divergences with some counterterms in the supergravity effective action. The finite part of the counterterm is what gives the ambiguity. We can also expect that such ambiguities will appear on more and more spins as we go further in the $1/c$ expansion. This means that ultimately the supergravity extrapolation of the CFT data at finite $N$ will be ambiguous for all spins, but the effects of this ambiguity will be apparent only at smaller and smaller values of $N$ as the spin is increased. As we have anticipated in the introduction (see \figurename~\ref{fig:BisectionReggeSingletComparisonINTRO}) this is indeed compatible with what we oberve, even though one could have expected the effects to be noticeable at much lower values of $N$.
\par
From an effective field theory perspective we can always estimate the value of these ambiguities because the cutoff of the theory is given by the single trace gap $\Delta_\mathrm{gap} = \lambda^{1/4}$. As explained in~\cite{Alday:2017vkk}, the gap cannot be taken bigger than $c^{1/4}$ in order for the double discontinuity $\mathrm{dDisc}[\CF]$ to remain bounded by 0 and 1. Thus we can take the worst case scenario and set $\lambda=c$. This, by dimensional analysis, implies a contribution at two loop of the order of $1/c^{7/2}$. In~\cite{Alday:2017vkk} it is also argued that such correction to the CFT data are, in any case, bounded by $1/c^{3/2}$, as a consequence of the inequality satisfied by $\mathrm{dDisc}[\CF]$ for theories with a large gap~\cite{Caron-Huot:2017vep}.\footnote{More generally, the bound for the ambiguities arising at spin $\ell$ is given by $1/c^{\frac{3+\ell}2}$.} This is clearly in accord with the effective field theory estimate, but it is not saturated.
\par
As the reader will see in the next section, the numerical results are actually quite far from the supergravity values, even by taking into account an error of the order of the estimates discussed above.

\section{Numerical results}
\label{sec:results}

\subsection[Bounds on the dimension as a function of \texorpdfstring{$N$}{N}]{Bounds on the dimension as a function of $\boldsymbol{N}$}

\begin{figure}[t!]
\centering
\subfloat[Mixed correlator system]{\includegraphics[scale=.65]{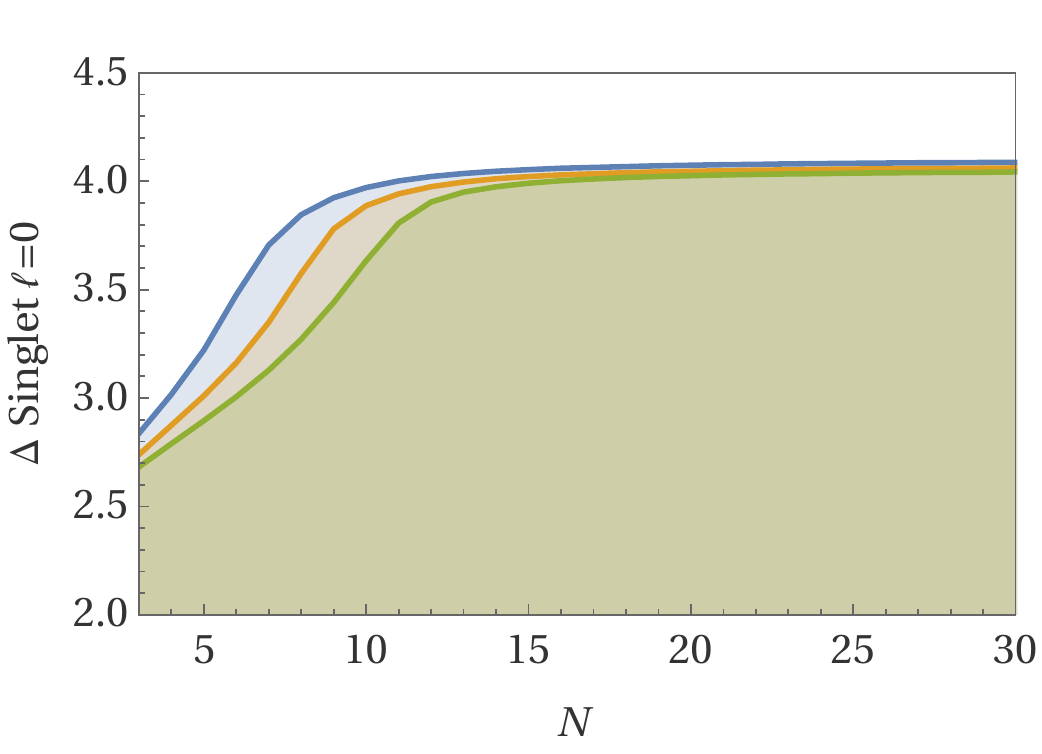}}\;
\subfloat[Only $\COp3$]{\includegraphics[scale=.65]{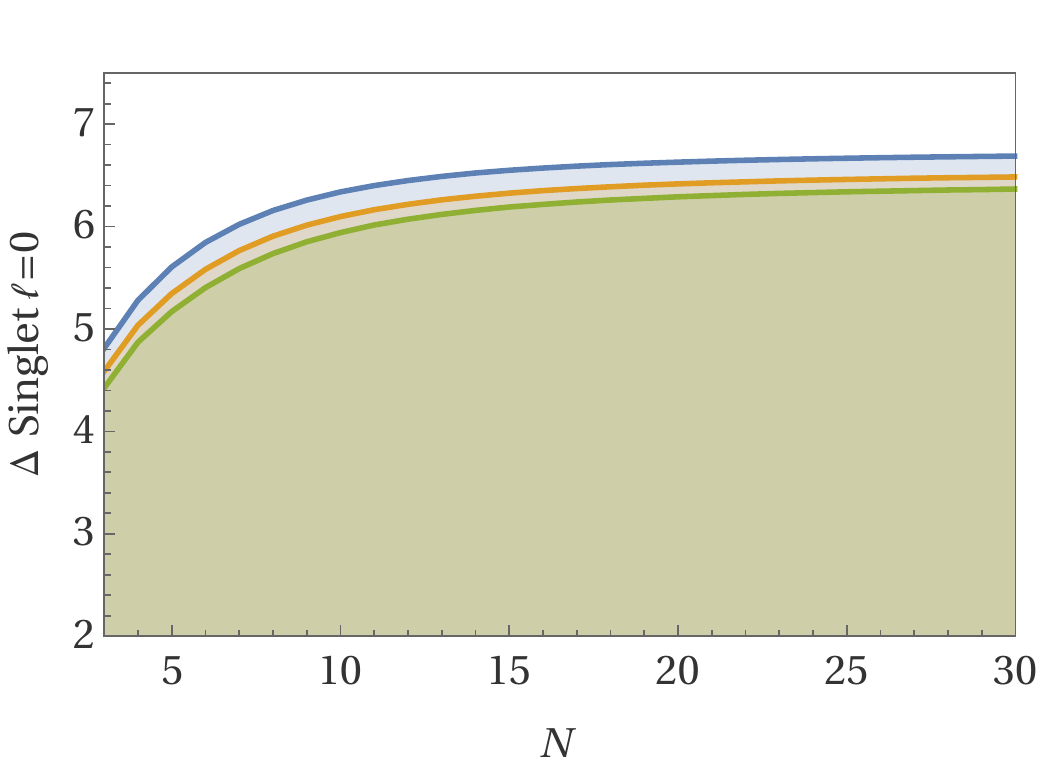}}
\caption{In the left we see the upper bound for the gap on the dimension of the scalar singlet as a function of $N$. The different curves are the different values of $\Lambda = 19,23,27$. The right shows the same plot using only $\langle\COp3\COp3\COp3\COp3\rangle$ (note the difference in the scale).}\label{fig:BisectionSingletMixed}
\end{figure}

\begin{figure}[t!]
\centering
\subfloat[Only $\COp2$\label{fig:BisectionRegge2222}]{\includegraphics[scale=.65]{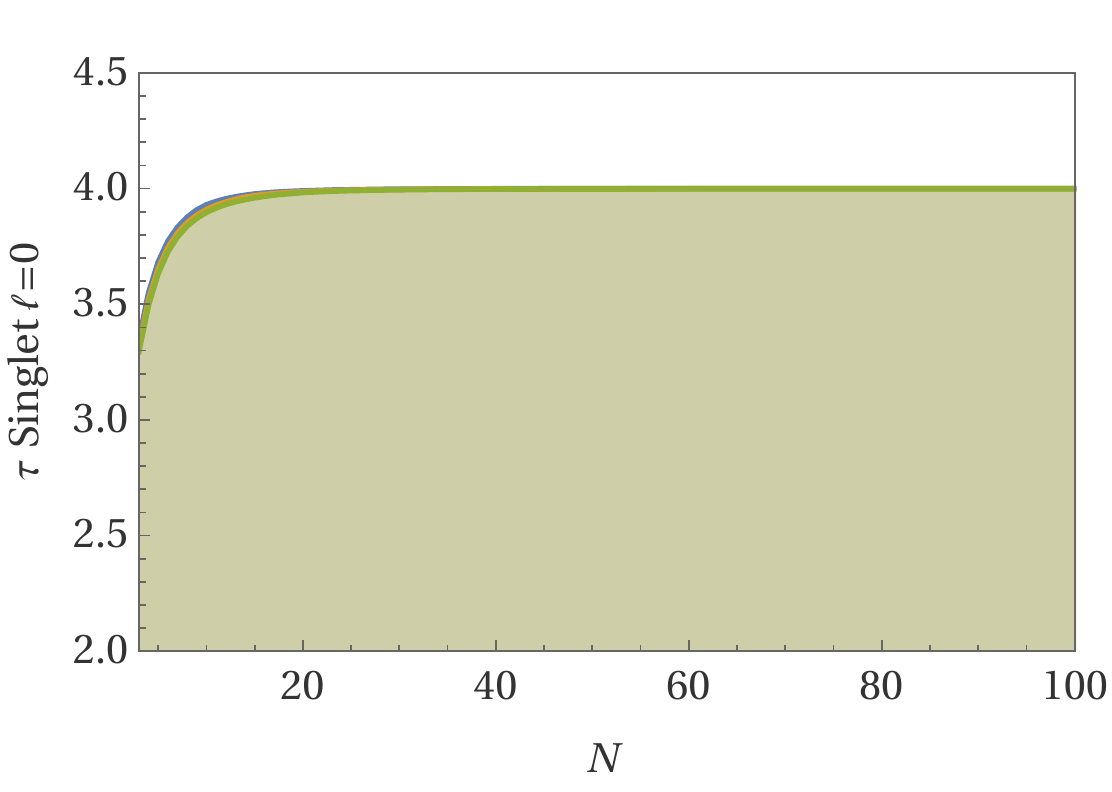}}\;
\subfloat[Only $\COp3$\label{fig:BisectionRegge3333}]{\includegraphics[scale=.65]{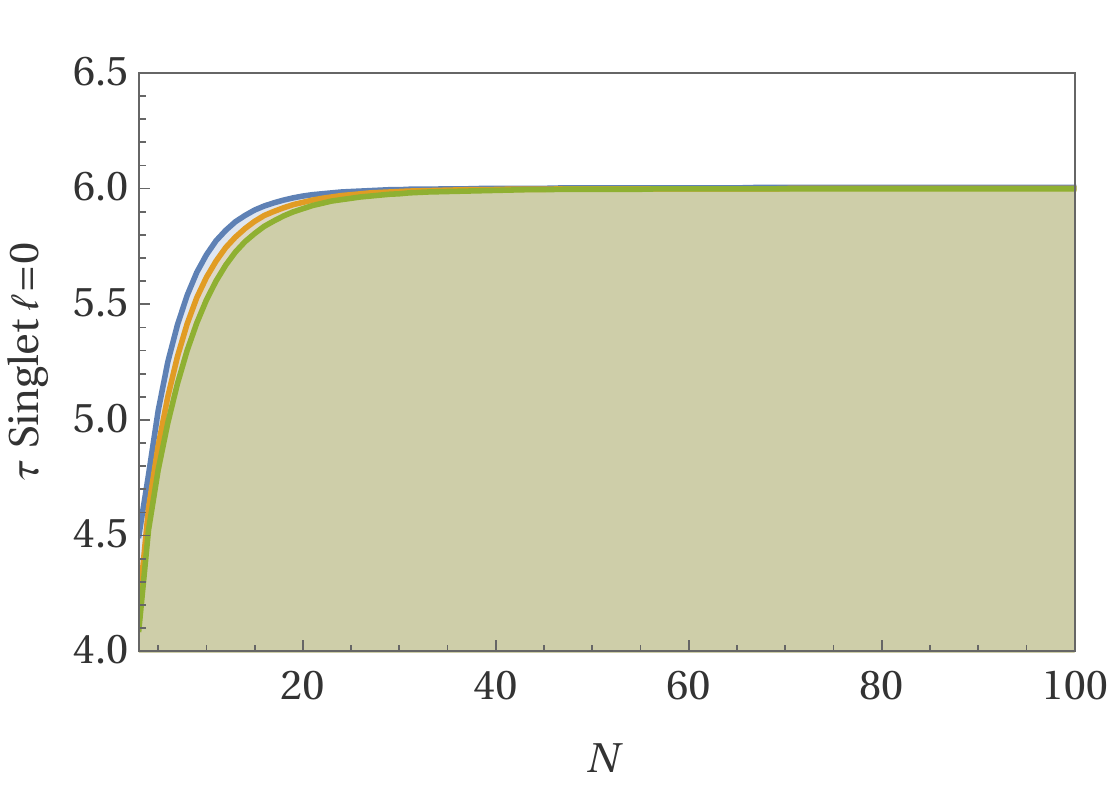}}
\caption{Upper bound for the gap on the twist of all singlets as a function of $N$. The different curves are the different values of $\Lambda = 19,23,27$. Here we pushed to much higher $N$ to show where it starts converging to $\tau = 4$.  The left plot uses only $\langle\COp2\COp2\COp2\COp2\rangle$ while the right plot uses only $\langle\COp3\COp3\COp3\COp3\rangle$ (note the difference in the scale).}\label{fig:BisectionReggeSingletMixed}

\end{figure}

\begin{figure}[t]
\centering
\includegraphics[scale=.60]{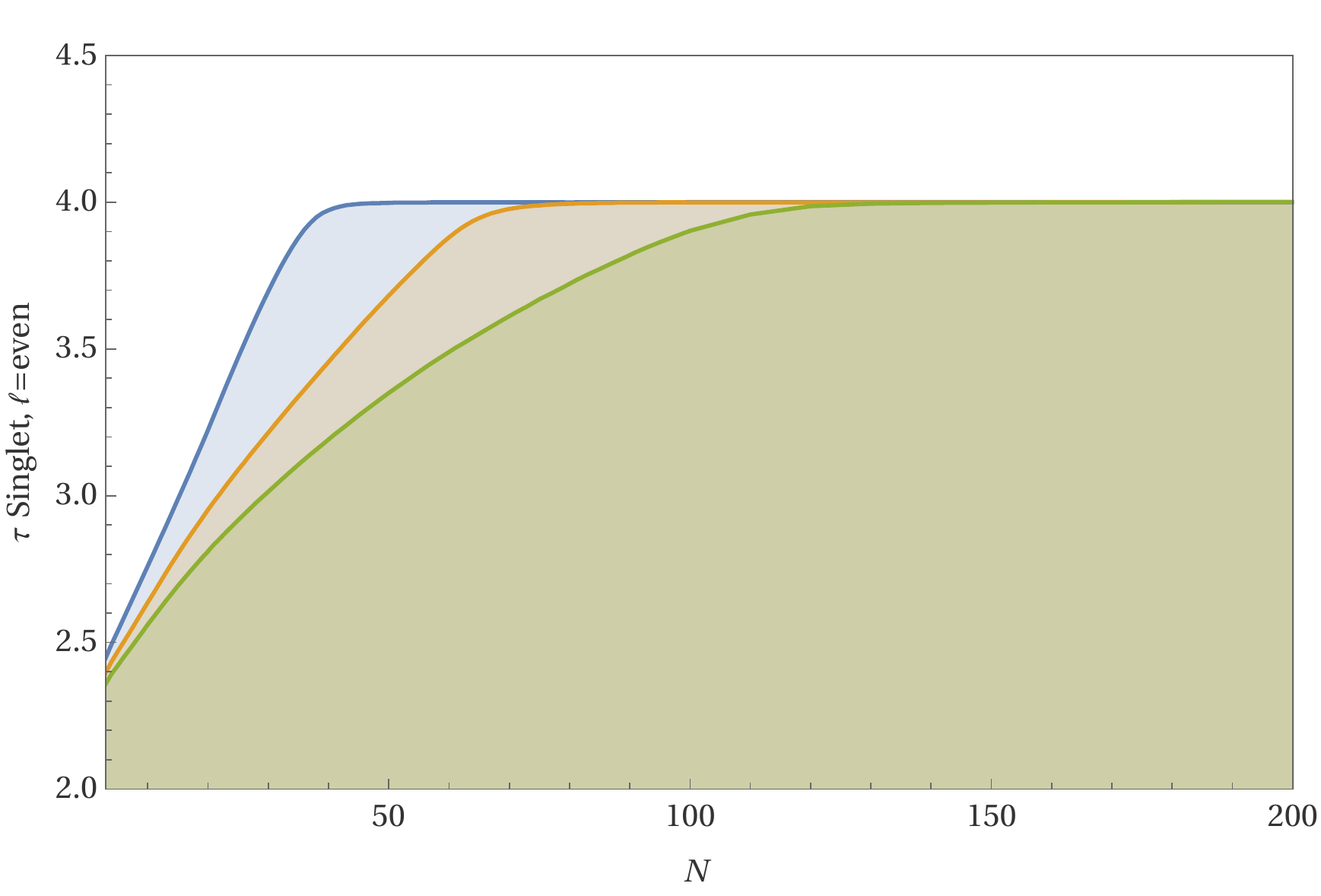}
\caption{Upper bound for the gap on the twist of all singlets as a function of $N$, using the full system of correlators. The different curves are the different values of $\Lambda = 19,23,27$.}\label{fig:BisectionReggeSingletMixeda}
\end{figure}

We start by showing some simple binary searches for the conformal dimension of the lightest operators in the various representations. The different lines are associated to a different value of $\Lambda$ (defined in the Appendix~\ref{app:numerical}) which is related to the maximal order of the Taylor expansion. The shaded areas are the allowed values for the gap imposed on the representation considered.
\par
In Figure~\ref{fig:BisectionSingletMixed} we show an upper bound for the gap on the conformal dimension of the scalar singlet sector. The left plot uses the mixed correlator system. We also looked at higher values of $N$, namely $200\leq N \leq 1000$ and the bound remains very close to $4$ for all the $N$'s we checked in that range. The correlator $\langle\COp3\COp3\COp3\COp3\rangle$ on the other hand (on the right) has a much weaker bound. It does not seem to asymptote to $\Delta_\mathrm{gap} = 6$ at large $N$, which would be the supergravity solution. It was conjectured in~\cite{Beem:2016wfs} that indeed the theory on the boundary in the limit $N\to \infty,\Lambda\to\infty$ should be planar supergravity. It seems that this plot does not follow this expected behavior. However we can try to make another kind of assumptions in our bootstrap problem. Instead of assuming a gap on $\Delta$ in a given channel we can impose a \emph{twist} gap in all spins of a given R-representation. This gives certainly more stringent bounds and, remarkably, they saturate the supergravity solution with very high precision. In Figure~\ref{fig:BisectionReggeSingletMixed} on the left we have the plot for the $\COp2$ single correlator which approaches $\Delta_\mathrm{gap} = 4$ at large $N$. On the right instead we have the $\COp3$ single correlator which also approaches its supergravity solution $\Delta_\mathrm{gap} = 6$. The bound can be observed to converge to $6$ even at intermediate values of $N$. However one would expect the bound to drop to $4$ because the OPE coefficient $\lambda_{\COp3\COp3[\COp2\COp2]_{0,0}}$ becomes non negligible. It seems that the crossing equations of the single correlator alone are not strong enough to require the presence of the twist four operator at finite $N$.
\par
Building on this idea of simultaneously gapping the twist of all operators in a given representation, we considered the entire system of mixed correlators. The result is given in~\figurename~\ref{fig:BisectionReggeSingletMixeda}. As we discussed in the introduction, the curve now drops more rapidly below the supergravity, large $N$, prediction of $\tau\sim4-O(1/N^2)$ and, in fact, it is in disagreement with it for somewhat large values of $N$ ($N\sim160$), see Figure~\ref{fig:BisectionReggeSingletComparisonINTRO}. Moreover, the numerics are not yet converged, and the tension will likely become more severe at higher values of $\Lambda$. In the introduction we also showed a bound on the twist of all operators with $\ell\geq2$, while leaving the scalars unconstrained. Some possible scenarios to resolve this puzzle were also discussed.

\begin{figure}
\centering
\subfloat[No operator at threshold]{\includegraphics[scale=.65]{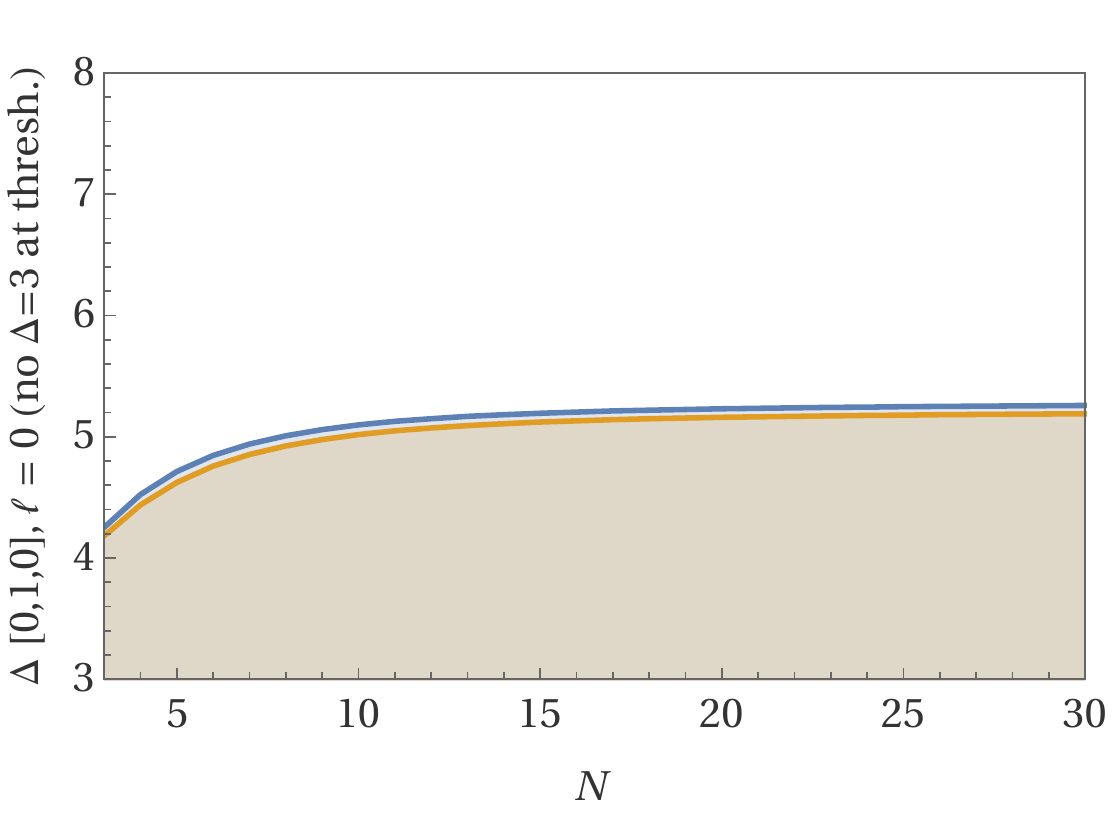}}\;
\subfloat[With operators at threshold]{\includegraphics[scale=.65]{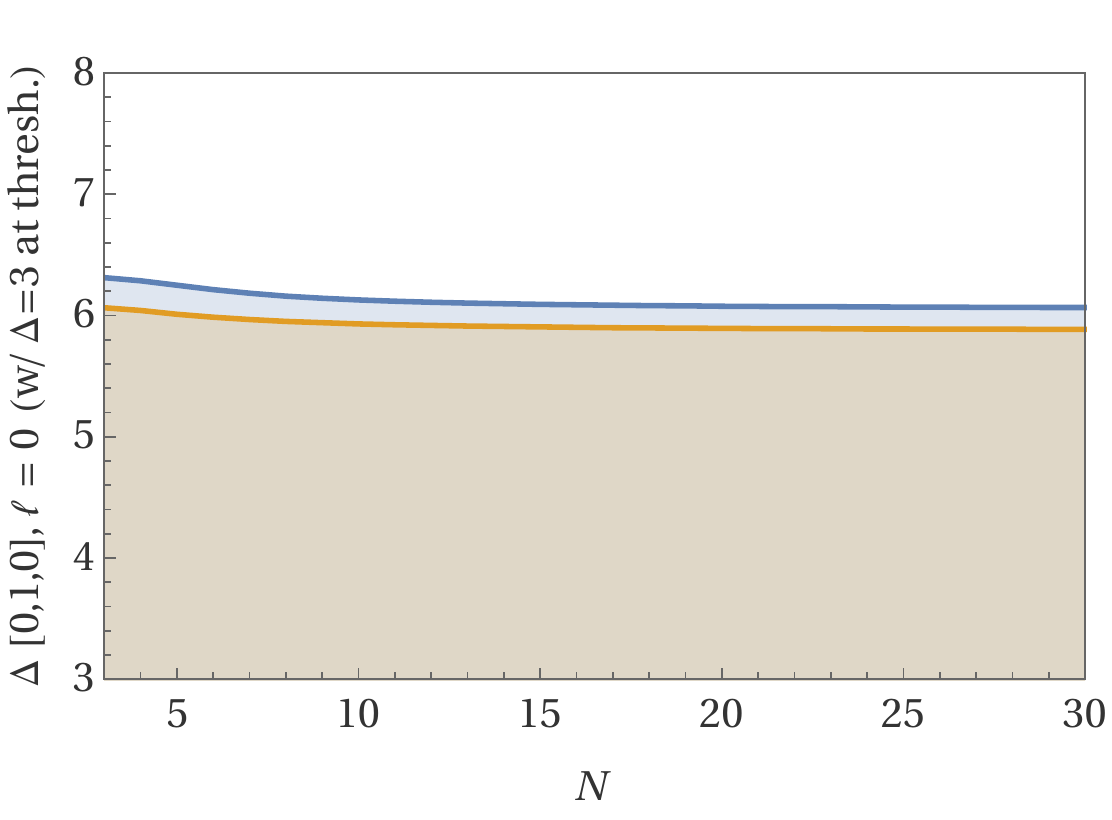}}
\caption{Upper bound for the gap on the dimension of the scalar transforming in the $[0,1,0]$ as a function of $N$. The different curves are the different values of $\Lambda = 19,23$. The first imposes a gap while the second allows for the presence of an operator sitting at $\Delta = 3$. This accounts for possible ``dynamically protected'' operators as discussed in Section~\ref{sec:recombination}.}\label{fig:Bisection010Mixed}
\end{figure}

\par We would also like to discuss the bounds on the representation $[0,1,0]$. We considered for simplicity just a scalar and put a gap on the lightest such operator. According to the discussion of Section~\ref{sec:recombination}, we cannot know the exact value of the dynamically protected OPE coefficient for a $[0,1,0]$ operator at twist $3$. Thus we made two plots in \figurename~\ref{fig:Bisection010Mixed}: one where we assume that there are no such dynamically protected operators at all (namely we cancel their contribution exactly in $F_\mathrm{short}$ and we do not require their presence in the bootstrap equations), and one where we allow for an operator at $\Delta = 3$ with positive but unknown OPE coefficient. We see that both cases are consistent and, as expected, the bounds for the latter case are less stringent.
\par
\begin{figure}
\centering
\subfloat[Singlet of spin $2$]{\includegraphics[scale=.65]{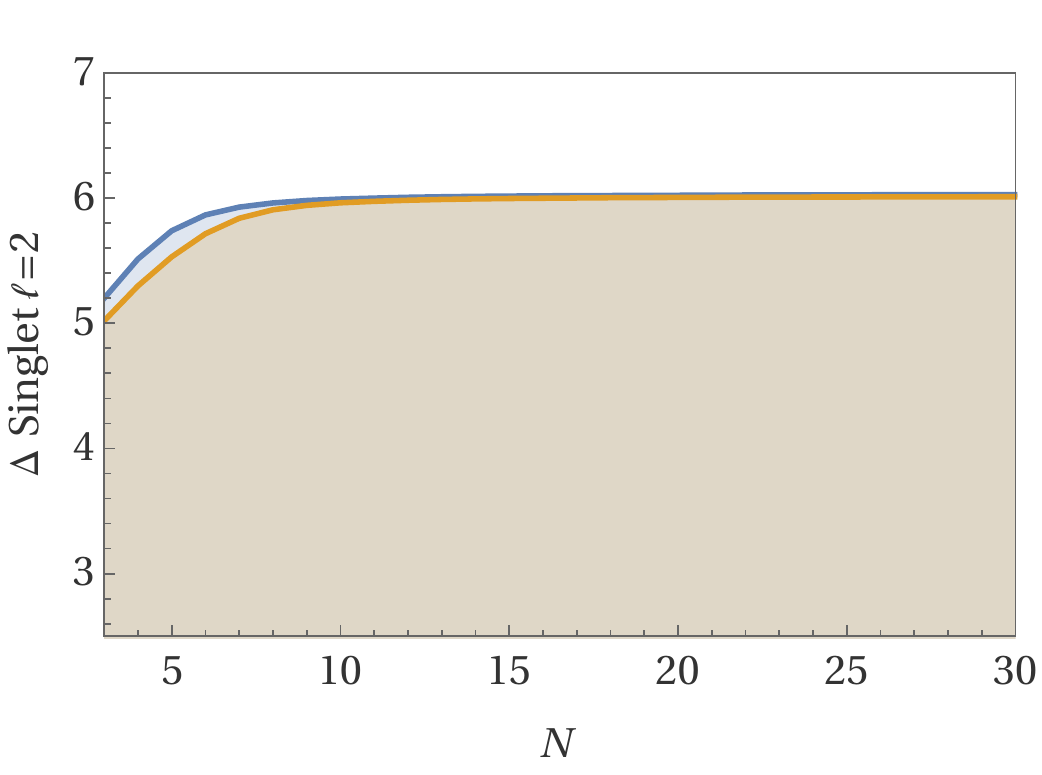}}\;
\subfloat[Singlet of spin $4$]{\includegraphics[scale=.65]{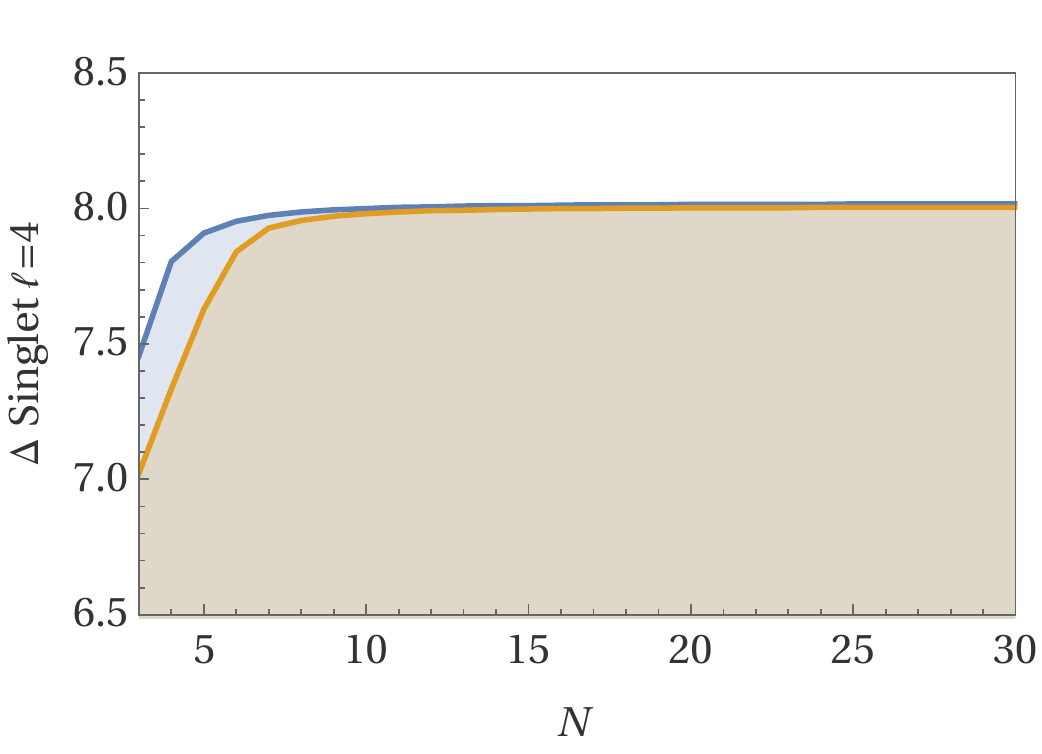}}\\
\subfloat[{$[1,0,1]$ of spin $1$}]{\includegraphics[scale=.65]{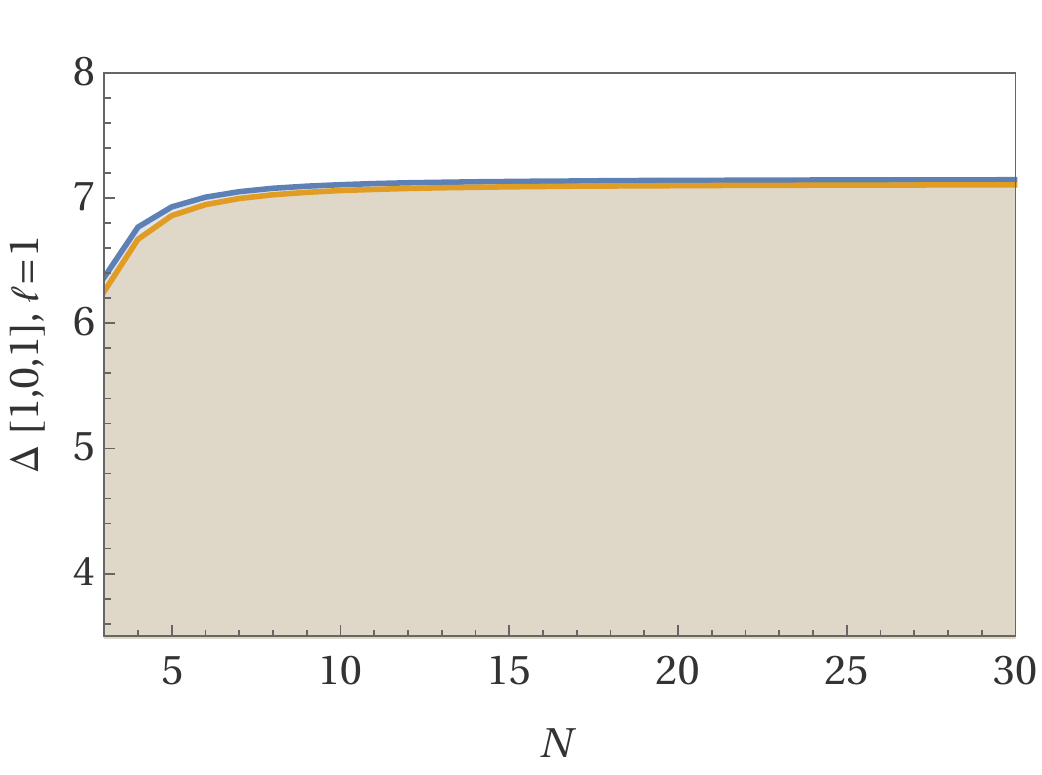}}\;
\subfloat[{$[0,2,0]$ of spin $0$}]{\includegraphics[scale=.65]{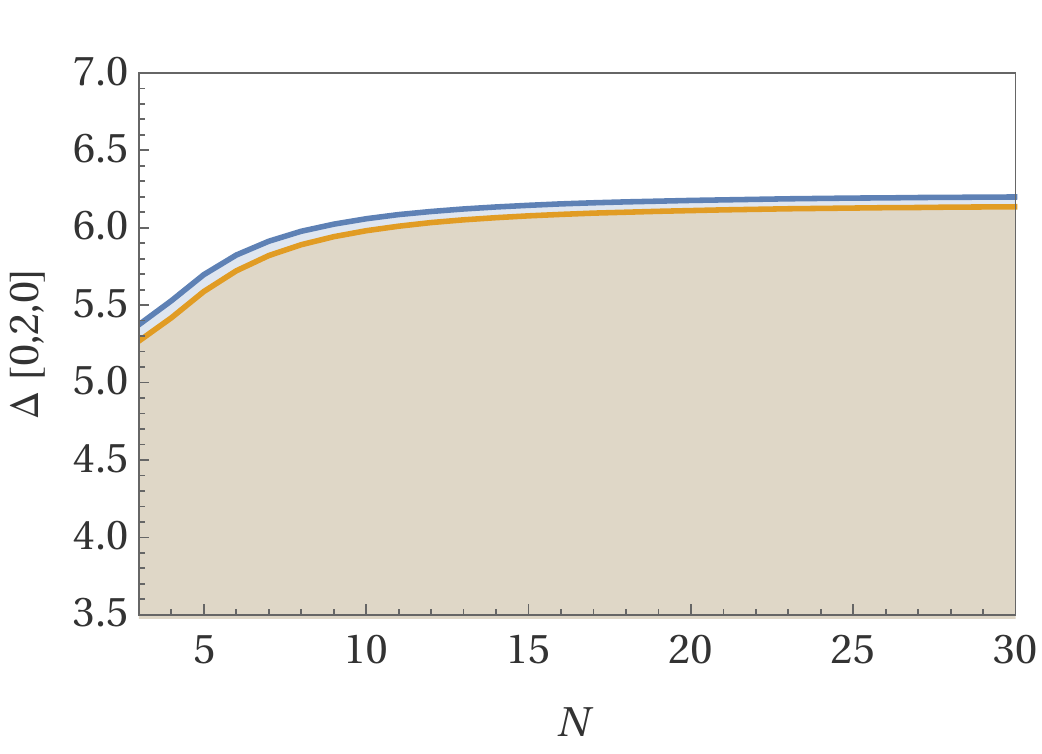}}
\caption{Bounds on the gap of the lightest operator for other representations of spin and R-symmetry with the mixed correlator. In order from the top left we have a singlet of spin $2$, spin $4$, a $[1,0,1]$ of spin $1$ and a $[0,2,0]$ of spin $0$.}\label{fig:otherreps}
\end{figure}
\begin{figure}[h]
\centering
\subfloat[{$[1,0,1]$ of spin odd}]{\includegraphics[scale=.65]{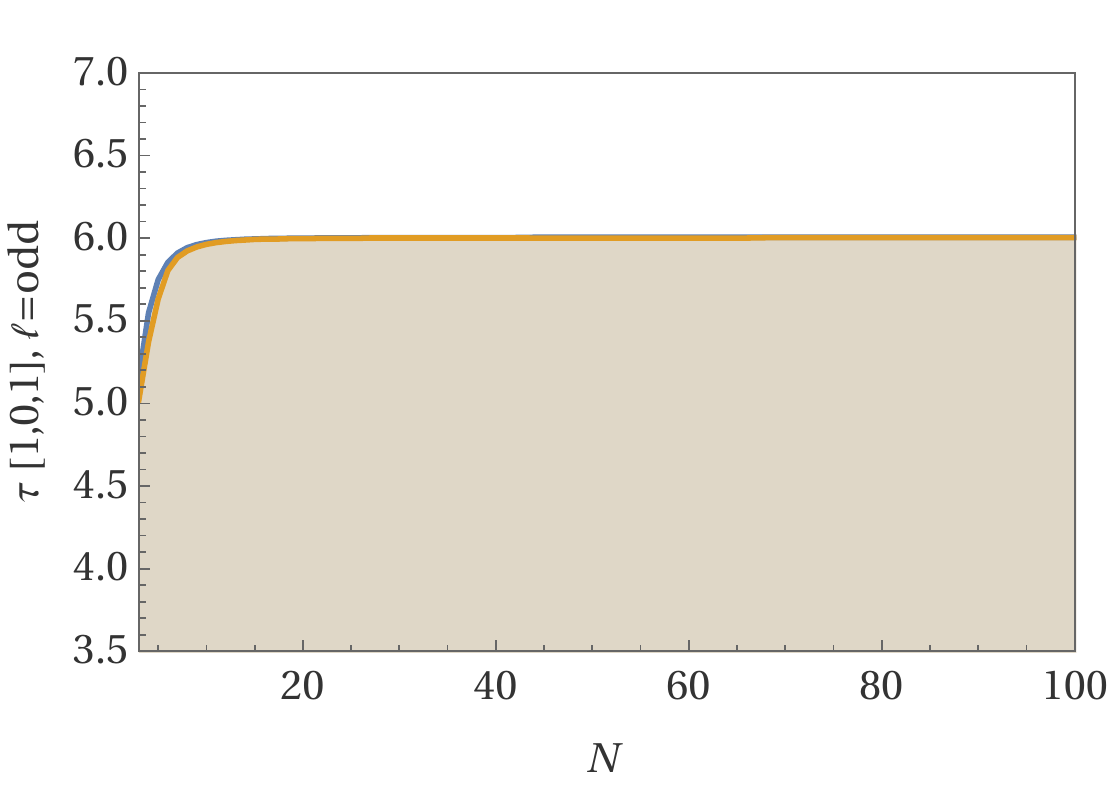}}\;
\subfloat[{$[0,2,0]$ of spin even}]{\includegraphics[scale=.65]{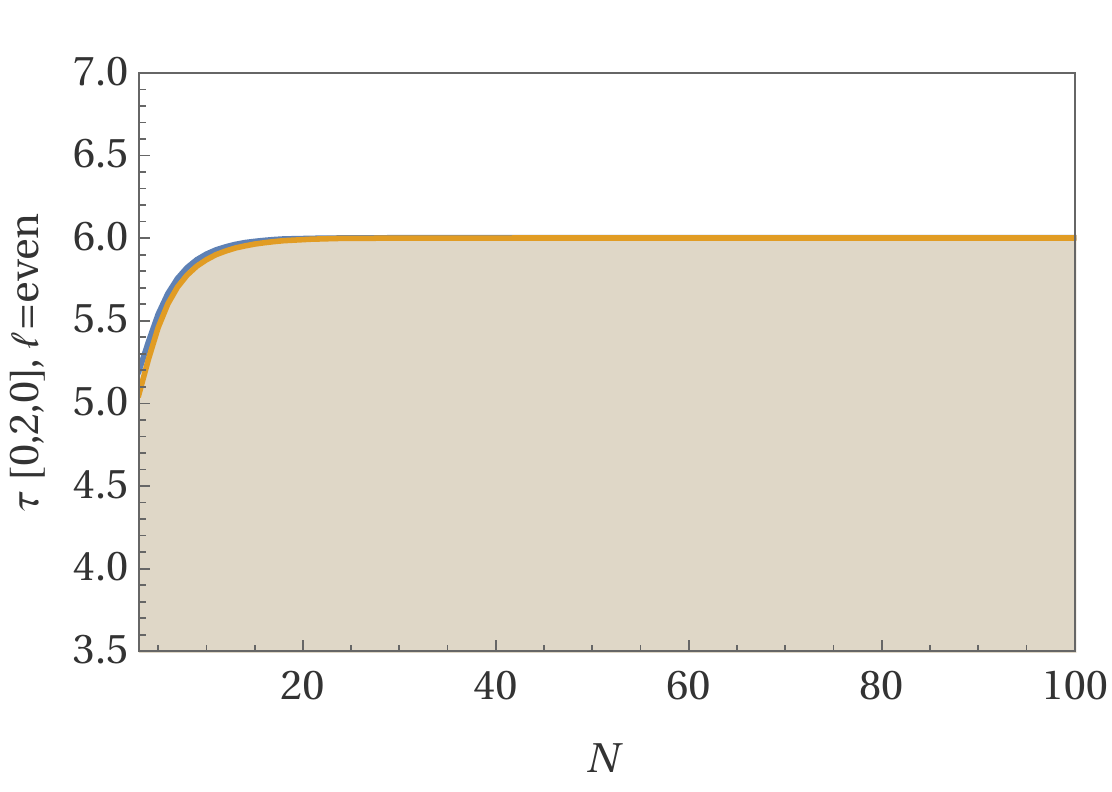}}
\caption{Bounds on the twist gap for other R-symmetry representations with the mixed correlator. We have a $[1,0,1]$ on the left and a $[0,2,0]$ on the right.}\label{fig:otherrepsRegge}
\end{figure}

In Figure~\ref{fig:otherreps} we show plots for some other values of the spin and the R-symmetry representation. In order from the top left we have a singlet of spin $2$, spin $4$, a $[1,0,1]$ of spin $1$ and a $[0,2,0]$ of spin $0$. Similarly, in Figure~\ref{fig:otherrepsRegge} we show bounds on the representations $[1,0,1]$ and $[0,2,0]$ but this time assuming gaps on the twist. All plots take into account the constraints of the entire system of mixed correlators.

\subsection[Behavior at large \texorpdfstring{$N$}{N}]{Behavior at large $\boldsymbol{N}$}

\begin{figure}[h]
\centering
\subfloat[{Twist gap in the singlet channel, with spin even, for the correlator of four $\COp2$.}\label{fig:fit2222}]{\includegraphics[scale=.68]{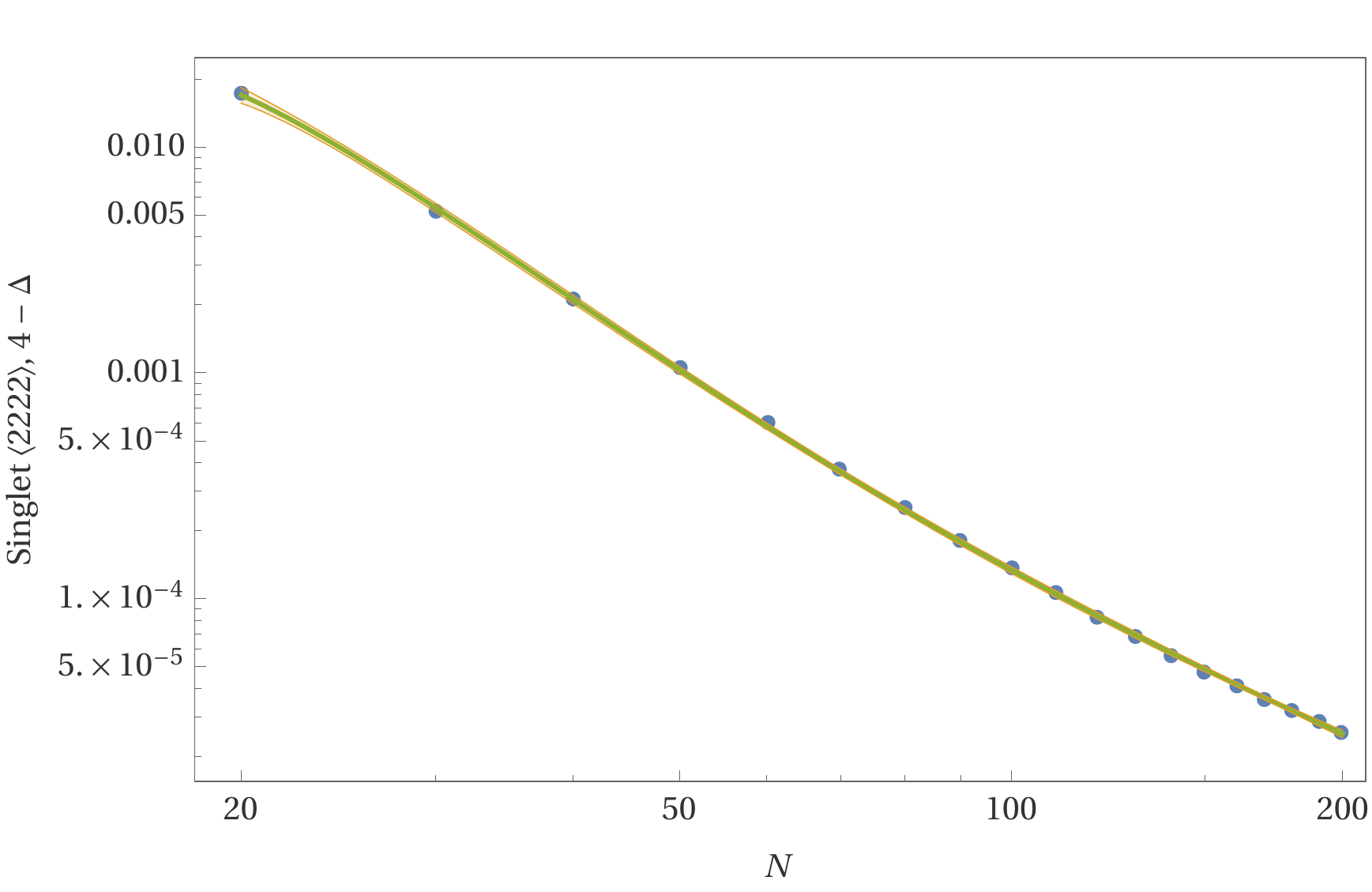}}\\
\subfloat[{Twist gap in the $[0,2,0]$ representation, with spin even, for the correlator of four $\COp3$.}\label{fig:fit3333020}]{\includegraphics[scale=.62]{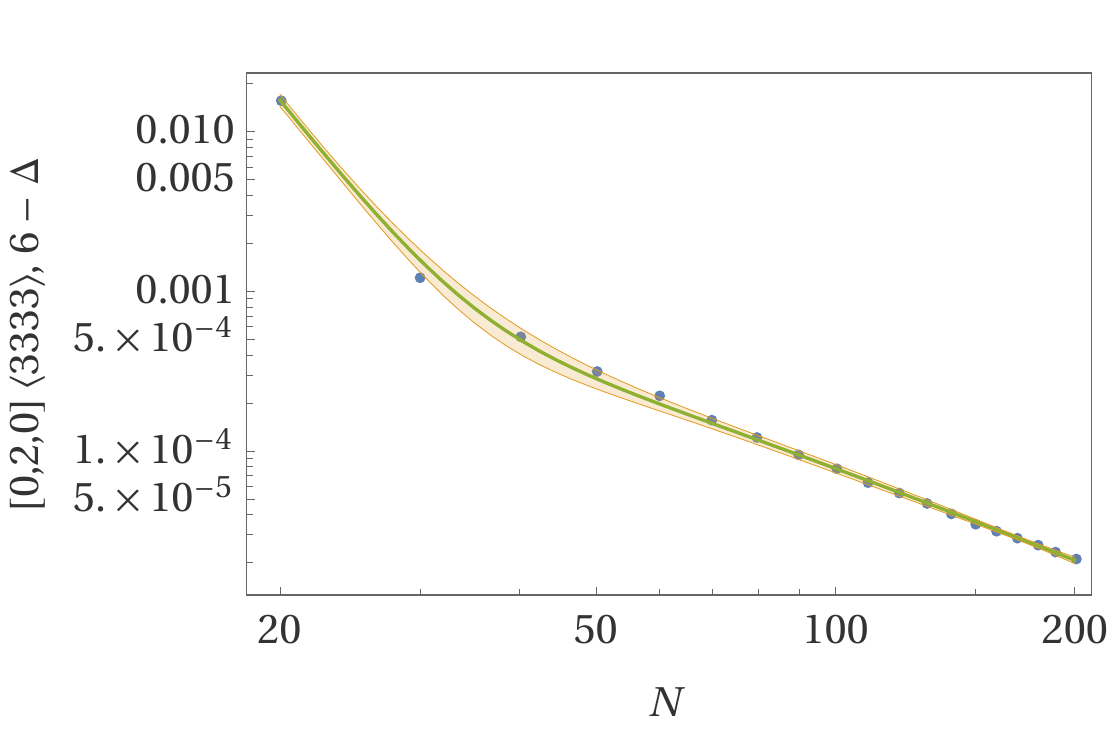}}\qquad
\subfloat[{Twist gap in the $[1,0,1]$ representation, with spin odd, for the correlator of four $\COp3$.}\label{fig:fit3333101}]{\includegraphics[scale=.62]{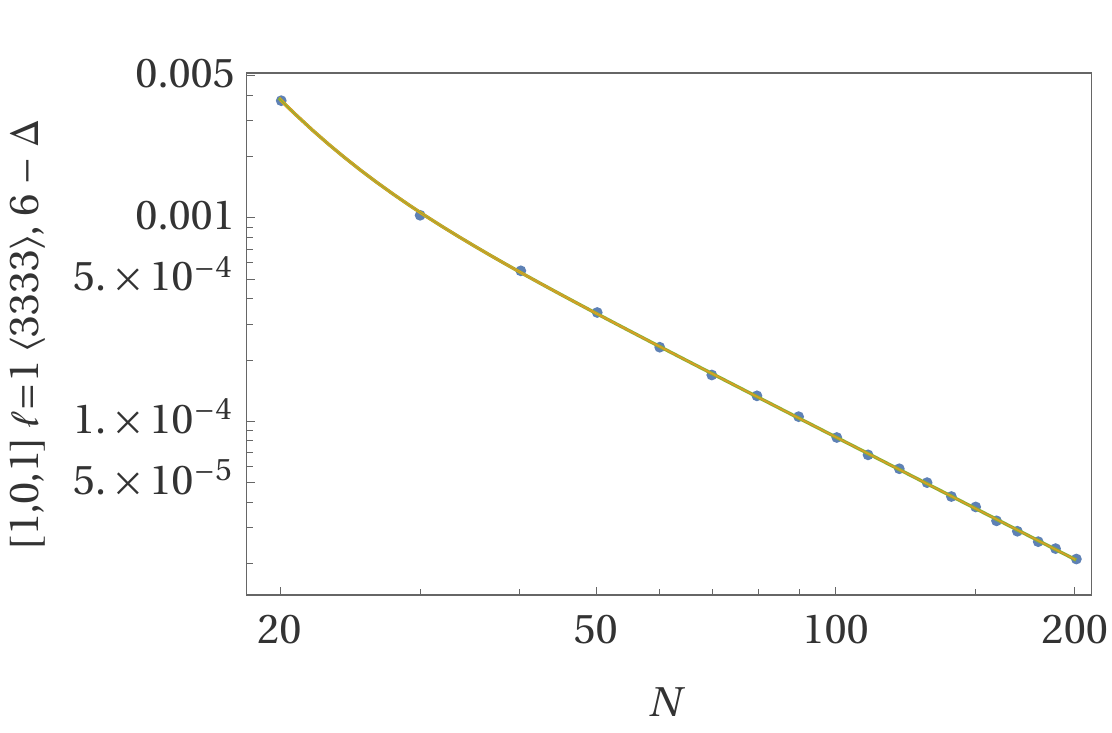}}
\caption{Numerical fit in logarithmic scale following \eqref{eqfit} of the smallest twist gap allowed in the respective channels. The numerical values of the curves in green is shown in equation \eqref{resfitSing}, \eqref{resfit020} and \eqref{resfit101} respectively. The yellow band is the error bar at $99\%$ confidence level.}
\end{figure}

\begin{figure}[h!]
\centering
\includegraphics[scale=.75]{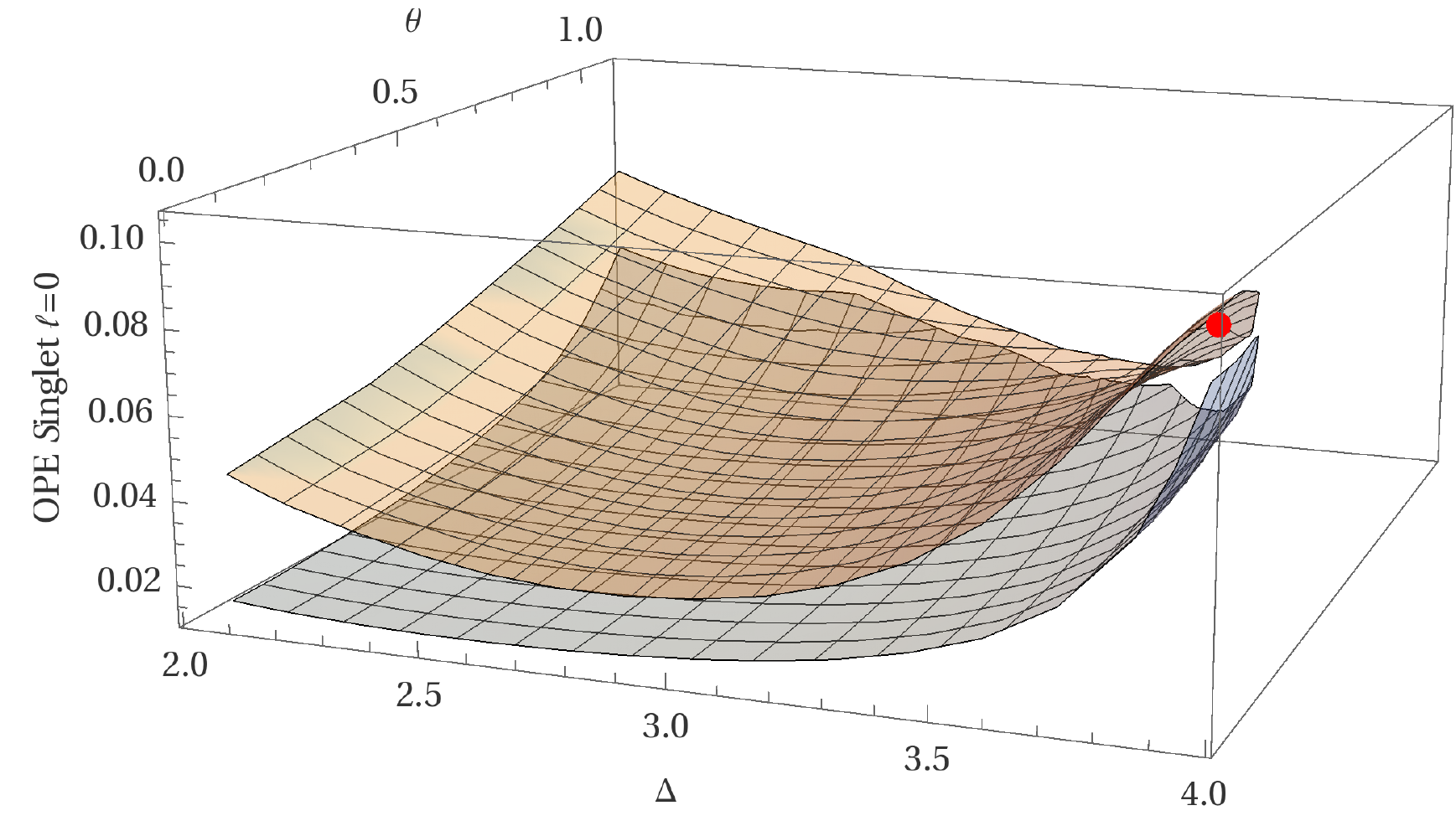}
\caption{Upper and lower bounds for the OPE coefficient $(\lambda_{22(\Delta,0)}^2 + \lambda_{33(\Delta,0)}^2)^{1/2}$ as a function of the dimension of the scalar singlet $\Delta$ and of the angle $\theta$ defined in \eqref{thetadef}. It is assumed a gap on the next scalar singlet of $4.2$. For this plot we fixed $N=20$. The angle $\theta$ ranges from $0$ to $1.1$ because for higher angles the theory is disallowed. The red dot marks the free theory lowest twist OPE coefficient~\opetwo (at $t=2$ in the free theory $\theta = 0$).}\label{fig:OPEMinMaxSinglet}
\bigskip

\subfloat[{Scalar transforming in the $[0,1,0]$. It is assumed a gap of 5.5 on the next scalar and a twist gap of 3.5 on the $[0,1,0]$ representation. The red dot marks the lowest twist free theory OPE coefficient~\opetwothree}]{\includegraphics[scale=.65]{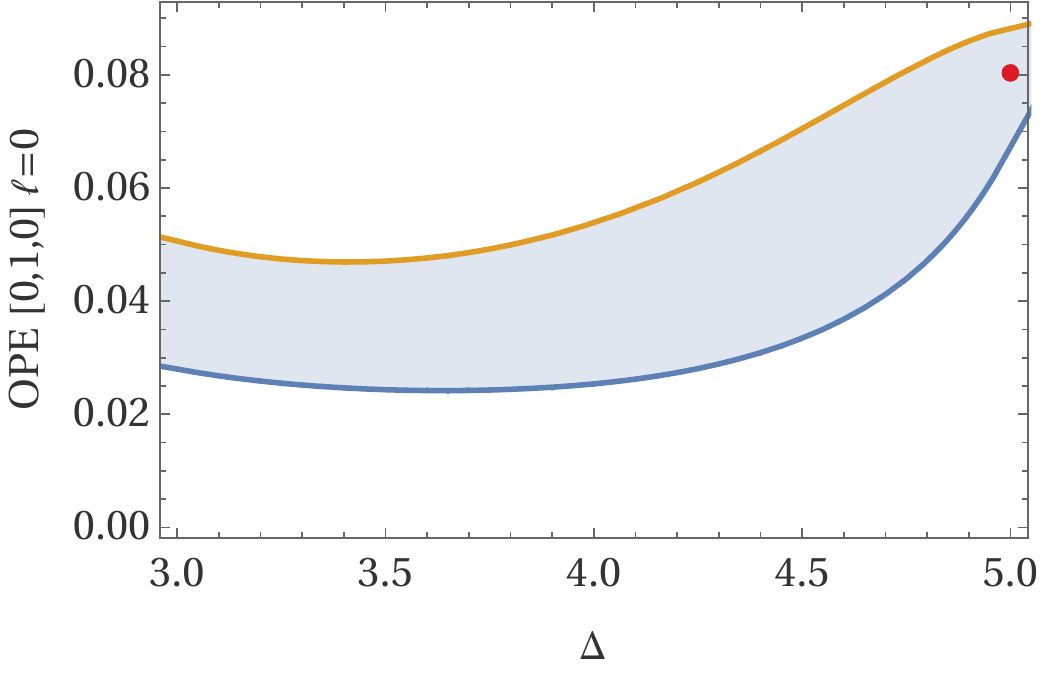}}\qquad
\subfloat[{Scalar transforming in the $[0,2,0]$. It is assumed a gap of 6.5 on the next scalar and a twist gap of 4.5 on the $[0,2,0]$ representation. The red dot marks the lowest twist free theory OPE coefficient~\opezerotwozero}]{\includegraphics[scale=.65]{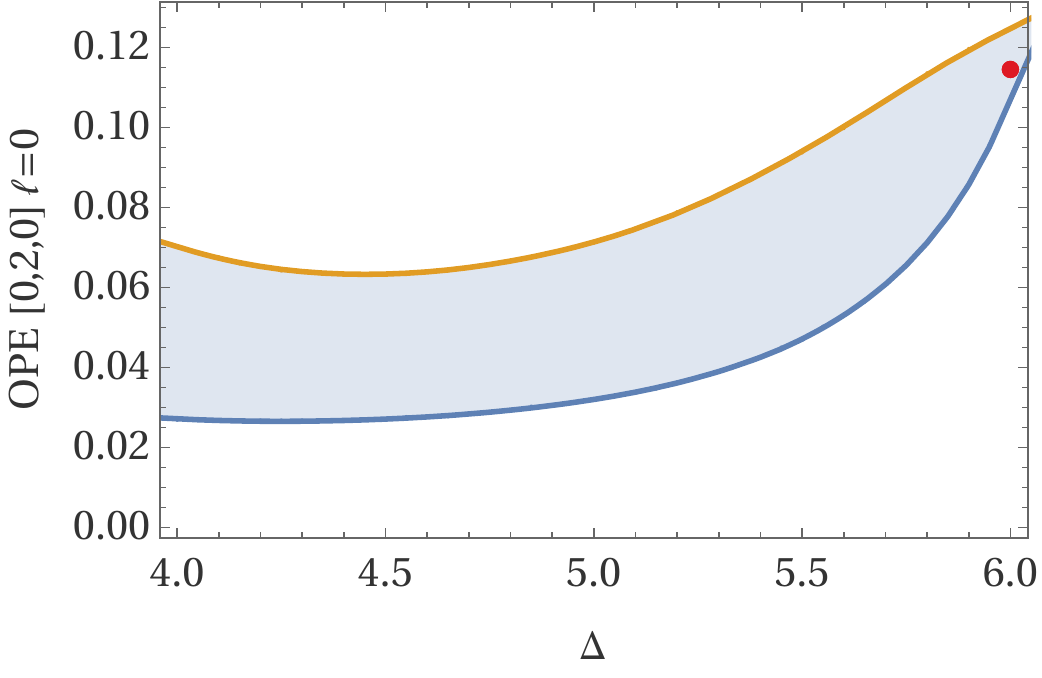}}
\caption{Upper and lower bounds of the OPE coefficient $a_{\Delta,\ell}$ of the lightest scalar in a given representation. We fixed $N=20$.}\label{fig:OPEMinMax}
\end{figure}

As we have observed in the previous subsection, by assuming gaps on the entire Regge trajectory we get very close to an integer value of $\Delta$. In particular the bound for the singlet in the $\langle\COp2\COp2\COp2\COp2\rangle$ gets very close to the supergravity solution $\Delta=4$. This was also observed in~\cite{Beem:2016wfs}. Here we provide a numerical fit for this case, plus other representations in the $\langle\COp3\COp3\COp3\COp3\rangle$ correlator. However, instead of assuming a gap on the dimension of the lightest operator in the OPE, we assume a twist gap.

In \figurename~\ref{fig:fit2222} we show the twist gap on the singlet sector in the correlator of four $\COp2$'s. In \figurename~\ref{fig:fit3333020} and \ref{fig:fit3333101} we instead show, respectively, the even spins trasforming in the $[0,2,0]$ and the odd spins transforming in the $[1,0,1]$, both in the correlator of four $\COp3$'s. Both axes are in logarithmic scale and the vertical one is translated by the offset value of the fit (called $\beta_0$ in \eqref{eqfit}). We looked for fitting functions of the form $\sum_a k_a N^{-2a}$ and observed that stopping at order $N^{-6}$ yielded a good approximation of the numerical data. We therefore chose the following parametrization
\eqn{
\tau_{\mathrm{max}} = \beta_0  + \frac{\beta_1}{N^2} + \frac{\beta_2}{N^4} + \frac{\beta_3}{N^6}\,.
}[eqfit]
With this definition, these are the result for, respectively, the singlet, the $[0,2,0]$ and the $[1,0,1]$ 
\begin{subequations}\label{resfit}\begin{align}
\mathrm{Singlet:}&&\beta_0 &= 4+9.696\times10^{-6}\,,\;& \beta_1 &= -0.8942\,,\;& \beta_2 &= -4541.\,,\;& \beta_3 &= 8.752\times10^5\,,
\label{resfitSing}\\*
\mathrm{[0,2,0]:}&&\beta_0 &= 6+5.441\times10^{-4}\,,\;& \beta_1 &= -0.8455\,,\;& \beta_2 &= 822.4\,,\;& \beta_3 &= -1.201\times10^6\,,
\label{resfit020}\\*
\mathrm{[1,0,1]:}&&\beta_0 &= 6+7.702\times10^{-4}\,,\;& \beta_1 &= -0.8367\,,\;& \beta_2 &= 28.25\,,\;& \beta_3 &= -1.212\times10^5\,.
\label{resfit101}
\end{align}
\end{subequations}
Unfortunately, at this value of $\Lambda$, the predictions for the large $N$ expansion coefficients are not reliable enough to make any conclusive statements. A possible improvement could be obtained by computing them with several values of $\Lambda$ and extrapolating to $\Lambda=\infty$, but we did not attempt to do this.

\subsection{Bounds on OPE coefficients}

\begin{figure}[t]
\centering
\subfloat[Singlet spectrum for $N=30$]{\includegraphics[scale=.55]{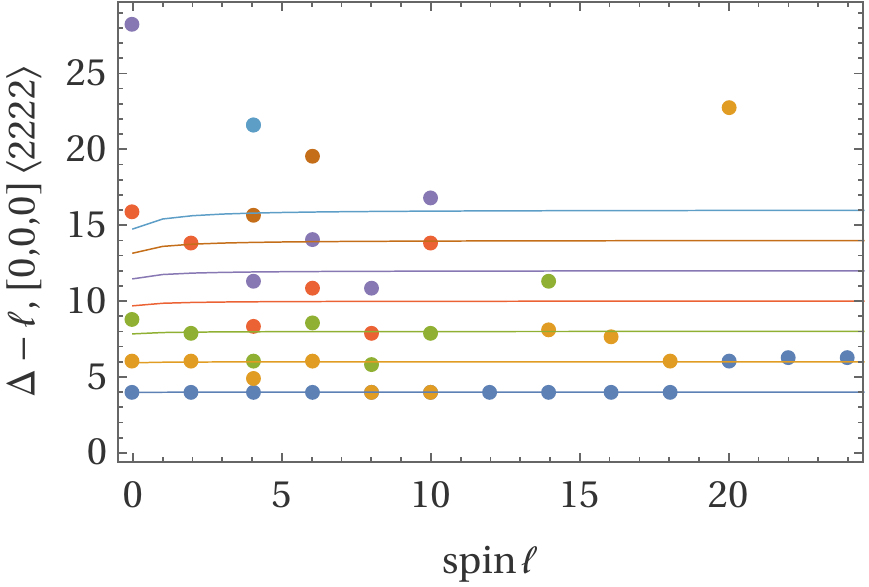}}\;
\subfloat[Singlet spectrum for $N=100$]{\includegraphics[scale=.55]{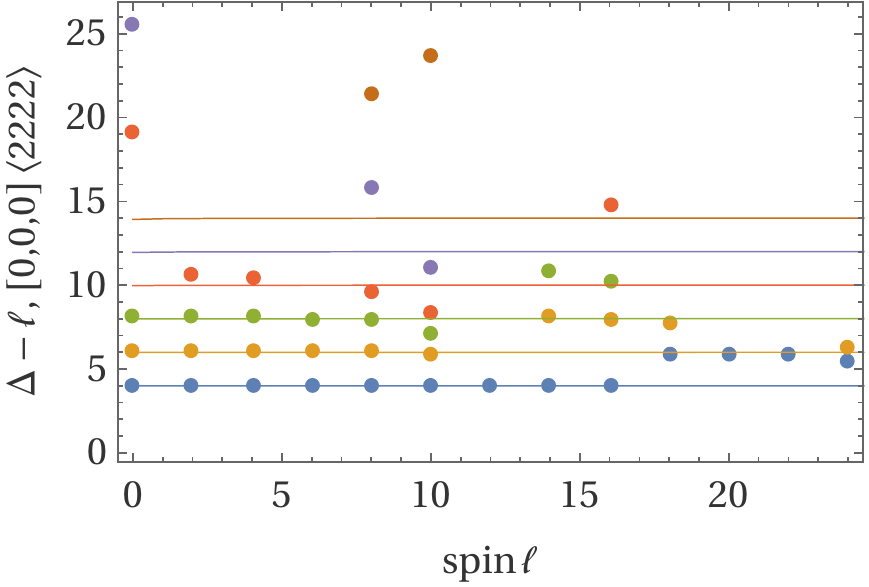}}\;
\subfloat[Singlet spectrum for $N=200$]{\includegraphics[scale=.55]{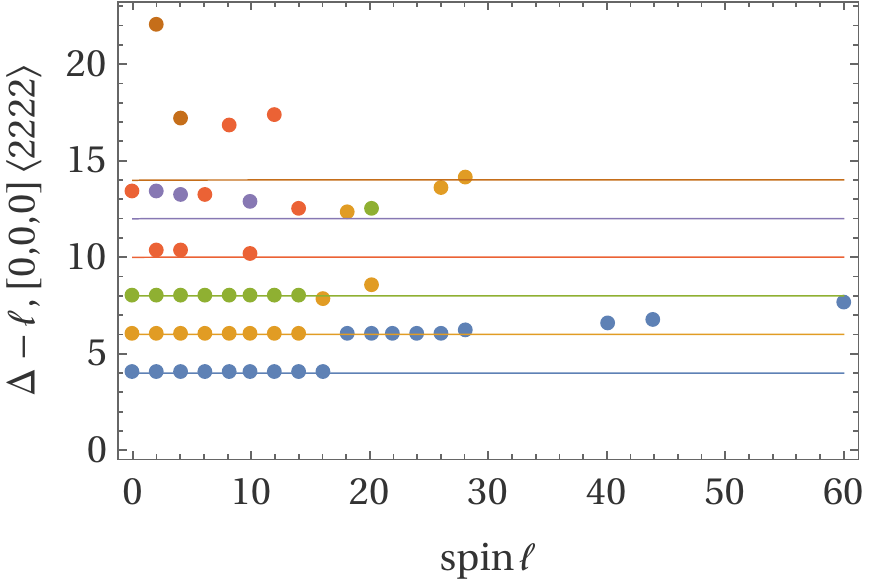}} \\
\subfloat[{$[1,0,1]$} spectrum for $N=30$]{\includegraphics[scale=.55]{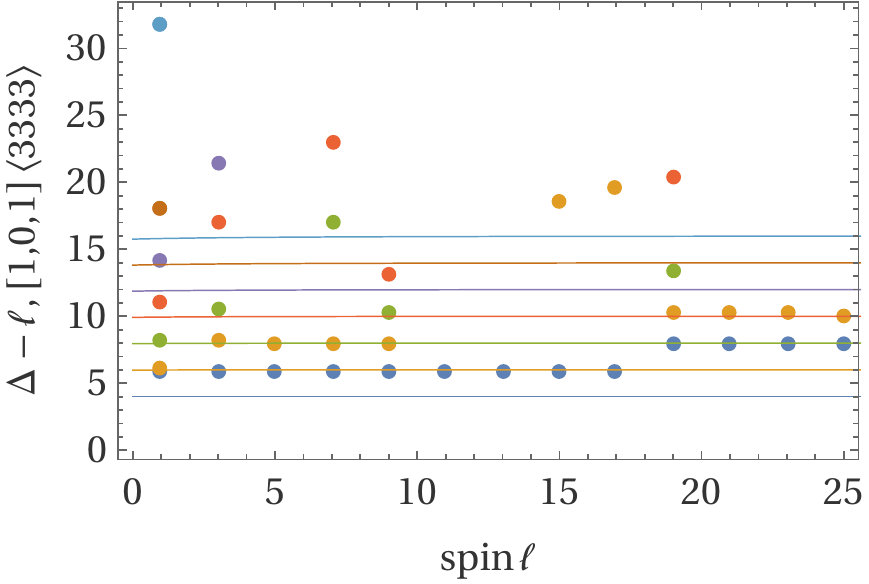}}\;
\subfloat[{$[1,0,1]$} spectrum for $N=100$]{\includegraphics[scale=.55]{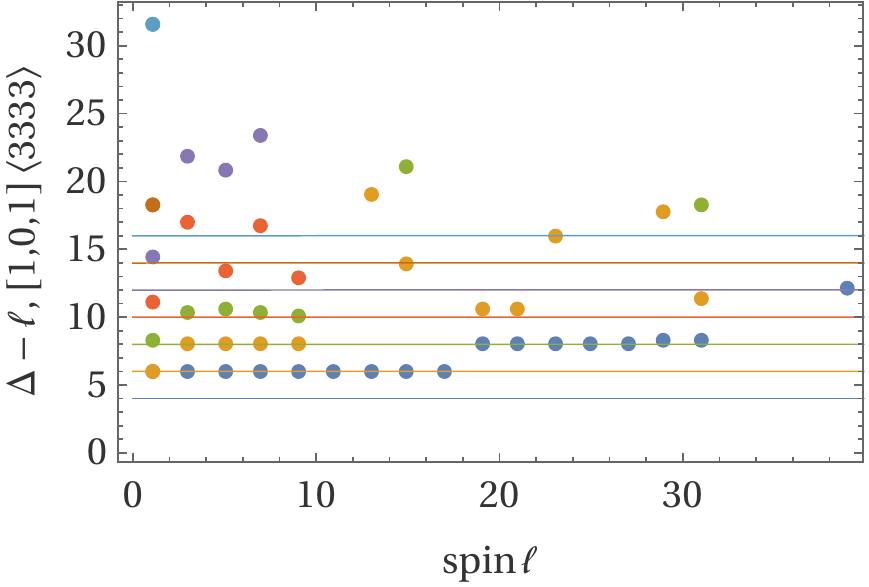}}\;
\subfloat[{$[1,0,1]$} spectrum for $N=200$]{\includegraphics[scale=.55]{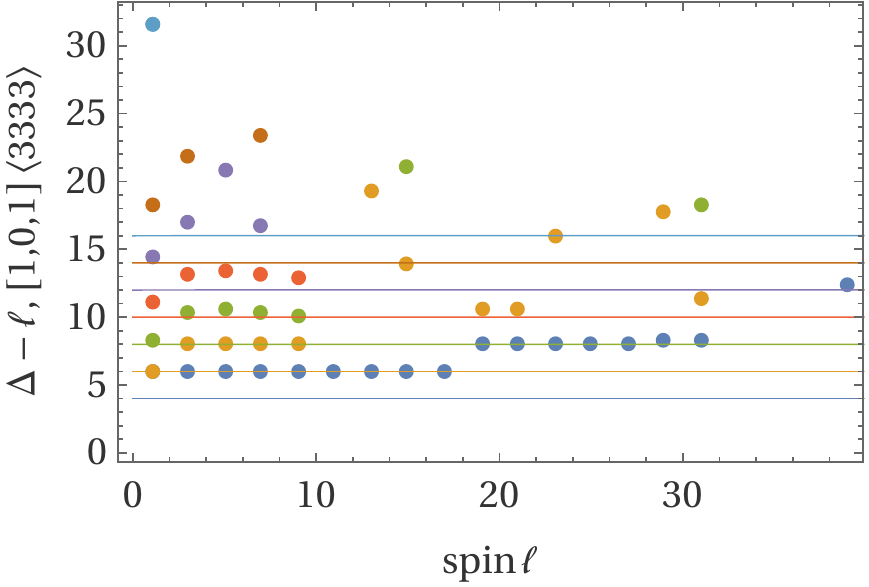}} \\
\subfloat[{$[0,2,0]$} spectrum for $N=30$]{\includegraphics[scale=.55]{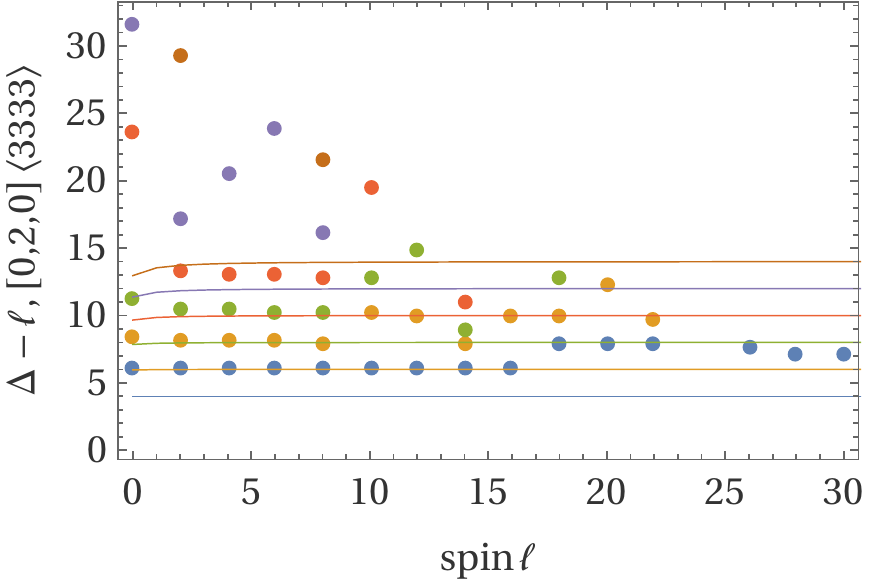}}\;
\subfloat[{$[0,2,0]$} spectrum for $N=100$]{\includegraphics[scale=.55]{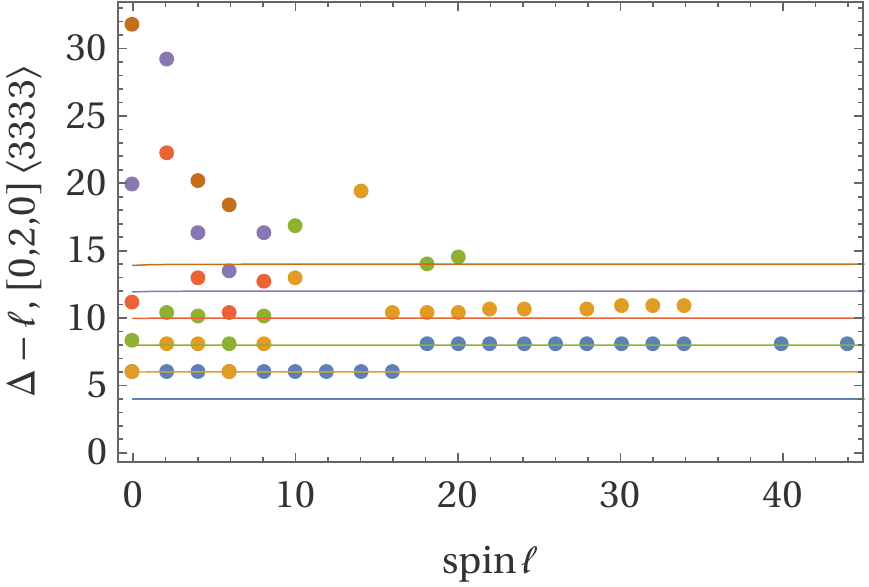}}\;
\subfloat[{$[0,2,0]$} spectrum for $N=200$]{\includegraphics[scale=.55]{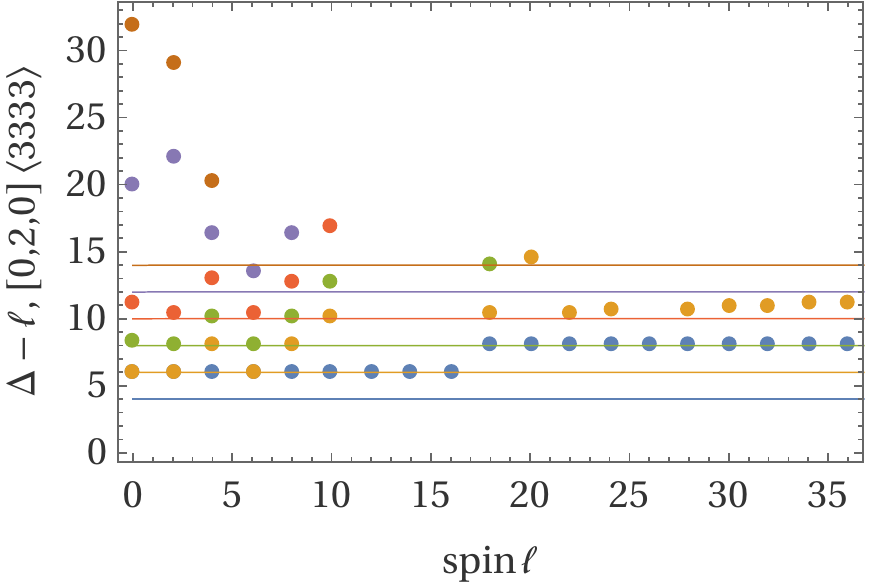}} \\
\caption{Zeros of the functionals at the boundary of the singlet twist gap exclusion plot. The first one uses only the $\langle\COp2\COp2\COp2\COp2\rangle$ which sits very close to $4$. Whereas the other two show the representations $[1,0,1]$ and $[0,2,0]$, respectively. They are obtained by bounding the singlet twist gap on $\langle\COp3\COp3\COp3\COp3\rangle$ which sits very close to $6$.
}\label{fig:spectra}
\end{figure}

Now we study the upper and lower bounds on the OPE coefficients. In order to scan over the possible values of the OPE coefficients of the singlet, we introduce $\theta$ defined by
\eqn{
\tan^{-1} \theta = \frac{\lambda_{33(\Delta,\ell)}}{\lambda_{22(\Delta,\ell)}}\,.
}[thetadef]
The first plot that we discuss is the one for the scalar singlet in \figurename~\ref{fig:OPEMinMaxSinglet}. The bound is on the norm of the vector $(\lambda_{22(\Delta,0)},\lambda_{33(\Delta,0)})$ where the angle $\theta$ in \eqref{thetadef} is varied.
\par
Next we show upper and lower bounds on the OPE coefficients of operators that transform in other representations. Those can be seen in \figurename~\ref{fig:OPEMinMax}. In this case there is no angle to scan over so the plots are two dimensional.
\par
The motivation behind the exclusion plot in \figurename~\ref{fig:OPEMinMaxSinglet} is to look for hints of a conformal manifold. Indeed fixing the value of $\Delta$ and $\lambda$ should, in principle, be enough to fix the value of the complex coupling $\tau$. Therefore one expects, in the limit of infinite numerical precision, a two-dimensional allowed region $(\Delta(\tau),\lambda(\tau))$. The plot obtained is compatible with this scenario, but we do not consider it to be a conclusive evidence.

\subsection{Spectrum extraction}

By studying the zeros of the functionals that exclude theories very close to the bound we can get information on the spectrum. In particular here we take the bound on the twist gap of the first scalar singlet using only the $\langle\COp2\COp2\COp2\COp2\rangle$ correlator (\figurename~\ref{fig:BisectionRegge2222}) and the bounds on the twist gap of the singlet using only the $\langle\COp3\COp3\COp3\COp3\rangle$ correlator (\figurename~\ref{fig:BisectionRegge3333}). In the latter we extract the spectrum in the other two representations $[1,0,1]$ and $[0,2,0]$. For both cases we consider $20 \leq N \leq 200$ and, for all $N$ in this range, the twist gap is very close to $4$ for the $\langle\COp2\COp2\COp2\COp2\rangle$ and to $6$ for $\langle\COp3\COp3\COp3\COp3\rangle$.

\begin{figure}[t]
\centering
\subfloat[Singlet spectrum for $N=20$]{\includegraphics[scale=.6]{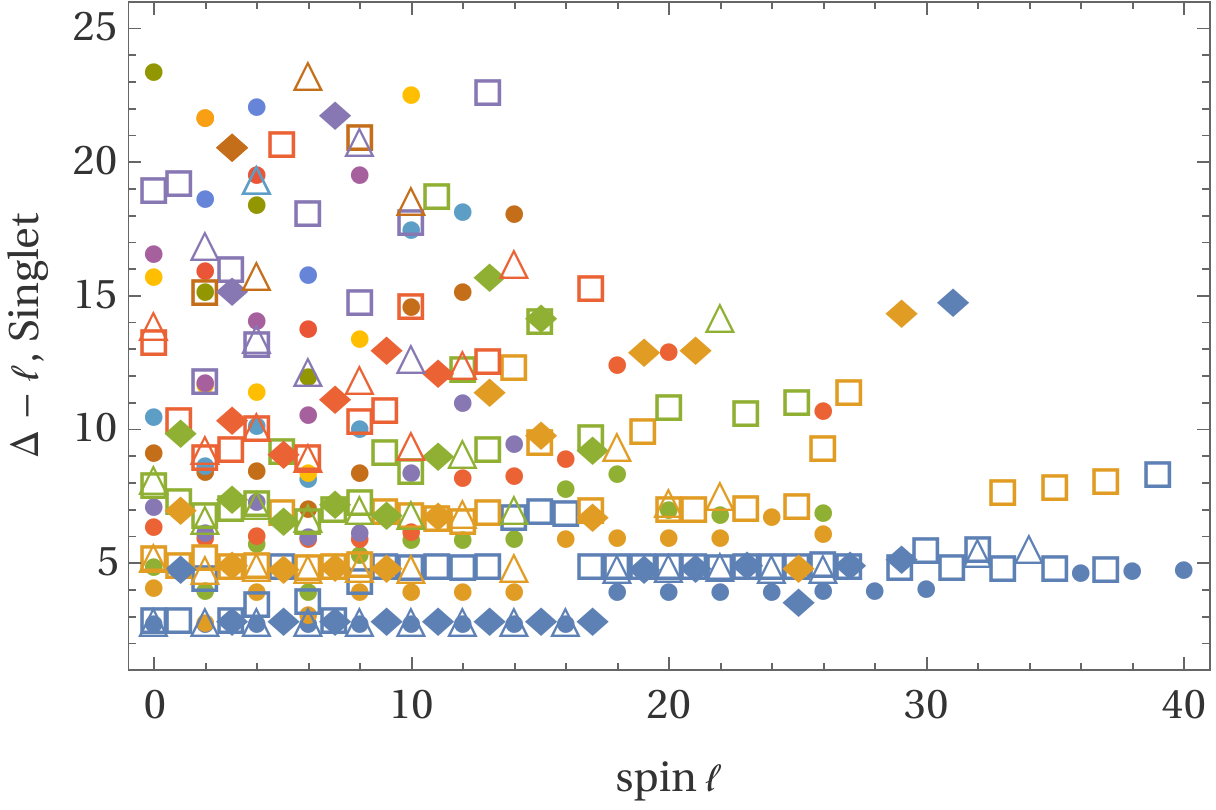}}\;
\subfloat[Singlet spectrum for $N=50$]{\includegraphics[scale=.6]{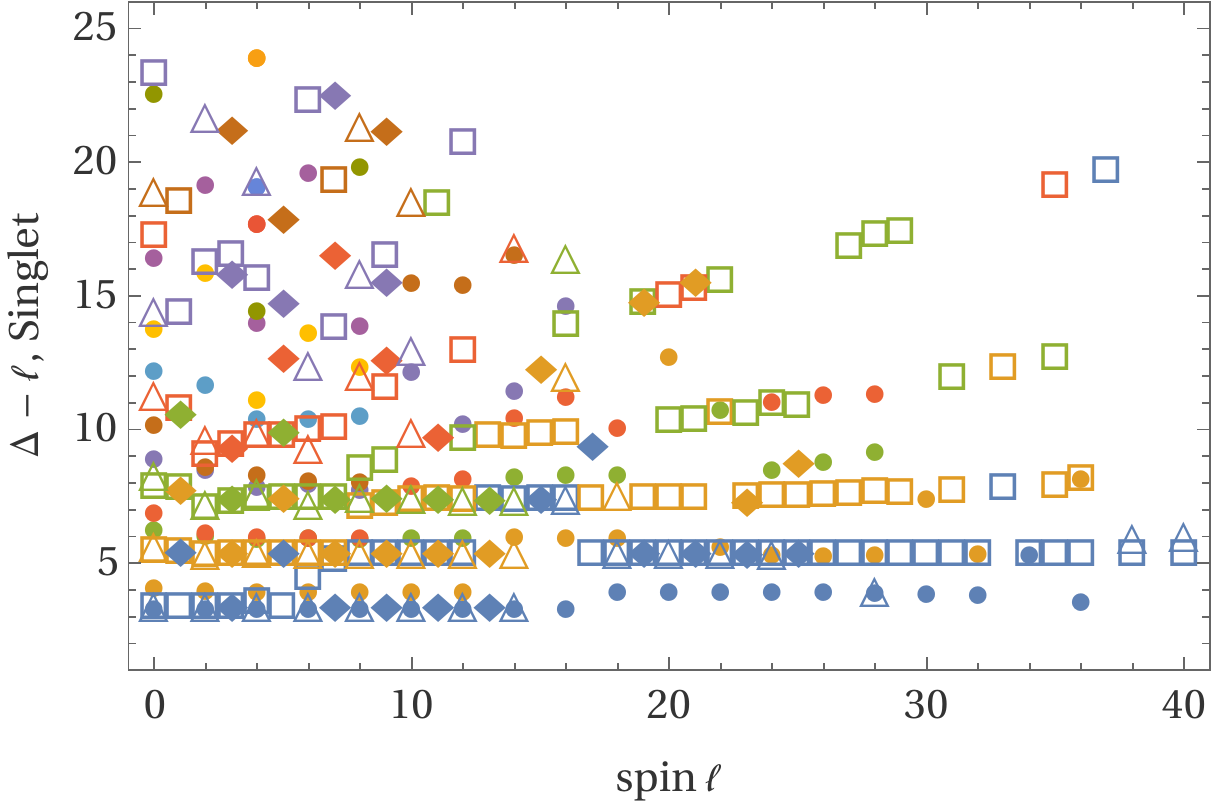}}\\
\subfloat[Singlet spectrum for $N=100$]{\includegraphics[scale=.6]{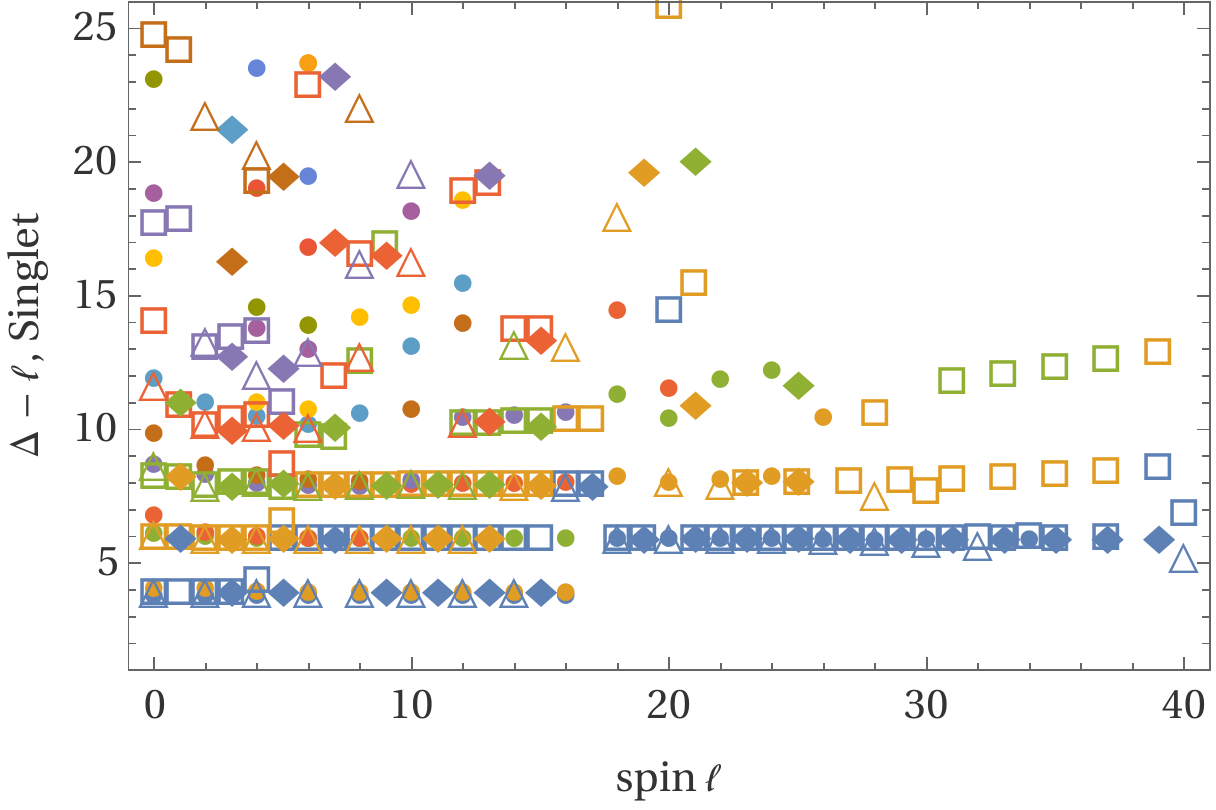}}\;
\subfloat[Singlet spectrum for $N=200$]{\includegraphics[scale=.6]{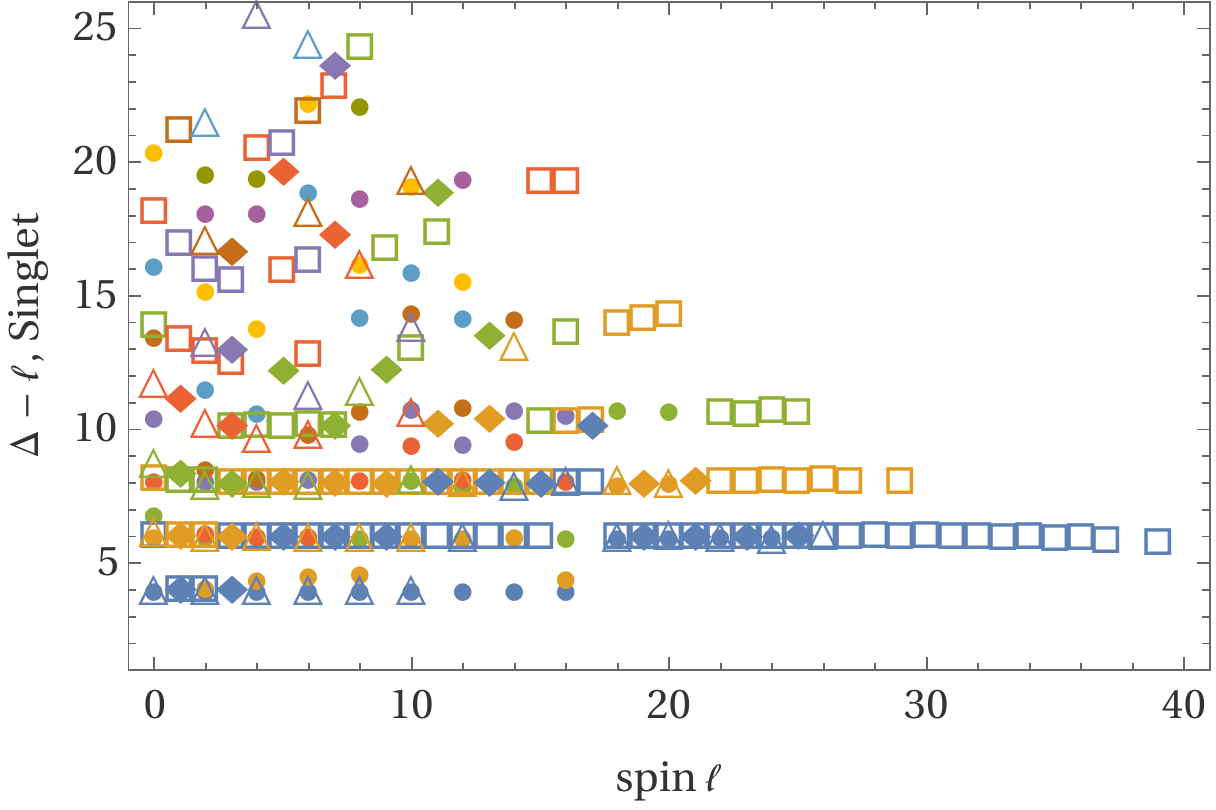}}\\
\caption[]{Zeros of the functionals at the boundary of the twist gap exclusion plot in \figurename~\ref{fig:BisectionReggeSingletMixeda}. The values of $N$ shown are, from the top left to the bottom right, $N=20,50,100,200$. We used the bound obtained with $\Lambda = 27$. The different markers are used to indicate the representation exchanged. They are defined as follows
\begin{minipage}{\linewidth}
\eqna{
\includegraphics[scale=5]{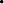} &\to \mathrm{Singlet}\,, &
\includegraphics[scale=4.2]{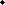} &\to {[1,0,1]}\,,\\
\includegraphics[scale=4.8]{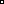} &\to {[0,1,0]}\,, & \includegraphics[scale=5.2]{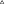} &\to {[0,2,0]}\,.}
\end{minipage}
}\label{fig:spectraMixed}
\end{figure}

In Figure~\ref{fig:spectra} we show the zeros of the functionals at the boundary for some values of $N$. The thin lines represent the expected Regge trajectories given in~\gammatl, \eqref{gamma101} and \eqref{gamma020}. The dots are colored according to how many zeros have been found in the same functional. For instance, a blue dot is the first zero while a purple dot is the fifth one.

Finally, we consider the mixed correlator. Specifically, we take some values of $N$ in the range $20\leq N \leq 200$ and extract the spectrum from the bound in \figurename~\ref{fig:BisectionReggeSingletMixeda}. The results are shown in \figurename~\ref{fig:spectraMixed}. Note that in this case the functionals are two-by-two matrices. The plotted points are the zeros of the determinant. In the plots in \figurename~\ref{fig:spectraMixed} we superimposed the different representations by using different markers (see the caption).


\acknowledgments
We would like to thank Fernando Alday, Simon Caron-Huot, Shai Chester, Leonardo Rastelli and Xinan Zhou for useful comments on the draft and Paul Heslop for discussions. We also thank Hugh Osborn for sharing the PhD thesis of Christopher Rayson before the publication and for several discussions. We are grateful to João Penedones for letting us use some of his CPU hours in the Fidis cluster at EPFL.
All numerical computations have been performed in the SCITAS high performance computing center at EPFL, which is supported by the SNSF grant PP00P2-163670. 
The work of AB and AM is supported by Knut and Alice Wallenberg Foundation under grant KAW 2016.0129 and by the VR grant 2018-04438. AM and AV are supported by the Swiss National Science Foundation under grant no. PP00P2-163670. AV is also supported by the European Research Council (ERC) Starting Grant no. 758903


\appendices

\section{Expansion of the free theory into atomic blocks}
\label{app:expansion}

\renewcommand{\arraystretch}{1.8}
\begin{table}[t]
\centering
\begin{tabular}{|r|ll|}
\hline
& Value & Parity \\
\Hhline
$b_{0,\ell}^{(2)}$& $\frac{((\ell +1)!)^2 }{(2 \ell +2)!}(\ell  (\ell +3)a_1-3 a_2 )$ & odd\\
\hline
$b_{1,\ell}^{(2)}$&$\frac{((\ell +1)!)^2}{(2 \ell +2)!}((\ell +1) (\ell +2)\lsp a_1+a_2)$ & even\\
\hline
$b^{(3)}_{0,\ell}$&$-\frac{((\ell +1)!)^2}{6 (2 \ell +2)!}\big((\ell -1) (\ell +4) (\ell  (\ell +3)b_1-8 b_2)-12 b_2+6 b_3\big)$
&odd\\
\hline
$b^{(3)}_{1,\ell}$&$-\frac{((\ell +1)!)^2 }{4 (2 \ell +2)!}\big((\ell +1) (\ell +2) ((\ell -1) (\ell +4)b_1-4 b_2)-8 b_2-4 b_3\big)$
&even\\
\hline
$b^{(3)}_{2,\ell}$&$-\frac{ ((\ell +1)!)^2 }{12 (2 \ell +2)!}(\ell +1) (\ell +2)(\ell  (\ell +3)b_1+4b_2)$
&odd\\
\hline
\multirow{2}{*}{$b^{(2,3,2,3)}_{0,\ell}$}&$\frac{(-1)^{\ell+1} (\ell+2)! (\ell+1)! }{3 (2 l+3)!}\times$
&\multirow{2}{*}{both}\\
&$\qquad\times \big(\ell (\ell+4) c_2-(-1)^\ell (\ell+2) (\ell (\ell+4)c_1-4 c_3)-2 c_3-3 c_4\big)$&\\
\hline
\multirow{2}{*}{$b^{(2,3,2,3)}_{1,\ell}$}&$\frac{(-1)^{\ell } (\ell +1)! (\ell +2)! }{6 (2 \ell +3)!}\times$
&\multirow{2}{*}{both}\\
&$\qquad\times\big(2 ( (\ell +1) (\ell +3)c_2+c_3)+(-1)^{\ell } (\ell +2) ((\ell +1) (\ell +3)c_1+2 c_3)\big)$&\\
\hline
$b^{(2,2,3,3)}_{0,\ell}$&$\frac{((\ell +1)!)^2 }{(2 \ell +2)!}( \ell(\ell +3)c_2-2c_3-c_4)$&odd\\
\hline
$b^{(2,2,3,3)}_{1,\ell}$&$\frac{((\ell +1)!)^2 }{ (2 \ell +2)!}((\ell+1)(\ell +2) c_2 +c_4)$&even\\
\hline
\end{tabular}
\caption{Coefficients $b_{n,\ell}$ for the expansion of $\hat{f}$ functions in atomic blocks. The last column refers to the parity of $\ell$. \label{tab:bnlcoeffs}}
\end{table}
\renewcommand{\arraystretch}{1}

The expansion in atomic blocks $g_\ell^{(0,0)}(x)$ has been carried in Appendix B of \cite{Dolan:2004iy}. Here we repeat the same computation for $g_\ell^{(1,1)}(x)$. For brevity and compatibility with their notation we define for this section $g_\ell^{(1,1)}(x)\equiv \tilde{g}_{\ell+2}(x)$.
\par 
Let us find the expansion coefficients for $x^n$ and ${x'}^n$, where $x' = \frac{x}{x-1}$.
\eqn{
x^{n+1}=\sum_{\ell=n}^\infty p_{n,\ell} \,\tilde{g}_{\ell+1}(x)\;,\qquad {x'}^{n+1}=\sum_{\ell=n}^\infty p'_{n,\ell}\, \tilde{g}_{\ell+1}(x)\,.
}[]
From the definition of $\tilde{g}_{\ell}(x)$ we see:
\eqn{
\tilde{g}_{\ell+1}(x) = (-)^{\ell+1}\sum_{n=\ell}^\infty \frac{(\ell+1)_{n-\ell}(\ell+2)_{n-\ell}}{(2\ell+3)_{n-\ell}} \frac{x^{n+1}}{(n-\ell)!}\,.
}[]
Define thus\footnote{The Pochhammer symbol $(a)_n$ is defined as $(a)_n = \frac{\Gamma(a+n)}{\Gamma(a)} = a(a+1)\ldots(a+n-1).$}
\eqn{
\alpha_{n,\ell}=\frac{(\ell+1)_{n-\ell}(\ell+2)_{n-\ell}}{(2\ell+3)_{n-\ell}} \frac{1}{(n-\ell)!}\,.
}[]
The problem now translates in finding a set of coefficients 
$\beta_{n,\ell}$ with the property
\eqn{
\sum_{k=\ell}^j \beta_{j,k}\alpha_{k,\ell} = \delta_{j\ell}\;,\quad \sum_{k=m}^n \alpha_{n,k}\beta_{k,m}=\delta_{nm}\,.
}[]
From this follows
\eqn{
\sum_{\ell = n}^\infty \beta_{\ell,n}\, \tilde{g}_{\ell+1}(x) = \sum_{\ell = n}^\infty\sum_{m =\ell}^\infty (-)^{\ell+1}\alpha_{m,\ell}\beta_{\ell,n}\, x^{n+1}\;,
}[]
and then the property of the coefficients $\beta_{\ell,n}$ together with the identity $\sum_{\ell=n}^\infty\sum_{m =\ell}^\infty \allowbreak= \sum_{m=n}^\infty \sum_{\ell = n}^m$ yields
\eqn{
x^{n+1}=\sum_{\ell=n}^\infty \underset{p_{n,\ell}}{\underbrace{(-)^{\ell+1} \beta_{\ell,n}}}\,\tilde{g}_{\ell+1}(x)\,. \label{pnl}
}[]
The coefficients $\beta_{\ell,n}$ can be found to be
\eqn{
\beta_{j,k}= (-)^{j-k}\frac{(k+2)_{j-k}(k+1)_{j-k}}{(2+j+k)_{j-k}}\frac{1}{(j-k)!}\,.
}[]
The case with $x'$ requires the use of the Pfaff transformation
\eqn{
\F(a,c-b;c;x) = \frac{1}{(1-x)^a}\F\left(a,b;c;\lifrac{x}{x-1}\right)\,.
}[]
This translates on $\tilde{g}_{\ell}(x)$ as
\eqna{
\tilde{g}_{\ell}(x')&= x^\ell \F(\ell,\ell;2\ell+1;x) =
(-)^{\ell+1}\sum_{n=\ell}^\infty \alpha'_{n,\ell}\, x^{n+1}\;,
}[]
where it has been defined, in analogy as before,
\eqn{
\alpha'_{n,\ell}=(-)^{\ell+1}\frac{\left[(\ell+1)_{n-\ell}\right]^2}{(2\ell+3)_{n-\ell}} \frac{1}{(n-\ell)!}\,.
}[]
Using the simple observation $x'' \equiv (x')' = x$ we can write
\eqn{
\tilde{g}_{\ell}(x)=
(-)^{\ell+1}\sum_{n=\ell}^\infty \alpha'_{n,\ell}\, {x'}^{n+1}\,.
}[]
Now by following the same steps as before, the coefficients $p'_{n,\ell}$ are given by
\eqn{
p'_{n,\ell} = (-)^{\ell+1} \beta'_{\ell,n}\;, \label{ppnl}
}[]
the inverse coefficients being
\eqn{
\beta'_{j,k}= (-)^{k-1}\frac{\left[(k+1)_{j-k}\right]^2}{(2+j+k)_{j-k}}\frac{1}{(j-k)!}\,.
}[]
Now now we can simply expand every $x$ and every $x'$ in the expression for $\hat{f}^{(2,3,2,3)}$ (which we report below) with the coefficients $p_{n,\ell}$ and $p'_{n,\ell}$ and add up the results. The list of coefficients for all cases is summarized in \tablename~\ref{tab:bnlcoeffs}. The expression for $\hat{f}$ is
\eqn{
\hat{f}^{(2,3,2,3)}(x,y) = \frac{y+1}{2}\left(c_1 x^2+c_2{x'}^2-c_3(x+x')\right)-\left(c_2{x'}^2 - (c_1+2c_3)x+(c_2+c_4)x'\right)\,.
}

\section{Short multiplets from the chiral algebra}
\label{app:chiral}
In this appendix we will derive the expression for $\hat{f}^{(3)}$ using the two dimensional chiral algebra described in \cite{Beem:2013sza}. The same computation has been done for $\hat{f}^{(2)}$ in \cite{Beem:2016wfs} with a different notation.
\par Two dimensional chiral algebras arise in theories with at least $\CN=2$ supersymmetry. It will be useful then to decompose our operators in terms of this smaller superalgebra. A representation of $\mathfrak{su}(4)$ decomposes into $\mathfrak{su}(2)_F \oplus \mathfrak{su}(2)_R \oplus \mathfrak{u}(1)_r$. The last two factors constitute the $\CN=2$ R-symmetry and the first is an additional flavor symmetry.
Now consider the cohomology of the nihilpotent supercharge $\bbQ$ given by this combination of the charges $\mathcal{Q}$ and $\mathcal{S}$:
\eqn{
\bbQ = \mathcal{Q}^1_- + \tilde{\mathcal{S}}^{2\dot{-}}\,.
}[]
A certain class of operators termed \emph{Schur operators} survives in the cohomology of $\bbQ$ and they will be the generators of the chiral algebra. The conditions for an operator to be of the Schur type can be expressed in $\CN=2$ language
\eqn{
\lifrac{1}{2}\left(\Delta - (j+\bar{\jmath})\right) - R = 0\,,\quad
r + (j-\bar{\jmath})=0\,.
}[]
where $\Delta$ is the dimension $(j,\bar{\jmath})$ the spin and $R,r$ are the charges under $\mathfrak{su}(2)_R \oplus \mathfrak{u}(1)_r$. Once we have identified the operators of this kind we can define the map $\chi$ that relates the four dimensional $\CN=2$ algebra to a $\CN = (0,4)$ two dimensional chiral algebra.
\eqna{
&\chi:  &4\mathrm{d}\; \CN=2\; &\longmapsto \;2\mathrm{d}\; \CN = (0,4)\\
&&\mathcal{O}(x)&\longmapsto \; \chi[\mathcal{O}(x)] =\mathcal{O}(z)\,.
}[]
This mapping is realized in three steps: first restrict $x^\mu$ to the plane $z = x^3 + i x^4$ and $\zb = x^3-ix^4$. Next consider a ``twisted translation'' of the operator to the point $z,\zb$. If we represent an operator in the spin $k$ representation of $\mathfrak{su}(2)_R$ as a symmetric tensor with $k$ indices $\mathcal{I}=1,2$, this is explicitly realized as
\eqn{
\CO(z,\zb) \equiv u_{\mathcal{I}_1}(\zb)\cdots u_{\mathcal{I}_k}(\zb)\,\CO^{\mathcal{I}_1\ldots\mathcal{I}_k}(z,\zb)\,,
}[]
where $u_\mathcal{I}(\zb) = (1,\zb)$. In the language of this paper the last two steps are equivalent to sending $\xb \to 1/\alphab$ as in \eqref{eq:WI},\footnote{The cross ratios $z,\,\zb$ in \eqref{eq:vars} should not be confused with the complex coordinates $z,\,\zb$ of this \appendixname.} but we will make this statement more precise in a moment. Finally consider the $\bbQ$ cohomology class of $\CO(z,\zb)$.
\eqn{
[\CO(z,\zb)]_{\bbQ} \equiv \CO(z)\,.
}[]
The result $\CO(z)$ is what we define to be $\chi[\CO(x)]$.
\par We proceed then by decomposing the $\COp3$ operator into $\CN=2$ language. We use the notation $(j_R,j_F)_r$ where $j_{R(F)}$ is the spin under the $\mathfrak{su}(2)_{R(F)}$ factor and $r$ the charge under $\mathfrak{u}(1)$.
\eqna{
\,[0,3,0]_{\mathfrak{su}(4)} \;\to\; &
\left(0,0\right)_{\pm 3}
\oplus
\left(0,0\right)_{\pm 1}
\oplus 
\left(\lifrac{1}{2},\lifrac{1}{2}\right)_{\pm 2}
\oplus
\left(\lifrac{1}{2},\lifrac{1}{2}\right)_{0}
\oplus 
\left(1,1\right)_{\pm 1}
\oplus
\left(\lifrac{3}{2},\lifrac{3}{2}\right)_{0}
}[]
The only one with Schur quantum numbers is $\left(\lifrac{3}{2},\lifrac{3}{2}\right)_{0}$, which in the two dimensional theory is an $h=\frac{3}{2}$ operator in a flavor four-plet. In $\CN=2$ language, $\CN=4$ Super Yang-Mills decomposes into a vector multiplet and two hypermultiplets with $\SU(2)_F$ flavor symmetry.
\eqn{
V = (\Phi, A_\mu, \lambda^{1,2}_\alpha, \ldots)\,,\quad H_i = (Q_i, \psi^i_\alpha,\ldots)\,,\quad i=1,2\,.
}[]
Accordingly, two of the six scalars contribute to the flavor singlet $\Phi$ and the other four to the flavor doublet $Q_{i}$. Recall that all these fields carry an implicit $\SU(N)$ adjoint index (the gauge group). In \cite{Beem:2013sza} the following mapping was shown
\eqn{
q_{i}^A(z) = \chi[Q_{i}^A]\,,\quad A = 1,\ldots, N^2-1\,.
}[]
where $A$ is the gauge group index. The field $q_{i}^A(z)$ has the following OPE
\eqn{
q_{i}^A(z) q_{j}^B(0) \sim \frac{\varepsilon_{ij}\delta^{AB}}{z}\,.
\label{eq:qOPE}
}[] 
Which implies that the holomorphic dimension of $q_i$ is $h=\frac{1}{2}$. Thus we can identify the Schur operator in the $\left(\lifrac{3}{2},\lifrac{3}{2}\right)_{0}$ as $B_{ijk}$ defined as
\eqn{
B_{ijk}(z) = \Tr :\!q_iq_jq_k(z)\!:\, = \chi[\Tr Q_iQ_jQ_k]\,,
}[]
where the trace is over $\SU(N)$. Using this definition we now want to compute the four-point function of four $B$'s
\eqn{
\langle B_{i_1j_1k_1}(z_1) B_{i_2j_2k_2}(z_2) B_{i_3j_3k_3}(z_3) B_{i_4j_4k_4}(z_4)\rangle\,.
}[]
The result is fixed by the singular part of the OPE, therefore the computation is done with the prescription of performing all Wick contractions between the $q$'s at different points. Moreover it will be convenient to contract all $\mathfrak{su}(2)_F$ indices $i,j,\ldots$ with commuting Weyl spinors $v^i$, e.g. 
\eqn{
B(z_1,v_1) \equiv v_1^{i_1}v_1^{j_1}v_1^{k_1}B_{i_1j_1k_1}(z_1)\,.
}[]
Before embarking on this computation however let us define the precise mapping between four dimensional and two dimensional correlators. We want to compare the chiral algebra result with $\mathcal{F}^{(3)}(x,\alpha,\xb,\alphab)$ restricted to the plane $x^1 = x^2 = 0$ and on the four dimensional subspace for the scalars $\phi^i$ corresponding to $t^5 = t^6 = 0$. If we introduce the complex coordinates mentioned earlier for the two space time dimensions $z = x^3 + i x^4$ and $\zb = x^3-ix^4$ we can write the cross ratios as
\eqn{
u = x \xb =\frac{|z_1-z_2|^2|z_3-z_4|^2}{|z_1-z_3|^2|z_2-z_4|^2}\,,\quad
v = (1-x)(1-\xb) = \frac{|z_1-z_4|^2|z_2-z_3|^2}{|z_1-z_3|^2|z_2-z_4|^2}\,.
}[]
Thus, clearly, the first entry in our dictionary is (calling $z_{ij} = z_i-z_j$)
\eqn{
x \to \frac{z_{12}\,z_{34}}{z_{13}\,z_{24}} \equiv \zeta\,,\quad \left(1-\zeta = \frac{z_{14} z_{23}}{z_{13}z_{24}}\right)\,.
\label{eq:zetadef}
}[]
Where we introduced the chiral cross ratio $\zeta$. Next we consider the $\mathrm{SO}(6)$ polarizations. Any four-vector can be split in bispinor notation
\eqn{
t^{\mu} = v^i \sigma^\mu_{i\mathcal{I}} \tilde{v}^{\mathcal{I}}\,,
}[]
with $\mu = 1,\ldots, 4$ an $\mathrm{SO}(4)$ index. Consistently with the previous notation upper case (lower case) indices correspond to the $\mathfrak{su}(2)_R$ ($\mathfrak{su}(2)_F$) subalgebra. Moreover $\sigma^\mu_{i \mathcal{I}} = ( i \unit, \vec{\sigma})_{i\mathcal{I}}$ are standard four dimensional (Euclidean) Pauli matrices. Following the convention of \cite{WessnBagger} (with minor changes due to the signature and $\alpha \to i,\,\dot{\alpha}\to \mathcal{I}$) one has
\eqn{
t_a \cdot t_b = 2\, v_a^i v_{bi}\; \tilde{v}_{a\mathcal{I}} \tilde{v}_b^{\mathcal{I}} \equiv 2\,v_{ab}\, \tilde{v}_{ab}\,.
}[]
where $a,b$ label the four-points in the correlator. This means that we can write the cross ratios as
\eqn{
\sigma = \alpha \alphab=\frac{v_{13} v_{24}}{v_{12}v_{34}}\frac{\tilde{v}_{13} \tilde{v}_{24}}{\tilde{v}_{12}\tilde{v}_{34}}\,,\qquad
\tau = (\alpha-1)(\alphab-1) = \frac{v_{14} v_{23}}{v_{12}v_{34}}\frac{\tilde{v}_{14} \tilde{v}_{23}}{\tilde{v}_{12}\tilde{v}_{34}}\,.
}[]
Two dimensional spinors satisfy the Schouten identity $v_{14}v_{23} = v_{13}v_{24} - v_{12}v_{34}$. This means that the second entry in our dictionary is
\eqn{
\alpha \to \frac{v_{13} v_{24}}{v_{12}v_{34}} \equiv \nu \,,\quad \left(\nu -1 = \frac{v_{14} v_{23}}{v_{12}v_{34}}\right)\,.
}[]
Where we introduced the $\mathfrak{su}(2)_F$ cross ratio $\nu$. According to the statement of the Ward identity \eqref{eq:WI}, in the limit $\xb = 1/\alphab$ the $\xb,\,\alphab$ dependence drops out and we end up with a function of $x$ and $\alpha$ only. Therefore it is consistent to require the matching
\eqn{
\mathcal{F}^{(3)}\left(x,\alpha,\lifrac{1}{\alphab},\alphab\right)\Big{|}_{\substack{
x \to \zeta \\ 
\alpha \to \nu}} = \mathfrak{f}^{(3)}(\zeta,\nu)\,,
}[]
where we defined
\eqn{
\langle B(z_1,v_1) B(z_2,v_2) B(z_3,v_3) B(z_4,v_4)\rangle = \left(\frac{v_{12}\,v_{34}}{z_{12}\,z_{34}}\right)^3 \mathfrak{f}^{(3)}(\zeta,\nu)\,.
}[]
\par As a warm up let us show this computation for $p=2$. The relevant four-point function is
\eqn{
\mathfrak{f}^{(2)}(\zeta,\nu)\equiv\left(\frac{z_{12}\,z_{34}}{v_{12}\,v_{34}}\right)^2\;\langle J(z_1,v_1) J(z_2,v_2) J(z_3,v_3) J(z_4,v_4)\rangle\,,
}[]
where
\eqn{
J(z,v) = \frac{1}{2 \sqrt{\kappa}} v^i v^j :\!q_i^A q_j^A(z)\!:\,,\quad \kappa =\frac{N^2-1}{2}\,.
}[]
The factor $1/\sqrt{\kappa}$ makes the OPE of $J$ canonically normalized, while the $1/2$ comes from the convention $\Tr t^At^B = \delta_{AB}/2$. The four-point function is readily computed
\eqn{
\mathfrak{f}^{(2)}(\zeta,\nu) = 1 + \nu^2\,\zeta^2 + (1-\nu)^2\,{\zeta'}^2 + \frac{2\nu (2-\nu)}{\kappa}\,\zeta + 
 \frac{2(1-\nu^2)}{\kappa}\, \zeta'\,,
}[]
where we defined $\zeta' = \zeta/(\zeta-1)$ with $\zeta$ as in \eqref{eq:zetadef}. According to \eqref{eq:F2freetheory} $\mathcal{F}^{(2)}$ is given by
\eqn{
\mathcal{F}^{(2)}(u,v,\sigma,\tau) = a_1\left(1 + (\sigma u)^2 +  \lifrac{(\tau u)^2}{v^2}\right) + a_2\left((\sigma u) + \lifrac{\tau u}{v} + (\sigma u) \lifrac{\tau u}{v}\right)\,.
}[]
A direct straightforward computation shows that $\mathfrak{f}^{(2)}(\zeta,\nu) = \mathcal{F}^{(2)}(\zeta,\nu,1/\alphab,\alphab)$ requires
\eqn{
a_1 = 1\,,\qquad a_2 = \frac{2}{\kappa}\,.
}[]
as in \eqref{eq:color2}. 
\par The case for $p=3$ follows the same logic, though the combinatorics is a bit heavier and one needs to use some trace identities for $\SU(N)$ matrices, which can be found in Appendix \ref{app:sun}. We performed this computation by implementing the OPE \eqref{eq:qOPE} and the $\SU(N)$ trace identities in \textit{Mathematica}. The result, modulo an overall factor which can be fixed by normalizing $B$, is 
\eqn{
\mathfrak{f}^{(3)}(\zeta,\nu) = 1+f_1\, \zeta^3 + f_1' \,{\zeta'}^3 +f_2 \,\zeta^2 + f_2'\, {\zeta'}^2+f_3 \,\zeta + f_3'\, \zeta'\,.
}[]
Where, as before, $\zeta'=\zeta/(\zeta-1)$ and the $f_k$ are defined as 
\eqna{
f_1 &= \nu^3\,,&\quad\; f_2 &= \frac{9 \nu ^2(2-\nu)}{N^2-1}\,,&\quad\; f_3 &= \frac{9 \nu  \left(N^2-4+(3N^2-28)(1-\nu)\right)}{\left(N^2-4\right) \left(N^2-1\right)}\,,\\
f_i'&= f_i\lsp\big{|}_{\nu \to 1-\nu}\,.
}[]
This has to be compared, as explained before, with
\eqna{
\mathcal{F}^{(3)}(u,v,\sigma,\tau) &= b_1 \left(1+(\sigma u)^3+\lifrac{(\tau u)^3}{v^3}\right)+b_3\,\sigma u \lifrac{\tau u}{v} \\&+b_2 \left((\sigma u)^2 \lifrac{\tau u}{v}+(\sigma u)^2+\sigma u
   \lifrac{(\tau u)^2}{v^2}+\sigma u+\lifrac{(\tau u)^2}{v^2}+\lifrac{\tau u}{v}\right)\,.
}[]
By matching $\mathfrak{f}^{(3)}(\zeta,\nu) = \mathcal{F}^{(3)}(\zeta,\nu,1/\alphab,\alphab)$ we can fix the coefficients $c_0,\,c_1$ and $c_2$:
\eqna{
b_1 &= 1\,,\qquad b_2 = \frac{9}{N^2-1}\,,\qquad b_3 = \frac{18\lsp(N^2-12)}{(N^2-4)(N^2-1)}\,.
}[]
And this result agrees with \eqref{eq:color3}.

\section{\texorpdfstring{$\boldsymbol{\SU(N)}$}{SU(N)} trace identities}
\label{app:sun}
We define the generators of $\SU(N)$ as $N^2-1$ hermitian traceless matrices labeled by $A,B,\ldots$ and normalized as
\eqn{
\Tr (t^A t^B) = \lifrac{1}{2}\delta_{AB}\,.
}[]
The commutation and anticommutation relations are
\eqn{
[t^A,t^B] = i f_{ABC} t^C\,,\quad
\{t^A,t^B\} = \frac{1}{N}\delta_{AB} \unit + d_{ABC}t^C\,.
}[]
Where $d_{ABC}$ is totally symmetric and $f_{ABC}$ totally antisymmetric. From this follows
\eqn{
\Tr (t^A t^B t^C) = \lifrac{1}{4}(d_{ABC} + i f_{ABC})\,.
}[]
Using the identity above the computation in Appendix \ref{app:chiral} for $p=3$ yields various contractions of the tensors $f_{ABC}$ and $d_{ABC}$. Some of the identities needed are the following (they can be found in \cite[\appendixname~B]{Cutler:1977qm}):
\eqna{
\delta_{AA} &= N^2-1\,,\\
f_{AAB} = d_{AAB} &=0\,,\\
d_{ABC}f_{ABD} &= 0 \,,\\
d_{ABC}d_{ABD} &= \frac{N^2-4}{N}\,\delta_{CD}\,,\\
f_{ABC}f_{ABD} &= N\,\delta_{CD}\,.
\label{eq:traces2}
}[]
For the contraction of four different tensors additional identities are needed. Those can be found in \cite[\appendixname~A]{Fadin:2005zj}. First let us define the matrices
\eqn{
(F^A)_{BC} = -i f_{ABC}\,,\quad
(D^A)_{BC} = d_{ABC}\,.
}[]
The possible traces of four $F$ or $D$ matrices are
\small
\eqna{
\Tr( F^A F^B F^C F^D) &= \delta_{AB}\delta_{CD} + \delta_{AD}\delta_{BC} + \lifrac{N}{4}\left(d_{ABE}d_{CDE}-d_{ACE}d_{BDE}+d_{ADE}d_{BCE}\right)\,,\\
\Tr( F^A F^B F^C D^D) &= \lifrac{iN}{4}\left(d_{ABE}f_{CDE} + f_{ABE}d_{CDE}\right)\,,\\
\Tr( F^A F^B D^C D^D) &= \left(\lifrac{N^2-8}{4N}\right)\left(d_{ABE}d_{CDE}-d_{ACE}d_{BDE}\right)+\\&\quad\; +\left(\lifrac{N^2-4}{N^2}\right)\left(\delta_{AB}\delta_{CD}-\delta_{AC}\delta_{BD}\right) +\lifrac{N}{4}d_{ADE}d_{BCE}
\,,\\
\Tr( F^A D^B F^C D^D) &= \lifrac{N}{4}\left(d_{ABE}d_{CDE}-d_{ACE}d_{BDE}+d_{ADE}d_{BCE}\right)\,,\\
\Tr( F^A D^B D^C D^D) &= \lifrac{2i}{N} f_{ADE}d_{BCE} + i \left(\lifrac{N^2-8}{4N}\right)f_{ABE}d_{CDE} + \lifrac{iN}{4}d_{ABE}f_{CDE}\,,\\
\Tr( D^A D^B D^C D^D) &= \left(\lifrac{N^2-16}{4N}\right)\left(d_{ABE}d_{CDE}+d_{ADE}d_{BCE}\right)+\\&\quad\;+\left(\lifrac{N^2-4}{N^2}\right)\left(\delta_{AB}\delta_{CD}+\delta_{AD}\delta_{BC}\right)-\lifrac{N}{4}d_{ACE}d_{BDE}\,.
\label{eq:traces4}
}[]
\normalsize
The contractions appearing in the correlator $\mathfrak{f}^{(3)}(\zeta,\nu)$ can be then obtained by contracting the indices $A,B,C,D$ in \eqref{eq:traces4} and, if needed, using again \eqref{eq:traces2}. E.g.
\eqn{
f_{AEF}f_{BFG}d_{AGH}d_{BHE} = - \Tr( F^A F^B D^A D^B)\,.
}[]

\section{Numerical implementation}\label{app:numerical}

The Taylor expansion of the superconformal blocks is obtained from the one of usual scalar blocks by shifting the value of $\Delta$ and inserting the factor of $u^{-2}$.\footnote{Recall that the usual scalar blocks are $G_{\Delta}^{(\ell)}(u,v;\Delta_{12},\Delta_{34})$ as in \eqref{confblockdef}, while the superconformal block is given by $u^{-2}G_{\Delta+4}^{(\ell)}(u,v;\Delta_{12},\Delta_{34})$.} We used the code in 
\href{https://gitlab.com/bootstrapcollaboration/scalar_blocks}{\texttt{gitlab.com/\hspace{0pt}bootstrap\hspace{0pt}collaboration/\hspace{0pt}scalar\_blocks}}. The code uses a recursion relation to generate a better and better approximation of the true function. If we solve the recursion relations up to order $N$ we obtain an expression which schematically reads
\eqn{
\partial^{n}_z\partial^m_\zb F_{\Delta,\ell}(z,\zb)\big|_{z=\zb=\frac12} \simeq r_0^\Delta\frac{P_N^{n,m}(\Delta)}{\prod_{k=0}^N (\Delta - \Delta^*_k)}\,,
}[eq:polesandres]
where $r_0 = 3-2\sqrt{2}$ is a constant and $P$ is a polynomial of order $N$. The set of poles $\Delta_k^*$ is fixed by representation theory and will be given later. In order to optimize the performance of \texttt{sdpb} while keeping a good precision in the approximation, we keep only the poles for $k \leq \kappa < N$. Then the residues in $P_N^{m,n}$ are shifted by means of a Padé-like approximation in order to compensate the effect of the missing poles. The denominator in \eqref{eq:polesandres} presents a peculiarity in even spacetime dimensions because there are double poles. This makes the recursion relations more involved~\cite{Kravchuk:unpublished}. Here are the values of $\Delta_k^*$ for the single poles $(\Delta-\Delta^*_k)$
\eqna{
\Delta^*_k \in \;&\{1-\ell-k\}_{k=\max\left(1,\left\lfloor\frac{\kappa-2\ell-2}{2}\right\rfloor+1\right)}^\kappa
\\&\cup \{3+\ell-k\}_{k=1}^{\min(\kappa,\ell)}\\&\cup \{3+\ell-k\}_{k=\ell+2}^{\min(\kappa,2\ell+2)}\,,
}[]
and for the double poles $(\Delta-\Delta^*_k)^2$
\eqn{
\Delta^*_k \in \{1-\ell-k\}_{k=1}^{\left\lfloor\frac{\kappa-2\ell-2}{2}\right\rfloor}\,.
}[]
We choose the size of the numerical problem by increasing the number of components of the vectors of Taylor coefficients, namely the range of values for $m$ and $n$ in $\partial^n_z\partial^m_\zb F$. All other parameters are adjusted accordingly. We keep the derivatives for which $n+m \leq \Lambda$ and vary $\Lambda$ as a parameter. Often in the literature an alternative definition is used: $\Lambda = 2\lsp\mathtt{nmax} -1$. Due to the symmetry $z \leftrightarrow \zb$ and to the symmetry or antsymmetry properties under $z,\zb \to 1-z,1-\zb$ the possible values that one needs to compute are those for $n > m$ and with $n+m$ even or odd according to whether $F(v,u) = \pm F(u,v)$.
\par
Now we discuss the parameters used for the plots presented in this paper. The parameter \texttt{Lmax} is the maximal value of the spin for any exchanged operator in any OPE channel, whereas $N$ and $\kappa$ have been defined previously. In Table~\ref{tab:param} we describe their values. Instead in Table~\ref{tab:sdpb} we describe the parameters for the \texttt{sdpb} jobs.
\par
All computations have been done using \texttt{sdpb 1} since the project started prior the release of \texttt{sdpb 2}~\cite{Landry:2019qug}. However, we double checked a few random points with the newer version and the results agree.

\begin{table}[H]
\centering
\begin{tabular}{|r|llll|}
\hline
$\Lambda$ & 11 & 19 & 23 & 27 \\
\Hhline
\texttt{Lmax} & 50 & 60 & 60 & 60 \\
$N$ & 50 & 60 & 60 & 60 \\
$\kappa$ & 10 & 14 & 14 & 18\\
\hline
\end{tabular}
\caption{Values of the various parameters for the bootstrap problem.}\label{tab:param}
\end{table}

\begin{table}[H]
\centering
\begin{tabular}{|l|c|c|}
\hline
Parameter & feasibility & OPE \\
\Hhline
\texttt{maxIterations} & $500$ & $500$ \\
\texttt{maxRuntime} &  $86400$&  $86400$ \\
\texttt{checkpointInterval} &  $3600$&  $3600$ \\
\texttt{noFinalCheckpoint} & \texttt{True} &  \texttt{False} \\
\texttt{findDualFeasible} & \texttt{True} &  \texttt{False} \\
\texttt{findPrimalFeasible} & \texttt{True} &  \texttt{False} \\
\texttt{detectDualFeasibleJump} & \texttt{True} &  \texttt{False} \\
\texttt{precision} &  $850$&  $850$ \\
\texttt{maxThreads} &  $28$&  $28$ \\
\texttt{dualityGapThreshold} &  $10^{-20}$&  $10^{-20}$ \\
\texttt{primalErrorThreshold} &  $10^{-60}$&  $10^{-60}$ \\
\texttt{dualErrorThreshold} &  $10^{-20}$&  $10^{-20}$ \\
\texttt{initialMatrixScalePrimal} &  $10^{20}$&  $10^{20}$ \\
\texttt{initialMatrixScaleDual} &  $10^{20}$&  $10^{20}$ \\
\texttt{feasibleCenteringParameter} &  $0.1$&  $0.1$ \\
\texttt{infeasibleCenteringParameter} &  $0.3$&  $0.3$ \\
\texttt{stepLengthReduction} &  $0.7$&  $0.7$ \\
\texttt{choleskyStabilizeThreshold} &  $10^{-70}$&  $10^{-70}$ \\
\texttt{maxComplementarity} &  $10^{200}$&  $10^{200}$ \\
\hline
\end{tabular}
\caption{Parameters used in \texttt{sdpb} for the feasibility problem used in the binary search or grid search (feasibility) and for the OPE coefficient maximization or minimization (OPE).}\label{tab:sdpb}
\end{table}

\bibliography{mixed}

\end{document}